\documentclass[a4paper,12pt]{article}
\pdfoutput=1
\usepackage{jcappub}
\usepackage[utf8]{inputenc}
\usepackage[table,dvipsnames]{xcolor}
\usepackage{multirow}
\usepackage{graphicx}
\usepackage{mathtools}
\usepackage{amsmath,amssymb,amsfonts,amsthm,latexsym,stmaryrd,physics,mathrsfs}
\allowdisplaybreaks[4]
\usepackage{marginnote}
\usepackage{color}
\usepackage{bm}
\usepackage{longtable}
\usepackage{tikz}
\usepackage{float}
\usepackage{nicefrac}
\usepackage{microtype} 
\usepackage{booktabs}
\usepackage[ruled,linesnumbered]{algorithm2e}

\usepackage{verbatim}
\usepackage{setspace}
\usepackage[T1]{fontenc}
\usepackage[utf8]{inputenc}
\usepackage{stfloats}
\usepackage{diagbox}
\usepackage{dsfont}
\usepackage[colorinlistoftodos,prependcaption,textsize=tiny]{todonotes}
\usepackage{epstopdf}
\usepackage{latexsym}
\usepackage{makecell}
\usepackage{xparse}
\usepackage[center]{subfigure}

\usetikzlibrary{positioning, shapes.geometric}

\usepackage{xargs}                      %
\usepackage[colorinlistoftodos,prependcaption,textsize=tiny]{todonotes}
	
\usepackage{ulem} %

 \renewcommand\arraystretch{2}
 \newcommand{\bq}{\begin{equation}}
 \newcommand{\eq}{\end{equation}}
 \newcommand{\bqn}{\begin{eqnarray}}
 \newcommand{\eqn}{\end{eqnarray}}

\NewDocumentCommand{\evalat}{sO{\big}mm}{%
  \IfBooleanTF{#1}
   {\mleft. #3 \mright|_{#4}}
   {#3#2|_{#4}}%
}

\newtheorem{lemma}{Lemma}

\newtheorem{corollary}{Corollary}
\def\be{\begin{eqnarray}}
\def\ee{\end{eqnarray}}

\newcommand{\lt}{\left}
\newcommand{\rt}{\right}

\newcommand{\rmd}{\mathrm d}

\newcommand{\ltbf}{\Xi}

\newcommand{\RN}[1]{%
\textup{\uppercase\expandafter{\romannumeral#1}}%
}

\newtheorem*{remark}{Remark}

\title{Regular black holes and their relationship to polymerized models and mimetic gravity}

\author[1]{Kristina Giesel}
\emailAdd{kristina.giesel@gravity.fau.de} 
\affiliation[1]{Department Physik, Institut f\"ur Quantengravitation, Theoretische Physik III, Friedrich-Alexander Universit\"at Erlangen-N\"urnberg, Staudtstr. 7/B2, 91058 Erlangen, Germany}

\author[1]{Hongguang Liu} 
\emailAdd{hongguang.liu@gravity.fau.de}

\author[2]{Parampreet Singh}
\emailAdd{psingh@lsu.edu}
\affiliation[2]{Department of Physics and Astronomy, Louisiana State University, Baton Rouge, LA 70803, USA}

\author[1]{Stefan Andreas Weigl} 
\emailAdd{stefan.weigl@gravity.fau.de}

\abstract{ 
We present further applications of the formalism introduced by the authors in \cite{Giesel:2023tsj}, which allows embedding of a broad class of generalized LTB models into effective spherically symmetric spacetimes. We focus on regular black hole models, where a broad class of models can be considered, including for example LQG-inspired models as well as the model with a regular center, e.g. of Bardeen and Hayward. For a certain class of regular black hole models, we can formulate a Birkhoff-like theorem in LTB coordinates. We further show that depending on the properties of the polymerization functions characterizing such regular black hole models in this formalism, the uniqueness of the effective spherically symmetric vacuum solutions might not be given in general in Schwarzschild-like coordinates. Furthermore, we introduce a reconstruction algorithm that allows for a subclass of this models to construct from a given metric in Schwarzschild-like coordinates the corresponding effective spherically symmetric model, its dynamics as an 1+1-dimensional field theory  as well as a corresponding covariant Lagrangian of extended mimetic gravity in four dimensions. Such a reconstruction allows us to obtain Lagrangians of extended mimetic gravity models for black holes  with a regular center, e.g. the Bardeen and Hayward metric as well as for effective LQG inspired models. Moreover, the reconstruction enables us to extend regular black hole models to general inhomogeneous dust collapse models. For the latter, within this formalism, we can investigate and look at the physical properties of the models such as the existence of weak shell-crossing singularities from a novel perspective.
}

\notoc
\begin{document}
\maketitle
\newpage
\tableofcontents

\section{Introduction}\label{sec:intro}
The generalization of black holes from classical general relativity  into the framework of modified gravity and quantum gravity is a topic that has a long history and has also gained extensive  interest in recent years. An interesting class of models in this context are regular black holes, for which the central singularity present in black hole solutions in GR is resolved, see \cite{Bambi:2023try} for an extensive and recent review on this topic and references therein. To obtain regular black hole solutions, one can either modify the gravitational sector (at the classical level or taking into account its quantization) or introduce (usually quite exotic) matter (see. e.g. \cite{Bronnikov:2022ofk, Giacchini:2023waa} for reviews) that resolves the central singularity.  Such regular solutions for black holes have been derived, for example, in effective models inspired by loop quantum gravity. In these models, the quantum gravity corrections are encoded in so-called polymerization functions and inverse triad corrections, which modify the classical dynamics and mimic to some extent the quantization used in loop quantum gravity in the form of holonomies and fluxes associated with symmetry-reduced models  \cite{Ashtekar:2005qt,Modesto:2005zm,Campiglia:2007pb,Gambini:2013ooa,Cartin:2006yv,Boehmer:2007ket,Chiou:2008nm,Corichi:2015xia,Olmedo:2017lvt,Cortez:2017alh,Yonika:2017qgo,Joe:2014tca,Chiou:2008eg,Brannlund:2008iw,Gambini:2013hna,Dadhich:2015ora,Ashtekar:2018lag,Ashtekar:2018cay,Ashtekar:2020ckv,Bodendorfer:2019xbp,Zhang:2021wex,Gambini:2020qhx,ElizagaNavascues:2022npm,Garcia-Quismondo:2022ler,ElizagaNavascues:2022rof,Han:2022rsx}
and see \cite{Ashtekar:2023cod} for a recent review. In these cases, quantum gravity effects are responsible for the resolution of the central singularity, and the corresponding effective metrics are usually derived as solutions of the underlying effective dynamical equations. Other very prominent examples of regular black holes are the Bardeen \cite{Bardeen68} and Hayward \cite{Hayward:2005gi} black holes which have a regular center. In both cases, one considers in a first step a modification of a spherically symmetric metric compared to the Schwarzschild metric for which the central singularity is removed, and then searches for possible Lagrangians of either exotic matter or modified gravity theories that have the Bardeen or Hayward black holes as solutions. The approach where we modify the underlying theory at either the classical or quantum level can be called a top-down approach and the one where we modify the classical metric and thus the solution can be called a bottom-up approach. Often these two strategies have been considered and studied separately due to difference in assumptions and approaches, but when explored collectively can provide valuable insights. 
~\\
~\\
In this work, we apply the formalism developed in \cite{Giesel:2023tsj}, which allows  embedding of a broad class of generalized Lemaitre-Tolmann-Bondi (LTB) models \cite{lemaitre,tolman,bondi} into effective spherically symmetric spacetimes. Here the effective spacetime description refers to inclusion of loop quantum gravity modifications in the classical constraints. In loop quantum cosmology, extensive numerical simulations show that the effective description has been extremely successful in capturing the underlying quantum dynamics \cite{Diener:2014mia} including in the presence of anisotropies \cite{Diener:2017lde}. In this work, we will assume the validity of the effective spacetime description for the entire spacetime. The usual LTB solutions in general relativity are spherically symmetric solutions with dust, and we can rediscover the Schwarzschild solution in the marginally bound models in  the special case where the dust energy density vanishes. Transferring this to the effective models considered in \cite{Giesel:2023tsj}, we find a  generic formalism in which one can study so-called polymerized vacuum solutions. These are solutions of the effective spherically symmetric model for a vanishing dust energy density. As we demonstrate, this formalism allows us to analyze the properties of regular black holes obtained with a top-down or bottom-up approach in a rather uniform way, and allows us to apply similar methods in both cases.
~\\
~\\
For the purpose of our investigation, we consider three different classes of effective models that parameterize the theory space following the representation in \cite{Giesel:2023tsj}. Class I comprises those effective spherically symmetric models for which a consistent reduction to their LTB sector exists. For models in class II, additionally the energy density is a conserved quantity, a requirement for the existence of polymerized vacuum solutions. Class III includes those effective models for which there is an underlying extended mimetic Lagrangian which establishes a connection to modified theories of gravity. In our former work in \cite{Giesel:2023hys} we applied the formalism from \cite{Giesel:2023tsj} mainly to class II models. There the goal  was to start with infinitely many decoupled LQC models along the radial direction and then construct the underlying spherically symmetric effective model, which, when restricted to its LTB sector, yields exactly the model with which one originally started. This allowed to compare existing effective LTB models from a novel perspective yielding useful insights. Continuing this class of models from \cite{Giesel:2023hys}, in the present manuscript  we study polymerized vacuum solutions in this setting. As we will show the insights on the polymerized vacuum solution in class II models will allow further insights on generic marginally bound LTB solutions for inhomogeneous dust profiles. 
~\\
~\\
The main topics and results of this work can be divided into two parts: in the first part we aim at investigating the Birkhoff's theorem \cite{+1927,jebsen} for effective models a question recently also analysed in \cite{Cafaro:2024vrw}. We show that using LTB coordinates one can formulate a Birkhoff-like theorem for effective models in class II. This result can be found in lemma \ref{thm:birkhoff-like} in Subsec. \ref{sec:Birkhoff_eff}. The detailed discussion in Sec. \ref{sec:effAnalysis} shows that effective models with bounded polymerization functions, as they occur in loop quantum gravity inspired effective models, are a bit more complicated to handle when compared to those effective models where the polymerization function is unbounded. Examples of the latter turn out to be the Bardeen and Hayward metric that can be embedded in the formalism here by choosing suitable unbounded polymerization functions. In order to compare our results to existing results in the literature and the the classical case  respectively where Schwarzschild-like and Schwarzschild coordinates are used, we further discuss a formulation of a Birkhoff-like theorem in Schwarzschild-like coordinates. In contrast to the classical theory where Birkhoff's theorem can be formulated in either LTB or Schwarzschild coordinates, this is no longer available at the effective level. As the corollary \ref{Cor:Birkhoff_Schwarzschild} in the subsection \ref{sec:sub_schwarzschildcoords} shows, uniqueness is generally given in LTB coordinates, but is no longer valid in Schwarzschild-like coordinates. Again, we note that the situation is different for bounded and unbounded polymerization functions or, more generally, bounded and unbounded phase space trajectories of the effective models. Further, other properties such as monotonicity as well as analyticity play an important role when dealing with the question of uniqueness. As the subclass $\rm{II\cap II}$ is contained in the class II the above discussed results on the Birkhoff-like theorem can be understood as Birkhoff-like theorems of extended mimetic gravity \cite{Chamseddine:2013kea,Sebastiani:2016ras,Takahashi:2017pje,Langlois:2018jdg} in this case.
~\\
~\\
In the second part we introduce a reconstruction algorithm for models in class II which provides a systematic procedure to construct a polymerized vacuum model from a given metric in Schwarzschild-like coordinates. In this work we focus on the reconstruction algorithm on models in the subclass $\rm{II\cap II}$ such that the construction of the underlying extended mimetic Lagrangian is part of the reconstruction. The results are summarized in lemma \ref{coro:reconstruction} in Subsec. \ref{sec:Reconstruction_Detail}. 
A further result that enters into the reconstruction algorithm is that we can extend the polymerized vacuum solution to the generic  marginally bound LTB solutions for inhomogeneous dust profiles, which can be found in lemma \ref{coro:reconstruction2} in Subsec. \ref{sec:Reconstruction_Detail}. This reconstruction algorithm plays a central role in establishing a connection to modified gravity theories in two ways. On one hand, it allows us to start with a modified metric in Schwarzschild-like coordinates compared to the Schwarzschild metric and then derive the effective dynamics and the underlying extended mimetic model from it, thus moving from a bottom-up model directly to its corresponding top-down model in the framework of modified gravity. We will apply this strategy to the Bardeen and Hayward metrics and explicitly derive the corresponding extended mimetic Lagrangians. On the other hand, our reconstruction algorithm allows us to establish a connection between loop quantum gravity inspired effective models and modified gravity models in the framework of extended mimetic models, and in this sense relate different top-down models to each other.
~\\
~\\
In general, it is not easy to consistently embed regular black holes, especially beyond the static case, in scalar-tensor theories. Our methods provide an example of such embedding with the following advantages: 1) We are able to embed static metrics of spherically symmetric black holes into the extended mimetic gravity, which is known to propagate only 2+1 degrees of freedom \cite{Takahashi:2017pje, Langlois:2018jdg}. 2) In our embedding methods, the mass of the static metric appears naturally as an integration constant of the equations of motion. The underlying extended mimetic gravity depends only on a parameter that can be related to the quantum gravity effect. 3) We naturally embed the associated inhomogeneous dust collapse solutions and Oppenheimer-Snyder models. Thus, the embedding is the associated 1+1d field theory instead of just the static solution. 4) We have a Birkhoff-like theorem for the underlying extended mimetic gravity. Moreover, our embedding provides a concrete example for exploring (regular) black hole solutions in extended mimetic gravity and related more general degenerate higher order scalar tensor (DHOST) theories \cite{Chamseddine:2019pux,BenAchour:2017ivq,Langlois:2018jdg,Langlois:2018dxi,Babichev:2023psy}.
~\\
~\\
Our analysis of the examples, which include effective models with a symmetric and asymmetric bounce as well as the Bardeen and Hayward metrics in section \ref{sec:examples}, shows that the formalism of \cite{Giesel:2023tsj} provides a systematic procedure to study physical properties of various regular black hole solutions that fall into the subclass $\rm{II\cap II}$ of effective models based on the properties of their corresponding polymerization functions or extended mimetic Lagrangians respectively. This includes, in particular, the form of the curvature invariants as well as addresses whether the corresponding marginally bound LTB solutions have shell crossing singularities or not. 
~\\
~\\
The paper is structured as follows: after the introduction in Sec. \ref{sec:intro} we briefly introduce the classical Hamiltonian formulation of spherically symmetric models in terms of Ashtekar-Barbero variables. In Subsec. \ref{sec:ClassTheory} we consider models with dust and discuss the form of the LTB solution in LTB coordinates. The special case of the vacuum solution is discussed in Subsec. \ref{sec:BirkhoffClass} where we further review Birkhoff's theorem in the classical theory in terms of LTB and Schwarzschild coordinates. This provides the necessary background and notation needed for Sec. \ref{sec:effAnalysis} in which these topics are revisited in the framework of effective models. In Subsec. \ref{sec:classification} we introduce a classification of the theory space of effective models relevant for the further analysis in this work. Then in Subsec. \ref{sec:BirkhofflikeThm} we focus on polymerized vacuum solutions and investigate the question whether the classical Birkhoff's theorem can be generalised at the level of effective models. Since most presentations of the classical Birkhoff's theorem are given in Schwarzschild coordinates,  afterwards in Subsec. \ref{sec:sub_schwarzschildcoords} we will discuss polymerized vacuum solutions in Schwarzschild-like coordinates and their relations to the solutions in LTB coordinates. In Subsec, \ref{sec:ReconstrAlg} we present a reconstruction algorithm that allows to construct an effective model from a broad class of metrics given in Schwarzschild-like coordinates. We further restrict the models to the subclass $\rm{II\cap III}$ for which an underlying Lagrangian of an extended mimetic model exists. At the level of the mimetic model 
we then observe a limiting curvature mechanism at the level of the mimetic model in the context of regular black holes which is discussed in Subsec. \ref{sec:LimCurvature}. To show the application of the reconstruction algorithm we discuss four examples in Sec. \ref{sec:examples}. Two of them, the symmetric and asymmetric bounce have bounded polymerization functions, whereas the two further examples are the Bardeen and Hayward metric which can be related to unbounded polymerization functions using the formalism introduced in this work. Finally, we summarize and conclude in Sec. \ref{sec:Concl}.

\section{Hamiltonian formulation of spherically symmetric models} \label{sec:HamFormSpherically}
Before we can formulate the effective model and discuss therein regular blackholes in Sec. \ref{sec:effAnalysis} and \ref{sec:examples}, we want to go in this section over some basic ingredients and definitions to introduce the reader to the main concepts of our framework. In the first part we will summarize the formulation of spherical symmetric gravity with dust in terms of Ashtekar-Barbero variables. Subsequently we will show how Lemaître-Tolman-Bondi (LTB) spacetimes written in Lemaître or LTB coordinates can be naturally embedded in this framework by implementing a suitable constraint. Finally we will derive the general vacuum solution and are able to rediscover Birkhoff's theorem.
\subsection{Spherically symmetric models with dust}
\label{sec:ClassTheory}
As a starting point for the upcoming analysis we first have to define a gravitational system with dust. In this work we choose non-rotational dust such that the Lagrangian density of our model is given by
\begin{equation}
    \mathcal{L}= \sqrt{\det(g)} \Big(R^{(4)} -\frac{\rho}{2}\big[g^{\mu\nu}(\nabla_\mu T)(\nabla_\nu T)+ 1\big]\Big)\,,
\end{equation}
where $g^{\mu\nu}$ is the four dimensional spacetime metric and $R^{(4)}$ the corresponding curvature scalar. Note that we use in this work as index convention greek letters for the spacetime indices $\mu,\nu,... = 0,\,...\,,3$ and latin letters for only the spatial dimensions $a,b,... = 1,2,3$. The second part of the action encodes the dynamics of the dust field $T$ which has the energy density $\rho$. It is also possible to work with other dust models like Gaussian \cite{Kuchar:1990vy} or Brown-Kucha\v{r} dust \cite{Kuchar:1995xn}. Although these dust types have more degrees of freedom, one can show that after performing the Dirac analysis of these systems in the LTB sector these additional degrees of freedom of the dust models actually reduce and in the end they are equivalent to the non-rotational dust model given above, for details we refer to \cite{Giesel:2023tsj}.

In loop quantum gravity \cite{Bengtsson_1990,Bojowald:1999eh,Bojowald:2004af} we are not working with the standard geometrodynamical variables \cite{Arnowitt:1960es}, but with so-called Ashtekar-Barbero variables which let us write gravity as a $SU(2)$ gauge theory. It turns out that this formulation is much more suited for a background independent canonical quantization of the theory \cite{Perez:2004hj,rovelli2004quantum,thiemann2008modern}. Furthermore we will impose spherical symmetry on the whole system. Assuming we have a global hyperbolic spacetime, then there exist a foliation into spatial hypersurfaces which we can chosen to be of the geometry $\mathbb{R}\times S^2$. The phase space is coordinatized by the Ashtekar connection $A_a^j$ and densitized triad variables $E^a_j$ which in the symmetry reduced case (after suitably fixing the Gau{\ss} constraint) can be written as
\begin{eqnarray*}
A_a^j \tau_j \mathrm{~d} X^a & = & 2\beta K_x(x) \tau_1 \mathrm{~d} x+\left(\beta K_\phi(x) \tau_2+\frac{\partial_x E^x(x)}{2E^\phi(x)} \tau_3\right) \mathrm{d} \theta \\
&&+\left(\beta K_\phi(x) \tau_3-\frac{\partial_x E^x(x)}{2E^\phi(x)} \tau_2\right) \sin (\theta) \mathrm{d} \phi+\cos (\theta) \tau_1 \mathrm{~d} \phi \\
E^a_j \tau^j \frac{\partial}{\partial X^a} & = &E^x(x) \sin (\theta) \tau_1 \partial_x+\left(E^\phi(x) \tau_2\right) \sin (\theta) \partial_\theta+\left(E^\phi(x) \tau_3\right) \partial_{\phi}\,,  
\end{eqnarray*}
where $X^a=(x,\theta,\phi)$ denote spherical coordinates, $\beta$ the Barbero-Immirzi parameter and $\tau_j=-\frac{1}{2}\sigma_j$ with $\sigma_j$ being the Pauli matrices. On one equal time slice the non-vanishing Poisson brackets are then given by \cite{Giesel:2023tsj}
\begin{eqnarray*}
 \{K_x(x), E^x(y)\} &=& G \,\delta(x,y),\quad  \{K_\phi(x), E^\phi(y)\} =G \,\delta(x,y),\quad\{T(x), P_T(y)\} =\delta(x,y) \,,
\end{eqnarray*}
where we introduced the canonical momentum of the dust field $P_T$. Before continuing our analysis we will implement at this point the dust time gauge $T=t$, i.e. we will use the value of the dust field $T$ as time variable. This is consistent with the fact that LTB coordinates can be seen as comoving observers with the dust, i.e. radially free falling observers move along $x=const$ geodesics and the coordinate time is equal to the proper time on these geodesics. In this setup the general spherical symmetric metric can be expressed in terms of densitized triad variables as
\be 
\label{eq:metricSphSymm}
\mathrm{d}s^2 = -\mathrm{d}t^2 + \frac{(E^{\phi})^2}{\abs{E^x}} (\mathrm{d}x + N^x \mathrm{d}t)^2 + \abs{E^x} \mathrm{d} \Omega^2\,.
\ee 
In our chosen comoving gauge the Hamiltonian or scalar constraint deparametrizes and the dynamical equations are generated from the primary Hamiltonian
\begin{eqnarray} \label{eq:defprimaryHamiltonian}
    H_P[N^x] = \int \mathrm{d}x \, \big[C + N^x C_x\big](x) \,,
\end{eqnarray}
where the shift vector $N^x$ is a Lagrangian multiplier to the spatial diffeomorphism constraint $C_x$ and $C$ is the gravitational part of the scalar constraint. Assuming a fixed orientation of the triad allows us to drop the absolute value of $E^x$ and the two contributions of the primary Hamiltonian have the form
\begin{align}\label{eq:defHamiltonianconstraintclassical}
    C(x)&= \frac{ E^{\phi}}{ 2 G \sqrt{E^x}}\bigg[-E^x\Big(\frac{4 K_x K_\phi}{E^{\phi}} +  \frac{(K_\phi)^2}{E^x}\Big) +  
\qty(\frac{  {{E^x}}'}{2{{E^{\phi}}} })^2 - 1    
+2\frac{E^x}{E^\phi}  \qty(\frac{   {{E^x}}'}{2{{E^{\phi}}} })'\bigg](x) \\
    C_x(x)&= \frac{1}{G}\left(E^{\phi} {K_{\phi}} ' - K_x {E^x}'\right)(x)\,.
\end{align}
Note that in above expression we already expressed the involved spin connection $\Gamma_\phi$ directly in terms of the triads and its derivatives according to $\Gamma_{\phi}=-\frac{\left(E^x\right)^{\prime}}{2 E^{\phi}}$.\\
\noindent
Next we want to specialize to the case that we have only dust, i.e. a perfect fluid with no pressure, in our gravitational system. The corresponding solutions to the Einstein equations are the Lemaître-Tolman-Bondi (LTB) spacetimes \cite{lemaitre, tolman, bondi}. At this point we will not implement any aerial gauge fixing (i.e. specifying the relation between radial coordinate $x$ and physical radius of shell $R(x)$ by gauge fixing the radial diffeomorphism constraint suitably) to our system, which would allow us to go to Gullstrand-Painlevé (GP) coordinates, as here the physical radius is used as radial variable. Although some authors argument that GP coordinates are classically very well suited for an initial value formulation \cite{Lasky:2006hq} and therefore a canonical description of such systems, it turns out that in our context of quantum corrected dust models LTB coordinates have several advantages in comparison to a formulation in (generalized) GP coordinates, see for example \cite{Fazzini:2023scu}\cite{Giesel:2023hys}. So for this analysis we will work with LTB coordinates. Coming back to the canonical formulation of a general spherical symmetric gravitational system, by comparing the metric in \eqref{eq:metricSphSymm} with the LTB metric written in LTB coordinates in the marginally bound case, we see that we first have to choose a vanishing shift vector $N^x=0$ and further we can deduce that the relation 
\begin{equation}\label{eq:ltbcondition_classical}
    C_{LTB}= (E^x)' - 2 E^\phi=0
\end{equation}
between the triad variables has to hold for this class of solutions. We will call this relation LTB condition from now on. It turns out that in the Hamiltonian formulation the condition \eqref{eq:ltbcondition_classical} adds another first class constraint to the spherical symmetric system \cite{Giesel:2023tsj}. The dynamical equations of the remaining pair of canonical variables are given by
\begin{eqnarray}\label{eq:dy_new}
        \partial_t E^x(x) = - 2 \sqrt{E^x}(x) K_\phi(x) \ , \qquad
        \partial_t K_{\phi}(x) =  \frac{(K_\phi)^2}{2 \sqrt{E^x}}(x) \,.
\end{eqnarray}
Note that in principle this reduction can also be done with respect to other choices for the remaining pair of canonical variables. We just have to use the vanishing of the diffeomorphism constraint and the LTB condition to eliminate the undesired two variables. However in the non-marginally bound system where the LTB condition can be seen as gauge fixing of the diffeomorphism constraint, above choice has the simplest Dirac bracket and subsequently the simplest dynamical equations. This last fact also holds in the marginally bound sector, which is why the variables $(K_\phi,E^x)$ are typically the favored choice. For more details we refer to the second section in \cite{Giesel:2023tsj}.~\\
~\\
As already specified in the text above, for this analysis we will only work in marginally bound LTB sector. In a moment it will become clear why this is fully sufficient for investigating the vacuum solutions. To begin with we are working in LTB coordinates that are describing the spacetime from the perspective of an observer comoving with the dust field. See for example in \eqref{eq:metricSphSymm} and \eqref{eq:defprimaryHamiltonian} that the lapse function $N$ is already fixed to the constant value one. Thus the LTB coordinates let us describe the dynamics of the dust field as the dynamics of infinitely many decoupled dust shells at each value of the radial coordinate $x_s$ with physical radius $R_s = \sqrt{E^x(x_s)}$. The so-called LTB function $\Xi(x)$, see for example \cite{Kiefer:2005tw}\cite{Bojowald:2009ih}, is a phase space independent function in $x$ which can be interpreted as the total energy of a dust shell \cite{Lasky:2006hq,Szekeres:1999}. The marginally bound sector is then defined by setting the total energy of each shell to exactly zero. This means while infinitely far away from any gravitational source the dust shell is at rest. However in a collapsing scenario the dust shell exactly gains the amount of kinetic energy it loses on the other side in terms of gravitational potential energy. The other parameter of a shell that we need to specify in order to describe all possible initial conditions is next to the energy its mass. As it turns out in the Hamiltonian formulation this is related to the conserved quantity $M(x)$, that can be interpreted as the accumulated dust mass distribution. i.e. it is a measure of the total gravitating mass inside the shell at radial coordinate $x$. Finally we can rediscover the vacuum solution in this setup by first setting the total energy of each shell to zero, thus we are in the marginally bound LTB sector, and in addition by setting the dust mass profile to zero which translates to assigning the same constant value for all $x$ to the conserved quantities $M(x)\equiv m$. In the following paragraph we will see in detail how the construction of the general vacuum solution in the marginally bound LTB sector can be carried out.
~\\
Based on the results of our previous work in \cite{Giesel:2023tsj}, we can write the gravitational contribution of the scalar constraint in the marginally bound LTB sector as 
\begin{align}\label{eq:LTBreducedHami}
    C(x)\eval_{LTB}= \partial_x \widetilde{H}(x), && \,-M(x) = \widetilde{H}(x) \coloneqq  -\frac{1}{2 G}\sqrt{E^x(x) } (K_\phi)^2(x)
\end{align}
where we defined the decoupled (not field theoretical) Hamiltonian $\widetilde{H}(x)$ at each $x$ that can be seen as the generator of dynamics with the Poisson bracket $\poissonbracket{K_\phi(x)}{E^x(x)}=-G$. Besides that we have introduced the conserved quantity $M(x)$ which is given by the negative value of the decoupled Hamiltonian itself. Note that this can be done for every classical mechanical system with a time independent Hamiltonian. With help of the equations of motions in \eqref{eq:dy_new} we can indeed verify that the temporal derivative of $M(x)$ is vanishing. As already explained these conserved quantities can be adopted to model arbitrary initial inhomogeneous dust mass distributions.\\ 
Note that we can also rewrite the dynamical equations \eqref{eq:dy_new} as a Friedmann equation by introducing $R(x)=\sqrt{E^x(x)}$ and using the conserved quantities to formulate
\begin{align}\label{Friedmann_class}
    \frac{\Dot{R}^2}{R^2}(x)=\frac{8\pi G}{3} \rho(x)\quad \quad\quad\mathrm{with}\quad\quad\rho(x)=\frac{3 }{4\pi }\frac{M}{R^3}(x)\,.
\end{align}
\subsection{Vaccuum solutions and Birkhoff's theorem}
\label{sec:BirkhoffClass}
Finally we want to derive the general vacuum solution in LTB coordinates. As explained beforehand this means first of all working in the marginally bound LTB sector, i.e. the total energy of each shell is vanishing, and additionally setting the dust mass profile to $M(x)\equiv m = const$. We can use the definition of the conserved quantities in \eqref{eq:LTBreducedHami} to express $K_\phi$ as a function of $E^x$, that is
\begin{equation}\label{eq:KphiClass}
K_\phi(x)= \frac{\sqrt{2Gm}}{(E^x)^{\frac{1}{4}}(x)} 
\end{equation}
and then using this rewrite the equations of motion in the marginally bound sector stated in \eqref{eq:dy_new} as
\begin{equation}
    \partial_t E^x(x) = - 2 \sqrt{2 G m} \,(E^x)^{\frac{1}{4}}(x)\,.
\end{equation}
Note that we can directly integrate this ordinary differential equation such that the general solution is given by
\begin{equation}\label{eq:solLTBclass}
    E^x = \Big[\frac{3}{2} \sqrt{2 G m} \qty(s(x) - t)\Big]^{\frac{4}{3}}\,.
\end{equation}
Without loss of generality we can set the integration constant to $s(x)=x$ since this is in the vacuum case just a rescaling of the radial coordinate. It is now not difficult to see that this solution is stationary (on the outside) since it has the timelike Killing vector $\partial_z = \partial_t + \partial_x = 0$ when plugging this solution back in the metric given \eqref{eq:metricSphSymm}. Using the LTB condition \eqref{eq:ltbcondition_classical} we have 
\begin{equation}
    E^\phi = \tfrac{1}{2} (E^x)' = \sqrt{2 G m} \Big[\frac{3}{2} \sqrt{2 G m} \qty(x - t)\Big]^{\frac{1}{3}}
\end{equation}
and further setting $N^x =0$, which is consistent with the gauge fixing of the diffeomorphism constraint with respect to the LTB condition in the non-marginally bound case, we realize that every function in the metric can be written as a function of $z=x-t$. Note that in the case of a radial dependent mass function $M(x)$, i.e. we have a non-trivial (inhomogeneous) dust mass distribution and not vacuum, from the solution in \eqref{eq:solLTBclass} we realize that $E^x$ cannot be written as a function of $z$ anymore. From the explicit solution of $E^x$ and the conservation of $m$ we can check that the extrinsic curvature components $K_\phi, K_x$ indeed go to zero for increasing values of $z$.\\
Summarizing these results we can rediscover Birkhoff's theorem in LTB coordinates as there exists a unique family of vacuum solutions which is labeled by $m$ and is stationary as well as asymptotically flat. We can verify that the conserved quantity $m$ is indeed equivalent to the Schwarzschild mass by transforming to Schwarzschild coordinates
\begin{eqnarray}
 \mathrm{d} s^2 = - ( 1-\mathcal{G}(r)^2) \mathrm{d} \tau^2 +\frac{1}{  \left( 1-\mathcal{G}(r)^2 \right)} \mathrm{d}r^2 + r^2 \mathrm{d} \Omega^2\,,
\end{eqnarray}
with $r=\sqrt{E^x}$ and the function $\mathcal{G}(r)^2$ can with above given solution of the equations of motion computed to be
\begin{align}
    \mathcal{G}(r)^2 := \frac{(\partial_x E^x)^2}{4  {E^x}} = \frac{r_s}{r}\,,
\end{align}
where we introduced $r_s = 2 G m$. A more detailed derivation of this transformation can be found in Sec. \ref{sec:sub_schwarzschildcoords} where this calculation is done in the effective model. However we can recover the classical case by setting $ {g}_{(\alpha)} \equiv1$. One important difference in the effective model is that depending on the dynamics this transformation might not be unique anymore.

\section{Investigation of different classes of effective spherically symmetric models}\label{sec:effAnalysis}
In this section, we will generalize the discussion of the last section by considering modifications of general relativity inspired by  underlying models of quantum gravity. The framework in which we analyze various models are so-called effective models, in which these modifications are encoded in so-called polymerization functions and inverse triad corrections. Both are inspired by the quantization techniques of loop quantum gravity, where connection variables are quantized as holonomy operators and inverse volume operators are formulated as commutators between holonomy and volume operators. 
In general effective models are expected to be derived from an underlying quantum model by computing the expectation values of the corresponding dynamical operators in suitable semiclassical states. The focus of this work is not to derive the form of the polymerization functions and inverse triad corrections in this way, instead we will start with a generic ansatz for an effective model and analyze which restrictions certain additional properties we require for the effective model will impose on the form of the polymerization functions and inverse triad corrections. In Subsec. \ref{sec:classification} we present a classification of the theory space of effective models relevant for our work. Then in Subsec. \ref{sec:BirkhofflikeThm} we focus on polymerized vacuum solutions and investigate the question whether we can formulate a Birkhoff-like theorem for these kind of effective models. Afterwards we will discuss in Subsec. \ref{sec:sub_schwarzschildcoords} polymerized vacuum solutions in Schwarzschild-like coordinates and their relations to the solutions in LTB coordinates. This will then provide the basis for introducing a reconstruction algorithm in Subsec. \ref{sec:ReconstrAlg} that allows to construct an effective model from a broad class of metrics given in Schwarzschild-like coordinates. For models in the subclass $\rm{II\cap III}$ an underlying Lagrangian of an extended mimetic model exists and we discuss in Subsec. \ref{sec:LimCurvature} how we can benefit from this in understanding a limiting curvature mechanism at the level of the mimetic model in the context of regular black holes.

\subsection{Classification of the effective models}\label{sec:classification}
The three main requirements for effective models that we will analyze can be divided into three different classes
\begin{itemize}
\item {\bfseries Class I:} effective models for which the possibility exists to do an LTB reduction of the effective spherical symmetric model to its effective LTB sector,
\item {\bfseries Class II:}  effective models for which there is an LTB reduction and which have a conserved energy density that allows the presence of a polymerized vacuum solution, and
\item {\bfseries Class III:} effective models for which an extended mimetic model as an underlying covariant Lagrangian exists. 
\end{itemize}
These classes are not disconnected from each other but, as discussed below, are useful to classify the theory space of effective models in the context of our investigation. The choice of these classes is motivated from the results found in our earlier works \cite{Giesel:2023tsj,Giesel:2023hys}. In \cite{Giesel:2023tsj} the implementation of the LTB reductions at the level of effective models was investigated in detail. As discussed in Subsec. \ref{sec:BirkhoffClass}, an important result in the classical theory is Birkhoff's theorem and the associated uniqueness of spherically symmetric stationary vacuum solutions. Therefore, class II includes effective models for which the analogue of a vacuum solution of the polymerized model exists, which we will refer to as polymerized vacuum solutions in our work. For effective models in class II we can then investigate the question whether a Birkhoff-like theorem also exists at the level of effective models. This corresponds to effective models for which the density of polymerized Hamiltonians, which we will refer to as $C^{\Delta}$ in the following, must vanish. Finally, 
for effective models that do not involve inverse triad corrections and polymerization functions compatible with the $\overline{\mu}$-scheme, an underlying extended mimetic Lagrangian can be formulated, and these types of models are included in class III.
~\\
~\\
Before discussing the properties of these three classes in more detail, we want to introduce the generic spherical symmetric polymerized model which we will take as a starting point. As in the classical case we will again work with a metric of the form \eqref{eq:metricSphSymm}. However the dynamics of the model will be different as the geometric part of the Hamiltonian constraint defined in \eqref{eq:defHamiltonianconstraintclassical} will be modified by generalized holonomy corrections $f(K_x,K_\phi,E^x,E^\phi)$ and inverse triad corrections $h_1(E^x),h_2(E^x)$ which are characterized by a polymerization parameter $\alpha$ such that geometric part of the effective Hamiltonian constraint is defined as
\begin{equation}\label{eq:defpolyhamiltonianconstraint}
    C^{(\alpha)}(x)= \frac{ E^{\phi}\sqrt{E^x}}{2 G }\left[ -{(1 + f) } \qty(\frac{4 K_x K_{\phi}}{E^{\phi}} + \frac{K_{\phi}^2}{E^x } ) + \frac{h_1 }{E^x}
\qty(\qty(\frac{  {{E^x}}'}{2{{E^{\phi}}} })^2 - 1   )+2\frac{h_2}{E^\phi} \qty(\frac{   {{E^x}}'}{2{{E^{\phi}}} })'\right](x)\,.
\end{equation}
In the semiclassical limit, realized by taking the polymerization parameter $\alpha\rightarrow0$, we have $f\rightarrow0$ and $h_1,h_2\rightarrow 1$. The polymerization parameter $\alpha$ encodes the magnitude of quantum gravity effects. In the case of loop quantized models it is related to the minimal area gap $\Delta$ such that we have $\alpha\propto\sqrt{\Delta}\propto {l}_\text{P}$ with $l_\text{P}$ denoting the Planck length. In this case we use the notation
\begin{equation}
C^\Delta(x):= C^{(\alpha=\alpha_\Delta)}(x), \quad{\rm with}\quad \alpha_\Delta:=\gamma \sqrt{\Delta}=\gamma 4\pi l_P,
\end{equation}
where $\gamma$ denotes the Barbero-Immirzi parameter.
However we want to keep the polymerization parameter $\alpha$ generic at this point, such that we can discuss loop quantized inspired effective and other modified gravity models together in this framework. Note that in our former work 
\cite{Giesel:2023tsj,Giesel:2023hys} such a generic polymerization parameter was included in principle as well but since we mainly applied our method to LQC inspired polymerizations we used the notation $C^\Delta$ through out the work in \cite{Giesel:2023tsj,Giesel:2023hys}. Similar to the framework introduced in \cite{Giesel:2023tsj} and which was applied in \cite{Giesel:2023hys} to effective LTB models, we will not consider modifications to the spatial diffeomorphism constraint. The reason for this, as discussed in detail in \cite{Giesel:2023tsj}, is that in this case the Poisson algebra of $C^{(\alpha)}$ and the spatial diffeomorphism constraint is simplified, while models with a polymerized spatial diffeomorphism constraint are more complex in this respect and would go beyond the scope of this article. Next, we discuss more detailed the properties of the individual classes and their corresponding restrictions on the polymerizations and inverse triad corrections respectively.
\subsubsection{Class I: models with an LTB reduction}\label{sec:ClassI}
The formalism that was introduced in our former work \cite{Giesel:2023tsj} allows to reduce a given effective spherical symmetric model to its LTB sector with the help of a so-called effective LTB condition. Based on the results in \cite{Giesel:2023tsj} the most general type of an effective LTB conditions allowing a consistent LTB reduction in the non-marginal bound case is of the form\footnote{Note that we have slightly changed the notation of the function $g_{(\alpha)}$ in this analysis compared to our previous work in \cite{Giesel:2023tsj}.}
\begin{equation}\label{eq:LTBconditioneffective}
  C_{LTB}^{(\alpha)}(x):= \Big[\frac{E^x{}'}{2 E^{\phi}} - \widetilde{g}_{(\alpha)}(K_\phi,E^x,\Xi)\Big](x) \,,
\end{equation}
where $\Xi(x)$ denotes the LTB function which is a measure of the total energy of a shell at radial coordinate $x$. Note compared to the notion in our former work \cite{Giesel:2023tsj} we as before label the LTB condition by a generic polymerization parameter $\alpha$ instead of $\Delta$ as done  \cite{Giesel:2023tsj}. Similar to the effective Hamiltonian we use the notation 
\begin{equation}
 \widetilde{g}_{\Delta}= \widetilde{g}_{(\alpha=\alpha_\Delta)},
\end{equation}
which will be used in the context of loop quantization inspired effective models. In this work we are mainly interested in the marginally bound LTB sector. Nevertheless
in our investigation we consider the more generic non-marginally bound case. The marginally bound case can then be rediscovered by choosing the LTB function to be a constant and without loss of generality this can be one, i.e. $\Xi(x)\equiv1$. This approach is motivated from the classical theory as here the same construction can be carried out. Note that the results in \cite{Giesel:2023tsj} show the existence of more general classes of LTB conditions in the effective models which are only implementable in the marginally bound case. Moreover restricting ourselves to LTB conditions where the dependence on the LTB function can be factored out, i.e.
\begin{equation}\label{eq:factorisedeffectiveLTBcondi}
\widetilde{g}_{(\alpha)}(K_\phi,E^x,\Xi) = { {g}}_{(\alpha)}(K_\phi,E^x)\Xi, 
\end{equation}
then corollary 2 in \cite{Giesel:2023tsj} tells us that the polymerization function $f$ can be expressed in terms of polymerization functions ${f}^{(1)}$ and ${f}^{(2)}$ in the following way
\begin{eqnarray}\label{eq:no_kx_f1}
    f = \frac{{f}^{(1)}(K_{\phi},E^x)- {f}^{(2)}(K_{\phi},E^x) K_{\phi}}{(K_{\phi}+4 {E^x} \widetilde{K}_x) K_{\phi}} + \frac{{f}^{(2)}(K_{\phi},E^x)}{ K_{\phi}} -1  \quad \text{with} \quad \widetilde{K}_x = \frac{K_x}{E^\phi}\,,
\end{eqnarray}
which means that no polymerization of $K_x$ is allowed. This can be easily seen by reinserting above polymerization function into the geometric part of the effective Hamiltonian written in \eqref{eq:defpolyhamiltonianconstraint} to get
\begin{align}\label{eq:defpolyhamiltonianconstraintwithoutKx}
    C^{(\alpha)}(x)= \frac{ E^{\phi}}{2 G \sqrt{{{E^x}}}}\Bigg[ -{ E^x}\bigg(&\frac{4 K_x {f}^{(2)}(K_{\phi},E^x)}{E^{\phi}} +  \frac{{f}^{(1)}(K_{\phi},E^x)}{E^x} \bigg)+\\  &+ h_1(E^x) 
\qty(\qty(\frac{  {{E^x}}'}{2{{E^{\phi}}} })^2 - 1   ) 
+2\frac{E^x}{E^\phi} h_2(E^x) \qty(\frac{   {{E^x}}'}{2{{E^{\phi}}} })'\Bigg](x)\,.\nonumber
\end{align}
The semiclassical limit of these new polymerization functions are given by
\begin{align}
    {f}^{(1)}(K_{\phi},E^x)\rightarrow (K_{\phi})^2\,, && {f}^{(2)}(K_{\phi},E^x)\rightarrow K_{\phi}\,, && h_1(E^x),h_2(E^x)\rightarrow 1\,.
\end{align}
In addition there are restrictions on the effective LTB condition. Firstly, the $K_\phi$ dependence of $g_{(\alpha)}$ has to vanish and secondly there is the differential equation
 \begin{eqnarray}    \label{eq:ltb_con_to_f1_f2}
    {g}_{(\alpha)}(K_{\phi},E^x) =  {g}_{(\alpha)}(E^x) \, ,\quad \,
    1-\frac{2{E^x} \partial_{E^x}  {g}_{(\alpha)}}{ {g}_{(\alpha)}} =\frac{-4 {E^x} \partial_{E^x}{f}^{(2)}(K_{\phi},E^x)+  \partial_{K_{\phi}}{f}^{(1)}(K_{\phi},E^x)}{2 {f}^{(2)}(K_{\phi},E^x)} =\text{Con}_{f}\nonumber \\
\end{eqnarray}
and both requirements have to be satisfied in order to perform a consistent LTB reduction in the effective model. If this is the case using the notion introduced in \cite{Giesel:2023tsj} we call such an effective LTB condition compatible. Note that the left hand side of the differential equation is only a function of $E^x$ and so this has to be the case for the right hand side too. This is highlighted in \eqref{eq:ltb_con_to_f1_f2} with the condition $\text{Con}_{f}$.
\subsubsection{Class II: models for which a polymerized vacuum solution exists}
\label{sec:ClassII}
Interestingly above class of compatible LTB conditions is closely related to the question whether $C^{(\alpha)}(x)$ are conserved quantities of the dynamics. Similar to the classical case this translates to a conservation of the mass for all individual dust shells at each radial coordinate $x$. Referring to the results given in Lemma 1 in \cite{Giesel:2023tsj} we are able to show this relation in detail. For an investigation of polymerized vacuum solutions this property is crucial, because it ensures that choosing $C^{(\alpha)}(x)$ to vanish entirely on one of the spatial hypersurfaces will then be preserved under the effective dynamics. Thus, only if this property is given we are in the situation that we can analyze polymerized vacuum solutions at all.
~\\
It turns out that we for one have to consider an effective model with holonomy polymerization of the type in equation \eqref{eq:no_kx_f1}, i.e. there is no $K_x$ polymerization and additionally the differential equation 
\be\label{eq:final_con_clo}
\frac{h_1(E^x) - 2 E^x \partial_{E^x} h_2(E^x)}{h_2(E^x)}= \frac{-4 {E^x} \partial_{E^x}{f}^{(2)}(K_{\phi},E^x)+  \partial_{K_{\phi}}{f}^{(1)}(K_{\phi},E^x)}{2 {f}^{(2)}(K_{\phi},E^x)} =\text{Con}_{f}\,.
\ee 
has to hold. We see that the same condition as in equation \eqref{eq:ltb_con_to_f1_f2} occurs, which allows us to embed a compatible LTB condition in an effective model in which $C^{(\alpha)}(x)$ are conserved quantities. This can be done by choosing the appropriate function $ {g}_{(\alpha)}(E^x)$ that satisfies differential equation \eqref{eq:ltb_con_to_f1_f2}, or vice versa.
~\\
In the case that we have a model with compatible effective LTB condition ${g}_{(\alpha)} =  {g}_{(\alpha)}(E^x)$ and additionally $C^{(\alpha)}$ being conserved quantities, the dynamical equations of the variables $K_\phi, E^x$ in the LTB sector are given according to corollary 4 in \cite{Giesel:2023tsj} by\footnote{Note that a global sign was added to the dynamical equation shown here compared to our former work in \cite{Giesel:2023tsj}. This was done to be consistent with the dynamical equation of the classical system \eqref{eq:dy_new} and merely corresponds to a redefinition of the direction of time.}
\begin{eqnarray}\label{eq:eff_dy_new}
        \partial_t E^x = -2 \sqrt{E^x} {f}^{(2)} \ , \qquad
        \partial_t K_{\phi} =  \frac{1}{2 \sqrt{E^x}} \left( f^{(1)}-   {g}_{(\alpha)}^2  \ltbf^2 (2 h_2 + 4 E^x \partial_{E^x} h_2 -h_1) + h_1 \right).
\end{eqnarray}
In addition we can define analogous to the classical case, see \eqref{eq:LTBreducedHami} the conserved quantities of the dynamics
\begin{align}
    \widetilde{H}^{(\alpha)}(x) := \frac{1}{2G}\qty[\frac{\sqrt{E^x}}{ {g}_{(\alpha)} \ltbf } \left( - F + h_2 \left(  {g}_{(\alpha)}^2 \ltbf^2  -1 \right) \right)](x)\,,&& C^{(\alpha)}(x)\eval_{LTB} = \partial_x \widetilde{H}^{(\alpha)}(x)\,,    \label{eq:Ham_non_marginal}
\end{align}
where we introduced 
\begin{equation}\label{eq:Def_F}
\partial_{K_{\phi}} F(K_{\phi},E^x) =2 {f}^{(2)}(K_{\phi},E^x)
\end{equation}
also satisfying the differential equation
\begin{eqnarray}\label{eq:closurebutwith}
        \frac{h_1 - 2 E^x \partial_{E^x} h_2}{h_2} = \frac{-2 E^x \partial_{E^x} F + f^{(1)} }{ F }\, .
    \end{eqnarray}
One important property of this class of models is that the effective dynamics decouples completely along the radial $x$ direction. Clearly $\widetilde{H}^{(\alpha)}$ is conserved under the effective dynamics. We will see later in Sec. \ref{sec:BirkhofflikeThm} that the conserved quantities are very useful for solving the equation of motion and subsequently constructing stationary solutions. We can also apply the procedure the other way around to derive from a given Schwarzschild-like metric the underlying effective model in terms of the polymerization functions and inverse triad corrections, for more details see Subsec. \ref{sec:sub_schwarzschildcoords}.
\subsubsection{Class III: models for which an underlying mimetic Lagrangian exists}\label{sec:ClassIII}
The last class of polymerized theories we want to consider is defined by the property of having an underlying extended mimetic Lagrangian \cite{Chamseddine:2016uef,BenAchour:2017ivq,Langlois:2018jdg,Han:2022rsx}. Here we will only very briefly discuss the Lagrangian in the case of spherically symmetric models and refer the reader to  \cite{Chamseddine:2016uef,BenAchour:2017ivq,Langlois:2018jdg,Han:2022rsx} for more details and to 
the appendix \ref{app:MimeticIntro} for a brief summary of the relevant formulas needed in this work.
~\\
The extended mimetic model is given by the following higher order Lagrangian
\begin{align}
\label{GA}
S[g_{\mu\nu},\phi,\lambda] =\frac{1}{8\pi G} \int_{\mathscr{M}_4} \rmd^4x \, \sqrt{-g} \, \left[ \frac{1}{2} \, {\mathcal{R}}^{(4)} + 
\, L_\phi(\phi,\chi_1,\cdots,\chi_n) \, + \, \frac{1}{2}\lambda(\phi_\mu\phi^\mu + 1)\right],
\end{align}
where we have an additional so-called mimetic field $\phi$ as well as the field $\lambda$ that acts after reduction of the corresponding primary constraint as a Lagrangian multiplier for the mimetic condition.
The higher order contributions are built from second order derivatives of $\phi$ and are encoded in $\chi_n$ with
\begin{align}
\label{chin}
\chi_n \equiv 
\operatorname{Tr}\left([\phi]^n\right) \equiv\sum_{\mu_1,\cdots,\mu_n} \phi_{\mu_1}^{\mu_2} \, \phi_{\mu_2}^{\mu_3} \cdots \phi_{\mu_{n-1}}^{\mu_n} \, \phi_{\mu_n}^{\mu_1}\,,\quad\phi_\mu=\nabla_\mu\phi\,,\quad  [\phi]_{\mu \nu} \equiv \phi_{\mu \nu}=\nabla_\mu\nabla_\nu\phi\,.
\end{align}
An analysis of this action shows that the theory only propagates $2+1$ degrees of freedom. 
In the spherically symmetric case it is sufficient to consider a mimetic potential of the form $L_\phi(\chi_1,\chi_2)$. We can introduce $\psi=\frac{1}{2}\ln(E^x)$, to write the action of the extended mimetic model as a two-dimensional (2D) action of the form
\begin{eqnarray}\label{eq:cov_action}
    S_2&=&\frac{1}{4G}\int_{\mathscr{M}_2} \rmd^2 x\, \det(e) \big\{ e^{2\psi}\left(R_{h}+2h^{ij}\partial_{i}\psi\partial_{j}\psi\right)+2 
    +\, e^{2\psi}\big[ L_{\phi}\left(X,Y\right)+\frac{1}{2}\lambda\left(\partial_j \phi \partial^j \phi+1\right)\big]\big\}\,, \nonumber
\end{eqnarray}
where $h_{ij}$ is the 2D metric, $\det(e)$ denotes the determinant of the 2D triad fields and $R_h$ is 2D scalar curvature. $L_{\phi}\left(X,Y\right)$ is again the mimetic potential and the quantities $X,Y$ consist of couplings between the metric $h_{ij}$ and derivatives of $\phi,\psi$ and are given by 
\be\label{eq:choiceofXandY}
X = - \Box_{h}\phi + Y \ , \quad  Y = - h^{ij}\partial_{i}\psi\partial_{j}\phi\ .
\ee
$L_{\phi}\left(X,Y\right)$ is actually defined on the cover space of $\chi_1,\chi_2$ as shown in (\ref{lift1}-\ref{lift2}). The four dimensional version $L_\phi(\chi_1,\chi_2)$ can be recovered from $L_{\phi}\left(X,Y\right)$ by choosing a lift, e.g. the lift with the branch \eqref{lift2} asymptotically such that it is compatible with the Schwarzschild geometry.
A choice of a specific mimetic potential can be carried over to certain choices of polymerization functions provided they satisfy certain assumptions as discussed below. Although a given choice of polymerization functions might not have a unique relation to some mimetic potential, it is very useful in the investigation of polymerized models to establish such a link because it provides the gauge-unfixed covariant extension for an effective model formulated in the canonical formalism by means of a Hamiltonian with respect to some choice of time. In the case in which the mimetic potential is absent, i.e. $L_{\phi}\left(X,Y\right)=0$, the action reduces to the standard action of general relativity coupled to some non-rotational dust field $\phi$ \cite{Brown:1994py} specialized to spherically symmetry. This suggests that the mimetic field can be used as a dynamical reference field for the Hamiltonian constraint \cite{Han:2022rsx}
and this will exactly be done in the work here. In case of a non-vanishing mimetic potential such a choice does however no longer corresponds to a dynamical reference field similar to non-rotational dust since the mimetic potential $L_{\phi}\left(X,Y\right)$ involves non-trivial higher derivative couplings between the mimetic field and the geometry. This can be summarized in the following remark
\begin{remark}
The mimetic field $\phi$ plays a twofold role: it is the dust clock that we use to deparametrize the theory, as well as a scalar field whose higher order derivatives introduce the polymerization effects, i.e. the quantum gravity corrections. In the classical limit, the mimetic condition $\frac{1}{2}\lambda(\partial_{\mu} 
\phi \partial_{\mu} \phi + 1)$ in the Lagrangian \eqref{GA} is nothing else but the non-rotational dust Lagrangian.   
\end{remark}
~\\
The equations of motion of the mimetic model are given by the pendant of Einstein's equations and the mimetic condition, which can be written as
\begin{eqnarray}\label{eq:cov_eq}
   G^{(\alpha)}_{\mu\nu} := G_{\mu\nu}- T^{\phi}_{\mu\nu}=-\lambda\partial_{\mu} \phi\partial_{\nu} \phi\,, \quad \partial_{\mu} \phi \partial^{\mu} \phi = -1 \,.
\end{eqnarray}
The right hand side of the modified Einstein equation is the energy momentum tensor of non-rotational dust, where $\lambda$ plays the role of the dust energy density. We call the case $\lambda = 0$ polymerized vacuum. One can check that this is in the dust time gauge equivalent to a vanishing effective energy density $C^{(\alpha)} = 0$. However in general the polymerized vacuum will deviate from the classical theory because compared to general relativity in spherically symmetry we still have a non-trivial mimetic field $\phi$ with some higher order derivative coupling which leads to a non-trivial $T^{\phi}_{\mu\nu}$ and thus we keep the quantum gravity or polymerization effects present in the effective model that corresponds to this extended mimetic model. 

~\\
~\\
Based on the works \cite{Chamseddine:2016uef,BenAchour:2017ivq,Langlois:2018jdg,Han:2022rsx} we can relate extended mimetic Lagrangians to effective models if the polymerization functions are compatible with a $\overline{\mu}$-scheme.  This means in terms of the polymerization function defined in \eqref{eq:defpolyhamiltonianconstraint} that they can be written in the form\footnote{In the literature polymerization functions are typically not defined in the "factorized" way as it is done in \eqref{eq:defpolyhamiltonianconstraint}, however since the classical Hamiltonian constraint already naturally obeys to the $\overline{\mu}$-scheme these definitions of the $\overline{\mu}$-scheme are equivalent.}
\begin{equation}\label{eq:mubarscheme}
    f(K_x,K_\phi,E^x,E^\phi)=f\big(\tfrac{\sqrt{E^x}}{E^\phi}K_x,b\big) ,\quad b:= \frac{K_{\phi}}{\sqrt{E^x}}.
\end{equation}
An additional requirement is that the inverse triad corrections are trivial, i.e. $h_1=h_2=1$. Moreover we can consider intersections with the two other classes of effective models described in the text above by first introducing the polymerization functions $\tilde{f}^{(1)},\tilde{f}^{(2)}$ which are related to the more general expressions defined in \eqref{eq:no_kx_f1} and \eqref{eq:defpolyhamiltonianconstraintwithoutKx} by the equations
\begin{eqnarray}\label{eq:mubar_mimetic_fs}
    f^{(1)}(K_{\phi},E^x) = {E^x} \tilde{f}^{(1)} (b) ,\quad f^{(2)}(K_{\phi},E^x) = \sqrt{E^x} \tilde{f}^{(2)} (b)\,.
\end{eqnarray}
In other words this implements a $\overline{\mu}$-scheme where only the $K_\phi$ extrinsic curvature component is polymerized. As we saw beforehand, only polymerizing $K_\phi$ is a necessary restriction in order to have $C^{(\alpha)}(x)$ as a conserved quantities. Analogously defining ${F}(K_{\phi}, E^x) =  E^x \tilde{F}\left(b\right)$ such that the equation $\tilde{F}{}'\left(b\right) = 2 \tilde{f}^{(2)}\left(b\right)$ holds, we can then write the pendant of condition \eqref{eq:closurebutwith} as
\begin{eqnarray}\label{eq:f1_and_F}
    \tilde{f}^{(1)} + b \tilde{F}' -3 \tilde{F} =0
\end{eqnarray}
in the mimetic sector. In this case, the corresponding mimetic potential is given by 
\begin{eqnarray}\label{eq:mimetic_L_from_F}
    L_{\phi}(X,Y) =2 X Y - Y^2 + \tilde{f}^{(1)} \qty[(\tilde{f}^{(2)})^{-1}\qty(-{Y})] + 2 X (\tilde{f}^{(2)})^{-1}\qty(-{Y}) .
\end{eqnarray}
In case $f^{(2)}$ is not a monotonic function where $(f^{(2)})^{-1}$ is not unique, this implies that the function  $L_{\phi}$ is actually defined in the cover space of $Y$. 
Due to the trivial inverse triad corrections, the only allowed compatible effective LTB condition in this subsector is according to equation \eqref{eq:closurebutwith} the classical one, i.e. $ {g}_{(\alpha)}=1$. In the more general case without having $C^{(\alpha)}(x)$ as conserved quantities, we can also have other compatible effective LTB conditions besides the classical one. The relevant equation is again \eqref{eq:ltb_con_to_f1_f2}, where the polymerization functions of the mimetic sector \eqref{eq:mubar_mimetic_fs} are inserted on the right-hand side.

\subsubsection{Brief summary of the properties of the classes} 
\begin{figure}[H]
\begin{center}
    \begin{tikzpicture}[thick,
                	set/.style = {circle,
                		minimum size = 3.2cm,
                		fill=MidnightBlue!50}]

                \begin{scope}
                \fill[Black!7](0,-0.23) ellipse(4.6cm and 3.0cm);
                    \fill[Dandelion!60](-0.9,0.55) circle(1.9cm);
                    \fill[RoyalBlue!37](0.9,0.55) circle(1.9cm);
                
                \end{scope}

                \begin{scope}
                    \clip (-0.5,0.55) circle(1.3cm);
                    \clip (-0.8,0.55) circle(1.9cm);
                    \fill[WildStrawberry!65](-0.9,0.55) circle(1.9cm);
                \end{scope}

                \begin{scope}
                	\clip (-0.9,0.55) circle(1.9cm);
                	\clip (0.9,0.55) circle(1.9cm);
                	\fill[ForestGreen!60](0.9,0.55) circle(1.9cm);
                \end{scope}

                \begin{scope}
                    \clip (-0.5,0.55) circle(1.3cm);
                    \clip (0.9,0.55) circle(1.9cm);
                    \fill[Purple!75](-0.9,0.55) circle(1.9cm);
                \end{scope}

                \draw (0,-0.23) ellipse (4.6cm and 3.0cm);
                \draw (-0.5,0.55) circle(1.3cm);
                \draw (-0.9,0.55) circle(1.9cm);
                \draw (0.9,0.55) circle(1.9cm);
               
                \node at (-2.32,0.55) {$\RN{1}$};
                \node at (1.85,0.55) {$\RN{3}$};
                
                \node at (-0.09,0.55) {$ \RN{2}\cap \RN{3}$};
                \node at (-1.4,0.55) {$\RN{2}$};
                \node at (0,-2.0) {\large{theory space of effective models}};
\end{tikzpicture}
\end{center}
   \caption{Classification of the theory space of effective models: class \RN{1} is the space of polymerized theories that admit to a compatible polymerized LTB condition, which means \eqref{eq:no_kx_f1} and \eqref{eq:ltb_con_to_f1_f2} have to hold, and class \RN{2} are effective models that in addition have a polymerized vacuum solution, which imposes condition \eqref{eq:final_con_clo}. Finally class \RN{3} are those effective models that have an extended mimetic theory as an underlying Lagrangian. In this case no inverse triad corrections are allowed and the polymerization has to be compatible with the $\overline{\mu}$-scheme, see \eqref{eq:mubarscheme}.}
       \label{fig:setsLagrangian_poly_mimetic}
\end{figure}
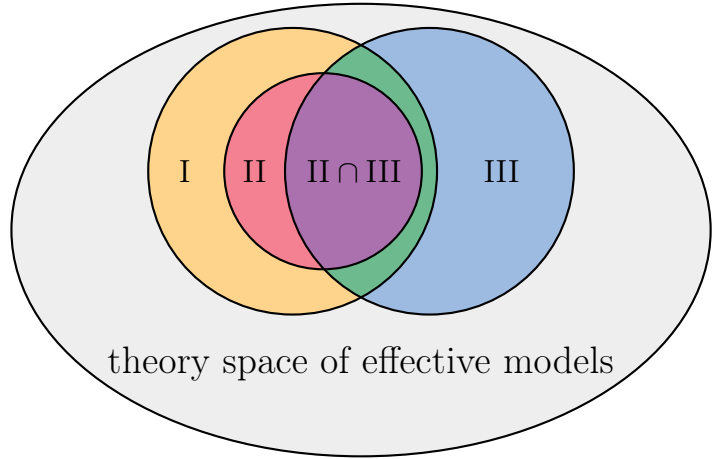

To summarize all the above presentation in a concise manner we want to refer at this point to  table \ref{tab:Classes} and Fig. \ref{fig:setsLagrangian_poly_mimetic}.  Of all possible effective models relevant for this work we are mostly interested  on one hand in those which have a compatible effective LTB condition (class I) and on the other hand in those which have an underlying extended mimetic Lagrangian (class III). Focusing on the table below this means that in order to have a compatible effective LTB conditions analogously to the classical case we have to restrict to polymerization functions only in the $K_\phi$ curvature component which can be written as \eqref{eq:no_kx_f1} and furthermore the differential equation found in \eqref{eq:ltb_con_to_f1_f2} between the polymerization functions and effective LTB condition has to hold. Specializing to the case that also the energy density $C^{(\alpha)}$ is conserved in the effective LTB sector we can define class II. This property poses the condition \eqref{eq:final_con_clo} on the form of the polymerization functions and the effective compatible LTB condition. On the other hand, we are not allowed to have any inverse triad corrections and also compelled to have a $\overline{\mu}$-scheme polymerization according to \eqref{eq:mubarscheme} in the mimetic sector.
Fig. \ref{fig:setsLagrangian_poly_mimetic} shows that the classes are not disconnected from each other. For instance models with polymerized vacuum solutions are naturally a subset of the models of class I denoted as class II in Fig. \ref{fig:setsLagrangian_poly_mimetic}. If we further consider in this subset only those polymerized vacuum solutions where inverse triad corrections are absent, then these models are also a subset of class III and are denoted by ${\rm II}\cap {\rm III}$ in Fig. \ref{fig:setsLagrangian_poly_mimetic}. Furthermore, the green region in Fig. \ref{fig:setsLagrangian_poly_mimetic} includes those effective models that have a compatible effective LTB condition, have no inverse triad corrections and the polymerization functions are compatible with the $\overline{\mu}$-scheme but have no polymerized vacuum solutions.

\begin{table}[H]
\begin{tabular}{c!{\vrule width 1.6pt}c}
\centering
\cellcolor{gray!23} Existence of  &  \cellcolor{gray!23} Conditions \\\noalign{\hrule height 1.6pt}
 \cellcolor{Dandelion!55} & \\[-2.3em]
 \cellcolor{Dandelion!55}\bgroup\def\arraystretch{1.05}\begin{tabular}{@{} c @{}}compatible LTB condition\\ $\widetilde{g}_{(\alpha)} =  {g}_{(\alpha)}\,\ltbf$ (class I)\end{tabular}\egroup & \bgroup \def\arraystretch{1.05}\begin{tabular}{@{} c @{}}
$f = \frac{{f}^{(1)}(K_{\phi},E^x)- {f}^{(2)}(K_{\phi},E^x) K_{\phi}}{(K_{\phi}+4 {E^x} \widetilde{K}_x) K_{\phi}} + \frac{{f}^{(2)}(K_{\phi},E^x)}{ K_{\phi}} -1$ \\
${g}_{(\alpha)}(K_{\phi},E^x) =  {g}_{(\alpha)}(E^x) \, ,\qquad
    1-\frac{2{E^x} \partial_{E^x}  {g}_{(\alpha)}}{ {g}_{(\alpha)}} =\frac{-4 {E^x} \partial_{E^x}{f}^{(2)}+  \partial_{K_{\phi}}{f}^{(1)}}{2 {f}^{(2)}} $
\end{tabular}\egroup    \\ 
     \cellcolor{Dandelion!55} & \\[-2.3em]\hline
 \cellcolor{WildStrawberry!47} \bgroup
\def\arraystretch{1.05}\begin{tabular}{@{} c @{}}compatible LTB condition and \\ conserved energy density (class II)\end{tabular} \egroup   & class I conditions \&   \scalebox{0.96}{$\frac{h_1 - 2 E^x \partial_{E^x} h_2}{h_2} = 1 - \frac{2{E^x} \partial_{E^x}  {g}_{(\alpha)}}{ {g}_{(\alpha)}} $}  \\\hline 
 
\cellcolor{RoyalBlue!44}  mimetic Lagrangian (class III) & $h_1=h_2=1$, $\overline{\mu}$-scheme: $f(K_x,K_\phi,E^x,E^\phi)=f\big(\tfrac{\sqrt{E^x}}{E^\phi}K_x,\tfrac{1}{\sqrt{E^x}}K_\phi\big)$\\ 
\end{tabular}
\caption{This table summarizes some of the results presented in the main text. On the left hand side the the different classes are listed and on the right hand side the corresponding conditions. For class I, i.e. having a compatible effective LTB condition, the necessary conditions are \eqref{eq:no_kx_f1} and \eqref{eq:ltb_con_to_f1_f2}, additionally having a conserved energy (class II) further imposes \eqref{eq:final_con_clo}. To have an underlying mimetic Lagrangian (class III) no inverse triad corrections are allowed and \eqref{eq:mubarscheme} has to hold.}
\label{tab:Classes}
\end{table}
\subsection{Effective dynamics, general solution and Birkhoff-like theorem}
\label{sec:BirkhofflikeThm}
\begin{figure}[h!]
    \centering
    \includegraphics[width=0.465\textwidth]{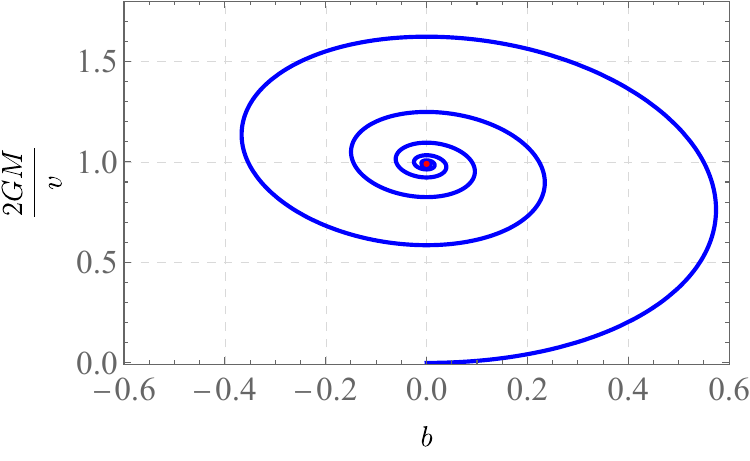}
    \includegraphics[width=0.45\textwidth]{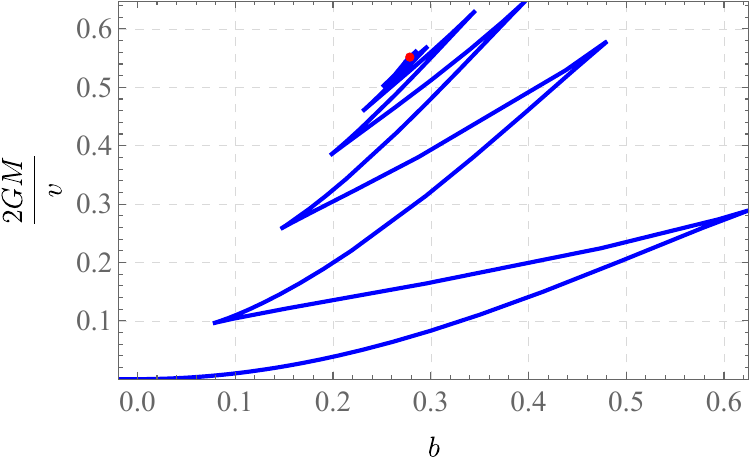}
    \includegraphics[width=0.45\textwidth]{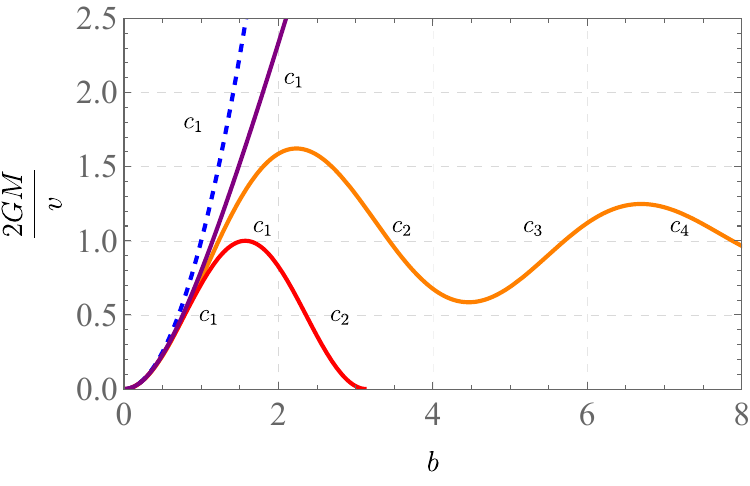}
    \caption{The possible examples of the phase space trajectories. In the upper two plots we show examples with an attractor (red points), while the lower plots are examples without attractors. The examples in the upper plots can happen in class II, however not in the intersection $\RN{2} \cap \RN{3}$. The blue dashed line marks the classical trajectory. As can be seen from the third plot, all effective models have the correct classical limit when $b$ tends to zero. The upper left plot is generated with $b(z)=\frac{5}{3} \sqrt{\frac{2}{11}} e^{-\frac{\sqrt{\frac{398}{11}}}{45 z}} \sin \left(\frac{\sqrt{22}}{5 z}\right),\frac{2GM}{v}(z)=e^{-\frac{\sqrt{\frac{398}{11}}}{45 z}} \cos \left(\frac{\sqrt{22}}{5 z}+\cos ^{-1}\left(-\frac{99}{100}\right)\right)+\frac{99}{100}$; The upper right plot is generated with the solution of the model presented in \cite{Han:2022rsx} with $\alpha=\alpha_{\Delta}=1,r_s=8 $; The bottom plot is generated with: red: $\tilde{F}=\sin^2(b)$, orange: $\tilde{F}(b)=e^{-\frac{1}{15} \sqrt{\frac{199}{22}} b} \cos \left(\frac{3}{5} \sqrt{\frac{11}{2}} b+\cos ^{-1}\left(-\frac{99}{100}\right)\right)+\frac{99}{100}$, purple: Hayward model \eqref{eq:F_Hayward} with $\alpha=1$.
    }
    \label{fig:phasespace}
\end{figure}

After the classification of effective models was discussed in Subsec. \ref{sec:classification}, we want to focus now on the marginally bound case and effective models in class II, i.e. effective models which have a compatible LTB condition and a conserved energy density $C^{(\alpha)}$. An interesting subclass therein is given by polymerized vacuum solutions with vanishing energy density. Note that the conservation of the energy density plays a twofold pivotal role for the polymerized vacuum solution here. On the one hand a conserved energy density is crucial because otherwise a polymerized vacuum solution would not be preserved under the effective dynamics. On the other hand similar to the classical case a conserved energy density does allow us to express the extrinsic curvature $K_\phi$ as a function of the triad $E^x$ which was a necessary step in the classical model to obtain a closed differential equation for $E^x$ that can be solved by simple integration. Although this step is in general more complicated for effective models, we will show that for the solutions of the modified Friedmann equations in class II, the steps performed in the classical model can be generalized to the effective models, and in the case of polymerized vacuum solutions, a Birkhoff-like theorem can be formulated, which we present in Subsec. \ref{sec:Birkhoff_eff}.
~\\
~\\
To further discuss the different types of effective models, it is convenient for our purpose to consider their trajectories in phase space, as this allows us to understand the main features of each type of effective model in a simple way. Moreover, it allows us to derive the general solution of the modified Friedmann equations for all effective models in class II. As already pointed out in section \ref{sec:ClassII} working in LTB coordinates and with respect to the comoving dust time, the dynamics of the effective models we consider are completely decoupled along the radial coordinate. 
In the marginally bound case, the effective dynamics are then according to \eqref{eq:Ham_non_marginal} completely determined by the following Hamiltonian $\widetilde{H}^{(\alpha)}(x)$ 
\begin{align}\label{eq:Ham_for_x}
    -M(x) = \widetilde{H}^{(\alpha)}(x) \coloneqq \frac{1}{2G}\qty[\frac{\sqrt{E^x}}{ {g}_{(\alpha)} } \left( - F+ h_2 \left(  {g}_{(\alpha)}^2  -1 \right) \right)](x)\,,
\end{align}
where similar to the classical case we denote the conserved quantity by $M(x)$. The Poisson bracket of the effective model for each radial coordinate $x$ is given by $\{K_{\phi}(x), E^x(x) \} = -G$. The corresponding equations of motion are those shown in \eqref{eq:eff_dy_new}. As can be seen from \eqref{eq:Ham_non_marginal}, we have $C^{(\alpha)}\eval_{LTB} = \partial_x \widetilde{H}^{(\alpha)}(x)$ and thus even if $M(x)$ is set to a non-vanishing constant, $C^{(\alpha)}\eval_{LTB}$ is vanishing as required for a polymerized vacuum solution. It is only important that the conserved quantities are set to the same constant value for all $x$ to have an polymerized vacuum solution.\\
 As discussed in our former work \cite{Giesel:2023tsj}, given the effective Hamiltonian \eqref{eq:Ham_for_x} of a decoupled model in class I according to the classification of effective models in Subsec. \ref{sec:ClassI}, one can reconstruct the underlying effective spherically symmetric model. As a special case this also includes polymerized vacuum solutions. Note that knowing the function $F$ which is related to the polymerization functions, the inverse triad corrections $h_2$ and the compatible effective LTB condition ${g}_{(\alpha)}$ in \eqref{eq:Ham_for_x} uniquely determines the underlying spherically symmetric model, since the conditions in \eqref{eq:ltb_con_to_f1_f2} and \eqref{eq:closurebutwith} will then fix $h_1$ and $f^{(1)}$ for given $F, h_2$ and ${g}_{(\alpha)}$. As a consequence the (unreduced) energy density $C^{(\alpha)}$ in \eqref{eq:defpolyhamiltonianconstraintwithoutKx} can be uniquely obtained. However, if we only know the form of $\widetilde{H}^{(\alpha)}(x) $ in the marginally bound case without the explicit form of $h_2$ and $ {g}_{(\alpha)}$, we cannot uniquely determine the corresponding Hamiltonian in the general non-marginally bound case \eqref{eq:Ham_non_marginal} which requires an explicit knowledge of $h_2$ and $ {g}_{(\alpha)}$. 
~\\ ~\\
If we compare the situation of the solution to the effective Friedmann equations with the classical solution, there are two aspects that become more complicated. First, it becomes more difficult to solve for $K_\phi$ if we consider the effective Hamiltonian $\widetilde{H}^{(\alpha)}(x)=-M(x)$ in \eqref{eq:Ham_for_x} and second, comparing the equation for $\partial_t E^x$ in \eqref{eq:eff_dy_new} with the equation for the classical model in \eqref{eq:dy_new}, we find that depending on the form of $f^{(2)}$, this equation also has a more complicated structure for generic effective models. To make progress we take advantage of the fact that $\widetilde{H}^{(\alpha)}(x)=-M(x)$ is a conserved quantity for each given $x$ and
will consider level sets of the effective Hamiltonian $\widetilde{H}^{(\alpha)}(x)$ for a given $x$. Hence, we can choose a value for $M(x)$ and consider the corresponding level sets in phase space as well as the corresponding phase space trajectories $c(K_{\phi},E^x):= \widetilde{H}^{(\alpha)}(x)=-M(x)$ defined by \eqref{eq:Ham_for_x}. However, since we only fix the asymptotic behavior of the system at $K_{\phi} \to 0$, the remaining part of the trajectory in phase space is determined by the polymerization functions $F$, the inverse triad corrections $h_2$ and the effective LTB condition  ${g}_{(\alpha)}$. Before we discuss some generic examples for phase space trajectories that effective models can have in Fig. \ref{fig:phasespace}, we briefly analyze the classical case. Considering the classical Hamiltonian in \eqref{eq:LTBreducedHami} and comparing it with \eqref{eq:Ham_for_x} we can directly read off $F_{\rm class}=K_\phi^2$ and $g_{(\alpha)}=1$. Thus, for a fixed value of $M(x)$ we can use \eqref{eq:Ham_for_x} to easily solve for $K_\phi$. The result is exactly \eqref{eq:KphiClass}. Since $F_{\rm class}$ is a monotonic function of $K_\phi$ for $K_\phi\geq 0$, a unique inverse function exists in this interval. Alternatively, instead of working with the variables $(K_\phi, E^x)$ we can use the variables $(b:=\frac{K_\phi}{\sqrt{E^x}},v:=(E^x)^{\frac{3}{2}})$, that will be useful for the particular examples discussed later, and consider the phase space trajectories $c(b,v)$. In the classical case we then have $\widetilde{H}(x)=-M(x)=-\frac{1}{2G}v\tilde{F}_{\rm class}(b)$ with $F_{\rm class}=E^x\tilde{F}_{\rm class}(b)$. Hence we can write $\tilde{F}_{\rm class}(b)=\frac{2GM(x)}{v}$ which is is plotted in the third plot in Fig. \ref{fig:phasespace} as a dashed blue line, where the phase space trajectories are plotted in the $b-\frac{2GM(x)}{v}$ plane. The plots in Fig. \ref{fig:phasespace} also include some more generic examples of phase space trajectories for effective models. Although not all of these trajectories might be physically favored, it turns out that already in the effective model with a symmetric bounce where no inverse triad corrections are present and $g_{(\alpha)}=1$ the phase space trajectory is not a monotonic function of $K_\phi$ and $b$ respectively. In fact as can be seen from the third plot in Fig. \ref{fig:phasespace}, we have $\tilde{F}(b)=\frac{\sin^2(\alpha_\Delta b)}{(\alpha_\Delta)^2}$ with $\alpha_\Delta:=\gamma\sqrt{\Delta}$ with $\gamma$ the Barbero-Immirzi parameter and $\Delta:=4\pi l_p$. As a consequence, the inverse function can only be constructed for each individual monotonic segments. However, when the Hamiltonian is analytic we will see that already from one monotonic segment we are able to get the full solution in LTB coordinates by analytical continuation.

\noindent Now in order to obtain the general solution of the modified Friedmann equations for an effective models and discuss the different types of solutions, we aim to solve \eqref{eq:Ham_for_x} for $K_{\phi}$  in terms of $E^x$ similar to the classical case. In contrast to the classical theory expressing $K_{\phi}$ as a function of $E^x$ for a given trajectory $c(K_{\phi},E^x)$ can only be done individually for each monotonic segment, we write the whole trajectory as the composition of the individual monotonic segments  $c = c_{1} \circ c_{2} \circ \cdots$. For each segment $c^{i}$, we have a one-to-one correspondence between $K_{\phi}$ and $E^x$ due to \eqref{eq:Ham_for_x}
\begin{align}
    K_{\phi}(t,x) = F^{-1}_{(i)} \left[ h_2 \left(  {g}_{(\alpha)}^2  -1 \right) -  2GM(x) \frac{ {g}_{(\alpha)} }{\sqrt{E^x}} , E^x \right] =: F^{-1}_{(i),M(x)}\left[E^X \right] \, .
\end{align}
Note that this procedure is possible since we work with effective models in class II that is a subset of class I and here $g_{(\alpha)}$ is a function of $E^x$ only. Similar to what happens in the classical theory, we can now substitute this into the equations of motion for $E^x$ in \eqref{eq:eff_dy_new} to obtain the modified Friedmann equation
\begin{align}\label{eq:eff_dy_new_rewritten}
    \frac{\partial_t E^x}{2 \sqrt{E^x} }(t,x) = - {f}^{(2)} \left[F^{-1}_{(i),M(x)}\left[E^X \right] , E^X \right] .
\end{align}
Integrating this equation over the temporal coordinate $t$ leads to the general solution
\begin{align}\label{eq:marginal_sol}
    E^x_{(i)}(t,x) = \mathcal{F}^{-1}_{(i)}(s(x) - t)\, ,
\end{align}
where $\mathcal{F}$ is a function given by
\begin{align}\label{eq:F_i}
    \mathcal{F}_{(i)} =  \int \frac{d E^x }{2 \sqrt{E^x} {f}^{(2)} \left[F^{-1}_{(i),M(x)}\left[E^x \right] , E^x \right] }\,.
\end{align}
The involved integration constant $s(x)$ is an arbitrary function of $x$. Hence, we can construct the solution for $E^x_{(i)}(t,x)$ of the modified Friedmann equation for each monotonic segment separately and the entire solution can be subsequently obtained by composing the individual ones $E^X= E^x_{(1)}(t,x)\circ  E^x_{(2)}(t,x)\circ \cdots$.
\begin{remark}
    The continuity of the solution $E^x_{(i)}(t,x)$ from the interval $i$ to $i+1$ can be ensured by a suitable choice of the integration constant $s(x)$. The modified Friedmann equation \eqref{eq:eff_dy_new_rewritten} implies that the continuity property of $\partial_t E^x$ in \eqref{eq:eff_dy_new} is determined by the continuity property of $F$ as $\partial_t E^x$ is $C^{n}$ if $F$ is $C^{n+1}$. 
\end{remark}
\noindent An important subclass of effective models is the intersection of class $\text{II} \cap \text{III}$, the latter are effective models that have an underlying mimetic Lagrangian and also admit a classical LTB condition. Since there are no inverse triad corrections in this case and the polymerization only affects the extrinsic curvature component $K_\phi$ and is performed using the $\overline{\mu}$-scheme, the polymerization takes the form of \eqref{eq:mubar_mimetic_fs}. In this subsector the effective decoupled cosmological Hamiltonians reduce to
\begin{eqnarray}\label{eq:ham_in_mimetic}
    -M(x) = \widetilde{H}^{(\alpha)}(x) = -\frac{1}{2G}\qty[v \tilde{F}(b) ](x)\,, \quad v:= (E^x)^{\frac{3}{2}}\,,
\end{eqnarray}
which means that the dependence on $v$ and $b$ completely factorizes in the Hamiltonians. As a result we cannot have trajectories with an attractor. The trajectories can only have multiple branches when $\tilde{F}$ is not monotonic, e.g., case three in Fig. \ref{fig:phasespace}. Moreover the trajectory in the $b$ -- $(\tilde{F}=2GM/v)$ plane is unique and independent of the integration constant $M$. The same happens also in the classical model where the Hamiltonian factorizes as can be seen in \eqref{eq:LTBreducedHami}.
~\\
~\\
The brief investigation of phase space trajectories for different effective models in Fig. \ref{fig:phasespace} also shows that models with unbounded and bounded phase space trajectories have very different properties. In the classical case we have an unbounded, monotonically increasing trajectory that can be identified with a polymerization function in the classical limit. Here the corresponding inverse function can be determined for all $b\geq 0$. The classical model has a singularity and no bounce. For the usual polymerizations used in LQC, the polymerization function and thus the phase space trajectory is bounded, not completely monotonic but analytical. For these models, the more elaborate procedure presented above must be applied and this will show that these models have a bounce and no central singularity. Moreover, there are unbounded phase space trajectories, that again can be related to unbounded polymerization functions, belonging to regular black hole solutions, which also do not contain a central singularity, but also no bounce. We will discuss their properties in detail in section \ref{sec:examples}, where we apply the framework presented here to specific examples of LQC-inspired polymerizations as well as to regular black hole solutions such as those of Bardeen \cite{Bardeen68} and Hayward \cite{Hayward:2005gi}.

\subsubsection{A Birkhoff-like theorem for polymerized vacuum solutions in class II}
\label{sec:Birkhoff_eff}
With the results obtained in the last section, we can proceed and specialize to the polymerized vacuum solutions and discuss their properties such as for instance their uniqueness. Similar to the case in general relativity, when we set $M(x) = m$, the $\mathcal{F}_{(i)}$ in \eqref{eq:F_i} will not contain an explicit dependence on $x$. Thus then the solution of $E^x$  in \eqref{eq:marginal_sol} has the symmetry $E^X = E^x(z)$ with $z := s(x)-t$. In such case it is straightforward to see that $s(x)$ is nothing else but the residual diffeomorphism that behaves as a redefinition of the $x$ coordinate, where w.l.g we can set $s(x)=x$. The fact the solution $E^x$ is only a function of $z$ implies that we have the Killing vector field $\partial_t +\partial_{x}=0$. This results in the polymerized vacuum solution for each segment $c_i$, which is clearly stationary. The result can be summarized in the lemma below as follows: 

\begin{lemma}\label{thm:birkhoff-like}
    (Birkhoff-like theorem) Effective models belonging to class II according to the classification in Subsec. \ref{sec:ClassII} admit a unique asymptotically flat polymerized vacuum solution characterized by a mass $m$ that is stationary. The solution is a continuous function and can be expressed as a piecewise function in LTB coordinates.
\end{lemma}
~\\
We note that when $F$ is analytic for both $E^x$ and $K_\phi$ or $v$ and $b$ for the trajectory $c(K_\phi,E^x)$ or $c(b,v)$ respectively, the different piecewise solutions $E^x_{(i)}(z)$ are related to each other by analytic continuation.
~\\
Making use of the Killing symmetry $\partial_t +\partial_{x}=0$ and the LTB condition \eqref{eq:LTBconditioneffective} we can write the polymerized vacuum metric in a similar from found in \eqref{eq:metricSphSymm} as
\begin{align}\label{eq:vacuum_metric_1}
    \mathrm{d}s^2 &= -\mathrm{d}t^2 + \frac{(\partial_x E^x)^2}{4  {g}_{(\alpha)}^2 {E^x}} \mathrm{d}x^2 + {E^x} \mathrm{d} \Omega^2=-\mathrm{d}t^2 + \mathcal{G}(z)^2 \mathrm{d}x^2 + {E^x}(z) \mathrm{d} \Omega^2 ,
\end{align}
where as expected the effective LTB condition ${g}_{(\alpha)}$ enters explicitly into the metric. Here we introduced
\begin{align}\label{eq:Gr_def}
        \mathcal{G}(z)^2&\coloneqq\frac{(\partial_x E^x)^2}{4  {g}_{(\alpha)}^2 {E^x}}(z)\,,\quad{\rm with}\quad z=x-t.
\end{align}
 At this point it should be mentioned that $\mathcal{G}(z)$ is generally a continuous function of $z$. Next let us define the coordinate transformation
\begin{eqnarray}
    && (t,x) \to (\tau,z=x-t), \quad \tau =  \tilde{\tau}(z)-t, \qquad \tilde{\tau}(z):=\int_{z_0}^z \mathrm{d}z' \frac{ \mathcal{G}(z')^2}{1- \mathcal{G}(z')^2 }. \nonumber
\end{eqnarray}
This enables us to transform the metric in \eqref{eq:vacuum_metric_1} to the form
\begin{eqnarray}
\label{metrictz1}
 \mathrm{d} s^2 = - ( 1-\mathcal{G}(z)^2) \mathrm{d} \tau^2 +\frac{\mathcal{G}(z)^2}{ 1-\mathcal{G}(z)^2} \mathrm{d}z^2 + E^x(z) \mathrm{d} \Omega^2\, .
\end{eqnarray}
From this form it is clear that the metric is asymptotically static. The trapped or anti-trapped surface is given by $1-\mathcal{G}(z)^2 =0$ and we have a static solution in the untrapped region where $1-\mathcal{G}(z)^2 >0$ and a homogeneous solution in the trapped or anti-trapped region where $1-\mathcal{G}(z)^2 <0$. This result agrees with the result calculated from the null expansion in the LTB coordinates, e.g. in \cite{Giesel:2022rxi}. As a result, the polymerized vacuum solution is static in the untrapped region and homogeneous in the trapped or anti-trapped region.
\subsection{Schwarzschild-like coordinates for polymerized vacuum solutions in class II and uniqueness}\label{sec:sub_schwarzschildcoords}
In this subsection we want to concentrate on the polymerized vacuum solution which we so far presented in LTB coordinates only. As a first step we want to explore its appearance in Schwarzschild-like coordinates and address the question of uniqueness also for these coordinates. Afterwards we will show that the formalism introduced in this work provides the possibility to reconstruct from a given effective metric in Schwarzschild-like coordinates the corresponding effective decoupled Hamiltonians $\widetilde{H}^{(\alpha)}$ and further the spherical symmetric Hamiltonian $C^{(\alpha)}$ from a given metric. This reconstruction procedure also enables us to embed regular black hole solutions like the one of Bardeen \cite{Bardeen68} and Hayward \cite{Hayward:2005gi} into our formalism as will be discussed in Sec. \ref{sec:examples}.
~\\
~\\
To obtain the metric in Schwarzschild-like coordinates we start from metric in the form of \eqref{metrictz1} and further define the physical radius as radial coordinate variable $r = \sqrt{E^x}(z)$. While the metric \eqref{metrictz1} is well defined for all $z$ in the domain of $\mathcal{G}(z)$, the coordinate transformation $r = \sqrt{E^x}(z)$ is only well defined for each piecewise segment $c_i$, which is also a piecewise monotonic segment of the solution $E^x$ in $t$. Moreover, for each piecewise segment $c_i$ we may in general have a different form of $\mathcal{G}_{(i)}(r)^2$ as a function of $r$. Therefore, the line element can be finally written as
\begin{eqnarray}
\label{metrictz2}
 \mathrm{d} s^2 = - ( 1-\mathcal{G}_{(i)}(r)^2) \mathrm{d} \tau^2 +\frac{1}{  {g}_{(\alpha)}(r)^2 \left( 1-\mathcal{G}_{(i)}(r)^2 \right)} \mathrm{d}r^2 + r^2 \mathrm{d} \Omega^2\,,
\end{eqnarray}
which gives the most general extension of the Schwarzschild metric as a stationary solution. Note that the $\mathcal{G}_{(i)}(r)^2$ are given by
\begin{align}\label{eq:Gr_def_F}
    \mathcal{G}_{(i)}(r)^2 := \frac{(\partial_x E^x_{(i)})^2}{4  {g}_{(\alpha)}^2 {E^x}_{(i)}} = \frac{\left({f}^{(2)} \left[F^{-1}_{(i),m}\left[E^x \right], E^x \right] \right)^2}{ {g}_{(\alpha)}^2} (r) \,,
\end{align}
where we used the equation of motion \eqref{eq:eff_dy_new_rewritten} and the Killing symmetry $\partial_t + \partial_x= 0$. ${g}_{(\alpha)}(r)$ here is understood as ${g}_{(\alpha)}(E^x(r))$.

\noindent
Taking lemma \ref{thm:birkhoff-like} into account, we can summarize these results in the following corollary 
\begin{corollary}\label{Cor:Birkhoff_Schwarzschild}
     The polymerized vacuum solution in lemma \ref{thm:birkhoff-like} expressed in terms of Schwarzschild-like coordinates may not be unique, but has countably (possibly infinitely) many solutions. Each Schwarzschild-like metric corresponds to a piecewise monotonic segment in $t$ of the unique polymerized vacuum solution written in LTB coordinates.
\end{corollary}
\noindent
Corollary \ref{Cor:Birkhoff_Schwarzschild} demonstrates that in contrast to the classical theory where the uniqueness can be obtained either in LTB or Schwarzschild coordinates, this is no longer the case for generic effective models. Note that the non-uniqueness of the polymerized vacuum solution in Schwarzschild-like coordinates is strongly related to the properties of the phase space trajectories determined by the effective Hamiltonian as discussed in Subsec. \ref{sec:BirkhofflikeThm}, more specifically, in Fig. \ref{fig:phasespace}. There we have already mentioned that effective models with unbounded monotonic trajectories and those with bounded trajectories have different properties. Applied to the discussion here, this means that every effective model with a monotonic trajectory implies a unique Schwarzschild-like solution. As discussed above a prominent example for an unbounded monotonic trajectory is the classical case. Besides that further interesting examples are the regular black bole solutions of Bardeen \cite{Bardeen68} and Hayward \cite{Hayward:2005gi}, that we will discuss more in detail in Sec. \ref{sec:examples}. 
~\\
Our analysis also shows that the appearance of the non-uniqueness is closely related to the non-uniqueness of the transformation between $r$ and $E^x$, in other words, the presence of bounces. When bounces occur, we always have $\partial_z E^x =-\partial_t E^x =\partial_x E^x =0 $. Considering that $E^{\phi} =\frac{\partial_x E^x}{ {g}_{(\alpha)}}$, we have $E^{\phi} =0$ unless ${g}_{(\alpha)}=0$ at the bounce. For the polymerized vacuum solution, this could be either a pure coordinate effect or a shell-crossing singularity. Precise knowledge about this can be gained by checking the curvature invariants, which we will calculate explicitly for all examples in section \ref{sec:examples}. 
~\\
A special case among the effective models with bounded phase space trajectories is the symmetric bounce where the uniqueness is also ensured as both symmetric segments yield the same Schwarzschild-like solution. Moreover for effective models in the subclass $\RN{2} \cap \RN{3}$, i.e. those where an underlying mimetic Lagrangian exist, the Schwarzschild-like coordinates are of the form
\begin{eqnarray}
\label{metrictz3}
 \mathrm{d} s^2 = - ( 1-\mathcal{G}_{(i)}(r)^2) \mathrm{d} \tau^2 +\frac{1}{ \left( 1-\mathcal{G}_{(i)}(r)^2 \right)} \mathrm{d}r^2 + r^2 \mathrm{d} \Omega^2\, ,
\end{eqnarray}
as we have $ {g}_{(\alpha)} =1$ for this subclass.
\subsection{Reconstruction algorithm for models in class II$\cap$III}
\label{sec:ReconstrAlg}
So far, the formalism introduced in \cite{Giesel:2023tsj,Giesel:2023hys} has mainly been applied to the following situations: One application is to start with an effective spherically symmetric model in the canonical framework and then perform its reduction to its effective LTB sector. A second application considers the class of effective models for which the LTB model is completely decoupled along the radial coordinates, then takes an effective cosmological model as a starting point and constructs the underlying effective spherically symmetric model and subsequently it was investigated under which assumptions an underlying extended mimetic Lagrangian exist. Now with the results discussed in Sec. \ref{sec:sub_schwarzschildcoords} on the Schwarschild-like coordinates we can go one step further and also investigate modified gravity theories such as regular black holes with our formalism if these are given in Schwarzschild-like coordinates. For a certain class of models this will even allow us to reconstruct the underlying mimetic Lagrangian as we will discuss below. This reconstruction is in detail discussed in Subsec. \ref{sec:Reconstruction_Detail} and a brief summary of the reconstruction algorithm can be found in Subsec. \ref{sec:Summ_Reconstruction}.

\subsubsection{Reconstructing the effective model from a given metric in Schwarzschild-like coordinates}
\label{sec:Reconstruction_Detail}
As the discussion in Sec. \ref{sec:sub_schwarzschildcoords} shows the solutions of the effective metric expressed in in Schwarzschild-like coordinates \eqref{metrictz2} are completely determined by the function $F^{-1}$ and ${g}_{(\alpha)}$. This allows us to reconstruct the decoupled Hamiltonians $\widetilde{H}^{(\alpha)}$ and further to associate a given polymerization function $F$ as well as a compatible LTB condition with an effective metric in Schwarzschild-like coordinates. Such an modified metric  may not have been motivated in the context of effective models in the canonical framework, but for example in the Lagrangian setup with phenomenological motivations.
Given $\widetilde{H}^{(\alpha)}$ our framework allows to further reconstruct the spherically symmetric Hamiltonian $C^{(\alpha)}$ from a given metric in Schwarzschild-like coordinates. 
~\\
As we discussed previously, in order to recover the spherical symmetric Hamiltonian $C^{(\alpha)}$, we need to determine the three different functions $F,h_2$ and $ {g}_{(\alpha)}$ appearing in the decoupled Hamiltonian $\widetilde{H}^{(\alpha)}$. As a result, this reconstruction is for theories in class $\RN{2}$ not unique, as we still have the freedom to choose the inverse triad corrections $h_2$. However, in the subclass $\RN{2} \cap \RN{3}$ no inverse triad corrections are present and for these effective models the reconstruction becomes unique. Moreover, since for these subclass of effective models an underlying covariant mimetic Lagrangian exist, we can reconstruct it from a given metric in the form of \eqref{metrictz3}. 
~\\
In detail this procedure goes as follows: We consider a  metric in Schwarzschild-like coordinates of the from in \eqref{metrictz3} 
as the starting point. Then we can immediately read off the form of $\mathcal{G}(r):=\mathcal{G}_{(i)}(r)$ for a fixed given $i$. As we consider models in the subclass $\RN{2} \cap \RN{3}$, we know that the inverse triad corrections are absent, that is $h_2=h_1=1$, the compatible LTB condition is the classical one, that is ${g}_{(\alpha)}=1$. Further, the polymerization functions need to be compatible with a $\overline{\mu}$-scheme polymerization and thus their generic form is given by the ansatz in \eqref{eq:mubar_mimetic_fs}. In addition we take advantage of the fact that the conserved quantity in the mimetic sector shown in \eqref{eq:ham_in_mimetic} takes a simpler form and factorizes concerning their dependence on the $(b,v=r^3)$ variables. Reinserting now the the ansatz of the polymerization function in \eqref{eq:mubar_mimetic_fs} into he definition of $\mathcal{G}(r)$ found in \eqref{eq:Gr_def_F} we realize that here the general expression for $\mathcal{G}(r)$ simplifies and can be written in terms of $\tilde{F}$ and its inverse respectively as well as a function of $\frac{2Gm}{v}=\frac{r_s}{r^3}$ due to the factorization
\begin{align}\label{eq:gr_def_in_mimetic}
    \mathcal{G}(r) = \frac{r}{2} \tilde{F}'\left[\tilde{F}^{-1}\left[\frac{r_s}{r^3}\right]\right] = - \frac{3 r_s }{2 r^3  \dv{}{r}\left(\tilde{F}^{-1}\left[\frac{r_s}{r^3}\right] \right)} ,
\end{align}
where we used the inverse function to rewrite the derivative of $\tilde{F}$ and defined $r_s\coloneqq 2 G m$. Integrating this equation we get
\begin{eqnarray}\label{eq:recon_f}
    \tilde{F}^{-1}\left[\frac{r_s}{r^3}\right]  = - \int\mathrm{d} r \frac{3 r_s }{2 r^3  \mathcal{G}(r)}  =  \int \mathrm{d}\left( \frac{r_s}{r^3}\right) \frac{r}{2 \mathcal{G}(r)}\, .
\end{eqnarray}
In this way we can explicitly reconstruct $\tilde{F}^{-1}\left[\frac{r_s}{r^3}\right]$ from a given function $\mathcal{G}(r)$. For the case of analytic $\tilde{F}$, we can extend the range of $\tilde{F}$ from the piecewise segment $c_i$ to the whole analytic domain of $\tilde{F}$.
~\\
~\\
In conclusion above described method provides a generic framework to construct alternative gravity models, which can have a regular, static, spherically symmetric metric as long as the right hand side of equation \eqref{eq:gr_def_in_mimetic} can be expressed as a function of $r_s/r^3$. In our classification these are models from the subclass II$\,\cap\,$III. From our perspective this subclass still involves interesting examples of the black holes with regular center such as for example the Bardeen \cite{Bardeen68} and the Hayward \cite{Hayward:2005gi} metric. It is known that these solutions can be given by general relativity coupled to nonlinear electrodynamics, e.g. in \cite{Bronnikov:2000vy}. Our framework is able to provide an explicit form of covariant Lagrangians in the context of scalar-tensor theories with (infinite) higher order derivative terms, which provides the class of black hole metrics with a regular center, e.g. the Bardeen and Hayward metric, as a solution. Moreover, it allows to compare them with the effective models with an LQC-inspired polymerization in the same framework and to gain new insights on how a chosen form of the polymerization function $\tilde{F}$ affects the properties of the effective model, which will be discussed in Sec \ref{sec:examples}.
~\\
The above results  can be summarized in Lemma \ref{coro:reconstruction} and the following remark:
\begin{lemma}\label{coro:reconstruction}
 For a Schwarzschild-like solution in the form of \eqref{metrictz3}, we can uniquely reconstruct a spherically symmetric Hamiltonian in the class II$\,\cap\,$III according to the classification in Subsec. \ref{sec:ClassI} and \ref{sec:ClassIII} and an underlying four dimensional mimetic Lagrangian having this solution as (one of) the polymerized vacuum solution. The reconstruction is given by \eqref{eq:recon_f},\eqref{eq:f1_and_F} and \eqref{eq:mimetic_L_from_F}. %
\end{lemma}
\noindent and
\begin{remark}
    Above lemma \ref{coro:reconstruction} %
    can be generalized to the Schwarzschild-like solution in the form of \eqref{metrictz2} if we fix an ansatz for the inverse triad correction $h_2$ of effective models in class II according to the classification in Subsec. \ref{sec:ClassII}. However, in such a case we cannot have an underlying mimetic Lagrangian. We leave the detailed reconstruction for this class to a future study.
\end{remark}
\noindent
We want to close that subsection with a discussion how we can use the results obtained so far to obtain the general marginally bound solution for generic $M(x)$, i.e. arbitrary dust profiles, from the solution of the modified Friedmann equation in for constant $M(x)=m$. As we can see from \eqref{eq:marginal_sol} the general marginally bound solution can be obtained from  knowing ${\cal F}$ that can again be computed once the polymerization function $F$ and the LTB function $g_{(\alpha)}$ are known. For models in II, we can compute ${\cal F}$ from a given ${\cal G}(r)$ and thus from a metric given in Schwarzschild-like coordinates.  For this purpose we use the definition of ${\cal G}(r)$ in \eqref{eq:Gr_def} and  \eqref{eq:Gr_def_F} that gives the 
the modified Friedmann equation in the marginally bound case by extending the constant $m$ to $M(x)$. As a result, we can integrate the equation to get the corresponding marginally bound solution for a given polymerized vacuum solution in Schwarzschild-like coordinates \footnote{Here $\widetilde{\mathcal{F}}$ is related to $\mathcal{F}$ in \eqref{eq:F_i} by identifying $E^x = R^2$}. 
\begin{eqnarray}\label{eq:marginal_sol_reconstruction_II}
    R(t,x) = \widetilde{\mathcal{F}}^{-1}(s(x)-t), \quad \widetilde{\mathcal{F}}(R) = \int_{R_0}^R \left. \frac{\mathrm{d}r}{\mathcal{G}(r)  {g}_{(\alpha)}(r)}\right|_{m=M(x)} .
\end{eqnarray}
The construction of the solution of the modified Friedmann equation is clearly unique for effective models of class II. Considering that we do not have a unique reconstruction of the Hamiltonian ${C}^{(\alpha)}$ in \eqref{eq:Ham_non_marginal} due to the occurrence of undetermined inverse triad corrections $h_2$ in this class, this implies that the effect of different ansatzes for $h_2$ is only distinguishable for non-marginally bound solutions. This also becomes apparent from the form of the decoupled Hamiltonian, where $h_2$ is directly coupled with the LTB function $\ltbf$.  In the marginally bound case we are able to redefine $F$ as $F_{\rm new}:=F-{f}_{\rm new}(v)$ and then reabsorb the additional ${f}_{\rm new}(v)$ into a redefinition of $h_2$ which no longer works in the non-marginally bound case where $\ltbf$ depends on $x$.
~\\
In the subclass $\RN{2} \cap \RN{3}$ the corresponding marginally bound solution is given by
\begin{eqnarray}\label{eq:marginal_sol_reconstruction}
    R(t,x) = \widetilde{\mathcal{F}}^{-1}(s(x)-t), \quad \widetilde{\mathcal{F}}(R) = \int_{R_0}^R \left. \frac{\mathrm{d}r}{\mathcal{G}(r)}\right|_{m=M(x)} .
\end{eqnarray}
As stated previously this class admits a unique reconstruction of the Hamiltonian ${C}^{(\alpha)}$ in \eqref{eq:Ham_non_marginal} and corresponding underlying mimetic gravity theory.
In this way we can for example derive the analytic solution of dust collapses with arbitrary dust matter profile $M(x)$ for a given Bardeen-like metric \eqref{metrictz3}. More generally, the construction of the analytic marginally bound solution works for all effective models in the subclass $\RN{2} \cap \RN{3}$.
~\\
~\\
We can summarize these findings by the following lemma:
\begin{lemma}\label{coro:reconstruction2}
    For a Schwarzschild-like solution in the form of \eqref{metrictz2}, we can directly construct the general marginally bound solution of the effective model in LTB coordinates with arbitrary dust profile beyond the vacuum solution according to \eqref{eq:marginal_sol_reconstruction_II}.
\end{lemma}
\noindent
Considering \eqref{eq:marginal_sol_reconstruction_II} we can apply this to a given Schwarzschild-like solution in the form of \eqref{metrictz2} with $\mathcal{G}(r):=\mathcal{G}_{(i)}(r)$ for a given monotonic segment index $i$ and obtain the corresponding marginally bound LTB solution. From the discussion above we know that different Schwarzschild-like solutions can be related to each other by the reconstruction procedure of the individual monotonic segments $c_i$ of the corresponding phase space trajectories. An interesting question is under which condition the class of $\{ \mathcal{G}_{(i)}(r) \}$ is an unique extension of certain $\mathcal{G}(r)$. Given the marginally bound LTB solution in \eqref{eq:marginal_sol_reconstruction_II} for models in class II we know that we have a decoupled dynamics along the radial coordinate in LTB coordinates. Thus we can construct the decoupled Hamiltonians $\widetilde{H}^{(\alpha)}(b,v)$ that yield the dynamics corresponding to the marginally bound solution. If this Hamiltonian is analytic in $b,v$ we know that is already determined from one monotonic segment and hence a given $\mathcal{G}_{(i)}(r)$ can be uniquely extended. Likewise if we reconstruct the marginally bound solution $R(t,x)$ for one given $\mathcal{G}_{(i)}(r)$ and thus one monotonic segment and this can be uniquely extended, the class of $\{ \mathcal{G}_{(i)}(r) \}$ is a unique extension of a certain $\mathcal{G}(r)=\mathcal{G}_{(i)}(r)$. We summarize this in corollary \ref{coro:g_i_extension}
\begin{corollary}\label{coro:g_i_extension}
For a given Schwarzschild-like solution in the form of \eqref{metrictz3} with $\mathcal{G}(r):=\mathcal{G}_{(i)}(r)$ for a given monotonic segment index $i$, different Schwarzschild-like solutions $\mathcal{G}_{(i)}(r)$ are related by the reconstruction procedure applied to each monotonic segment of the corresponding phase space trajectories. The class of $\{ \mathcal{G}_{(i)}(r) \}$ can be understood as the unique extension of a certain $\mathcal{G}(r)$ if 
\begin{itemize}
\item the effective Hamiltonian $\widetilde{H}(b,v)$ corresponding the reconstructed marginally bound LTB solution is analytic in $v,b$
\item or there exist a unique extension for the marginally LTB solution $R(t,x)$ in \eqref{eq:marginal_sol_reconstruction_II} from one monotonic segment to another.
\end{itemize}
\end{corollary}

\subsubsection{Summary of the reconstruction algorithm for subclass II $\cap$ III}
\label{sec:Summ_Reconstruction}
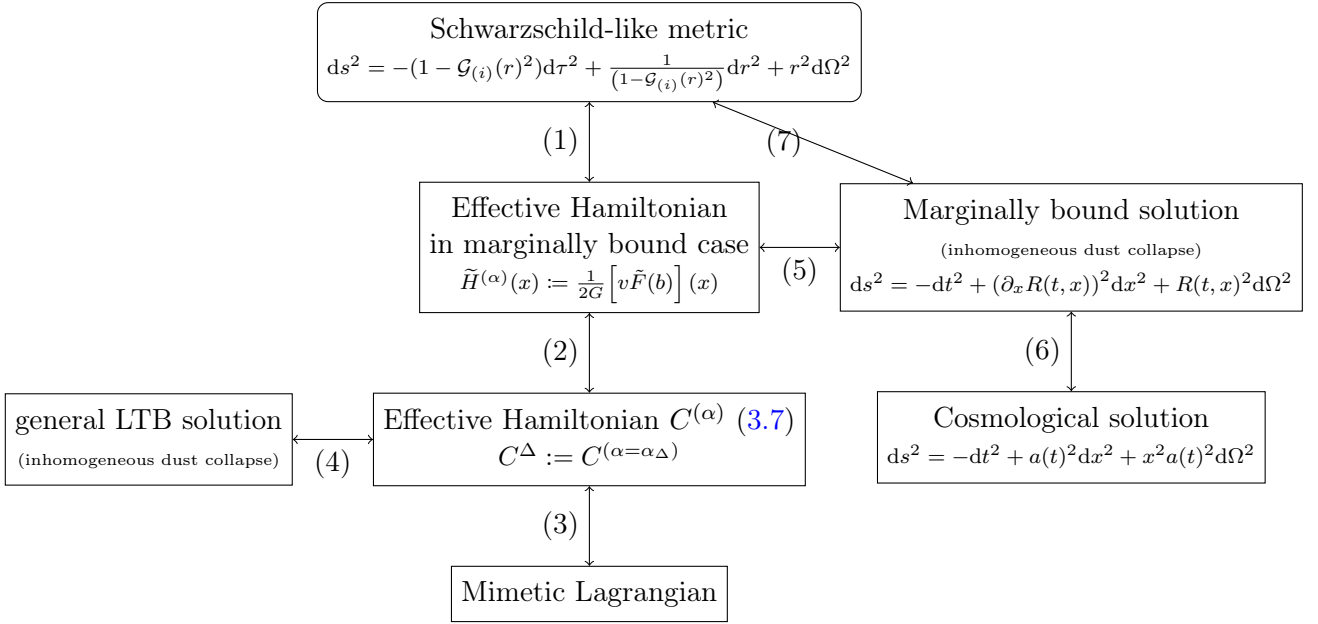
\begin{figure}[h!]
    \centering
    \begin{tikzpicture}[node distance=30pt]
    \small
  \node[draw, rounded corners]                        (metric)   {\makecell{Schwarzschild-like metric\\{\scriptsize $\scriptsize \mathrm{d} s^2 = - ( 1-\mathcal{G}_{(i)}(r)^2) \mathrm{d} \tau^2 +\frac{1}{   \left( 1-\mathcal{G}_{(i)}(r)^2 \right)} \mathrm{d}r^2 + r^2 \mathrm{d} \Omega^2$}}};
  \node[draw, below=of metric]                         (Finv)  {\makecell{Effective Hamiltonian\\
  in marginally bound case\\{\scriptsize $\widetilde{H}^{(\alpha)}(x) \coloneqq \frac{1}{2G}\qty[v \tilde{F}(b) ](x)$}}};
  \node[draw, below=of Finv]                         (Hsp)  {\makecell{Effective Hamiltonian $C^{(\alpha)}$ \eqref{eq:defpolyhamiltonianconstraintwithoutKx}\\
  {\footnotesize $C^\Delta:= C^{(\alpha=\alpha_\Delta)}$}}};

  \node[draw, below=of Hsp]                         (mimetic)  {\makecell{Mimetic Lagrangian\\
  }};

  \node[draw, right=of Finv]                         (marginal)  {\makecell{Marginally bound solution\\{\tiny(inhomogeneous dust collapse)}\\{\scriptsize $\scriptsize \mathrm{d} s^2 = - \mathrm{d} t^2 +{\left(\partial_x R(t,x) \right)^2}\mathrm{d}x^2 + R(t,x)^2 \mathrm{d} \Omega^2$}}};

  \node[draw, below=of marginal]                         (cosmo)  {\makecell{Cosmological solution\\{\scriptsize $\scriptsize \mathrm{d} s^2 = - \mathrm{d} t^2 +a(t)^2\mathrm{d}x^2 + x^2 a(t)^2 \mathrm{d} \Omega^2$}}};

  \node[draw, left=of Hsp]                         (nomarginal)  {\makecell{general LTB solution\\{\tiny(inhomogeneous dust collapse)}\\
  }};
 
  \draw[<->] (metric) -- node[left]  {(1)} (Finv);

  \draw[<->] (metric) -- node[left]  {(7)} (marginal);

  \draw[<->] (Finv) -- node[left]  {(2)} (Hsp);

  \draw[<->] (Hsp) -- node[left]  {(3)} (mimetic);

    \draw[<->] (Finv) -- node[below]  {(5)} (marginal);

    \draw[<->] (Hsp) -- node[below]  {(4)} (nomarginal);

    \draw[<->] (marginal) -- node[left]  {(6)} (cosmo);
\end{tikzpicture}
    \caption{Reconstruction algorithm} %
    \label{fig:reconstruction}
\end{figure}

In this subsection we restrict our study to subclass II $\cap$ III, in which the effective models have a unique reconstruction of the effective Hamiltonian $C^{\alpha}$ in \eqref{eq:Ham_non_marginal} and a reconstruction of the corresponding mimetic Lagrangian can be performed. In this subclass we have ${g}_{\alpha}=1$ which means that the classical LTB condition is a compatible one. 
As a consequence, a direct further symmetry reduction of the LTB metric into the standard flat FLRW cosmological case with the choice of $R(t,x) = x a(t)$ is possible. 
~\\
Here we summarize the in terms of a reconstruction algorithm correspondence  and relations between the Schwarzschild-like solutions and the Birkhoff-like theorem, the polymerized Hamiltonian and the mimetic Lagrangian as well as
the marginally bound solutions with a generic mass function for theories in the subclass II $\cap$ III described in section \ref{sec:sub_schwarzschildcoords}. We provide a graphical illustration of the construction algorithm in Fig. \ref{fig:reconstruction} and in addition list the individual steps numbered from one to seven in table below.
\begin{algorithm*}[h!]
\SetAlgorithmName{Reconstruction algorithm}{}{}
\caption{}
\label{Table:Algorithm}
\small
Reconstruction of $\tilde{F}(b)$: $\tilde{F}^{-1}\left[\frac{r_s}{r^3}\right]  = - \int\mathrm{d} r \frac{3 r_s }{2 r^3  \mathcal{G}(r)}  =  \int \mathrm{d}\left( \frac{r_s}{r^3}\right) \frac{r}{2 \mathcal{G}(r)}$\;
Reconstruction of $\tilde{f}^{(1)}(b),\tilde{f}^{(2)}(b)$: $\tilde{f}^{(2)}(b)=\tilde{F}'(b)/2, \; \tilde{f}^{(1)}(b) = 3 \tilde{F}(b)- b \tilde{F}'(b) $\;
Reconstruction of mimetic potential $L_{\phi}(X,Y)$: $L_{\phi}(X,Y) =2 X Y - Y^2 + \tilde{f}^{(1)} \qty[(\tilde{f}^{(2)})^{-1}\qty(-{Y})] + 2 X (\tilde{f}^{(2)})^{-1}\qty(-{Y})$ \;
Modified Friedmann equation:  $\frac{\dot{R}}{R} = -\frac{1}{2} \tilde{F}'\left[\tilde{F}^{-1}_{(i)}\left[\frac{2 G M(x) - \left(\ltbf(x)^2-1\right) R }{R^3}\right]\right]$ \; %
Solution of the modified Friedmann equation with $\ltbf(x)=1$:  $R(t,x) = \mathcal{F}^{-1}_{(i)}(s(x) - t), \quad \mathcal{F}_{(i)} =\int_{R_0}^R  \frac{2 \mathrm{d}r}{r \tilde{F}'\left[\tilde{F}^{-1}\left[\frac{2 G M(x)}{r^3}\right]\right]}$ \;
Cosmological reduction:  $R(t,x) = x^2 a(t)^2$  \; %
Extension to marginally bound solution:  $R(t,x) = \mathcal{F}^{-1}(s(x)-t), \quad \mathcal{F}(R) = \int_{R_0}^R \left. \frac{\mathrm{d}r}{\mathcal{G}(r)}\right|_{m=M(x)}$  \;
\end{algorithm*}
~\\
~\\
 As we mentioned before, there is a so-called polymerization parameter $\alpha$ that we introduced for the effective models which is included in the modifications in the effective models is defined such a way that when $\alpha=0$ we rediscover classical general relativity. Here such a modification could result from quantum gravity effects, e.g. if we identify the parameter $\alpha=\alpha_{\Delta} \propto \sqrt{\Delta} \propto l_{P} $ with $\Delta$ the minimal area gap in loop quantized models. Note the reconstruction algorithm is independent of the choice of polymerization parameters, so in principle more parameters or even no parameter are also allowed options. Moreover, in our construction from the static Schwarzschild-like metric, the mass $m$ naturally appears as the integration constant of the equations of motion. This is ensured by \eqref{eq:ham_in_mimetic}, where we identify the integration constant $M(x)$ with $m$. As a result, the polymerization function $\tilde{F}$, the 1+1d Hamiltonian $C^{(\alpha)}$ and the underlying mimetic model are independent of $m$ and contain only $\alpha$ as a parameter.
\subsection{Regular black holes and limiting curvature mechanism in the subclass II $\cap$ III}\label{sec:curvature_limiting}
\label{sec:LimCurvature}
 Using the reconstruction algorithm introduced above, we can take advantage of the fact that for models in the subclass II $\cap$ III the underlying extended mimetic models exits. Given this we can now analyze the conditions for having regular black holes and extend the study to the most general case of the inhomogeneous dust collapse by considering the extension to the marginally bounded solutions provided by the reconstruction algorithm.
 ~\\
 \noindent
 In effective cosmological model a common way to obtain regular solutions is to have a bounded polymerization function for the effective Hamiltonian such that the energy density is bounded. In this case a bouncing solution appears. Following our analysis for  effective spherically symmetric models included in the subclass II $\cap$ III we have a decoupled Hamiltonian \eqref{eq:Ham_for_x} along the radial coordinate with a polymerization function $\tilde{F}$, so we can still have bouncing solutions in the sense that bounces happens at $\dot{R}(t,x) = 0$ in these models.
 However, in this case the boundedness of the curvature scalars is not directly related to the boundedness of the polymerization functions of the decoupled Hamiltonian, especially in the case of the vacuum where $C^{(\alpha)} =0$. Note that in the marginally bound case in LTB coordinates the three volume  is given by $V = R^2 R'$. So the dust energy density given by $\rho_{dust} = \frac{C^{(\alpha)}}{ R^2 R'}$ and for models of class II $C^{(\alpha)}$ is conserved quantity. As a result, at least for the bouncing solution, the boundedness of the dust energy density for non-vacuum solutions is related to whether $R'=0$. This is nothing else than the condition for shell-crossing singularities in LTB coordinates, where in this case $\det g =0$. However, this could be a pure coordinate effect and the divergence only implies an inconvenient choice of coordinates at this point. Moreover, this does not explain the case where we have an unbounded polymerization function but a regular solutions such as for the Bardeen and Hayward case.
~\\
~\\
Taken the corresponding extended mimetic model into account, we have a simpler explanation from a limiting curvature mechanism \cite{Babichev:2016rlq,Babichev:2016fbg,BenAchour:2017ivq}. As can be seen from \eqref{eq:mimetic_L_from_F} the reconstructed mimetic potential is given by $X$ and the functions $\tilde{f}^{(1)}$ and $(\tilde{f}^{(2)})^{-1}$ which depend on $Y$, where in the unitary gauge $\phi=t$, $X$ and $Y$ are related to the extrinsic curvatures of the slices determined by $\phi=t$ as in \eqref{eq:choiceofXandY}. 
In the case of a non-trivial mimetic potential, it is possible that the functions $\tilde{f}^{(1)}$ and $(\tilde{f}^{(2)})^{-1}$ impose a condition on $Y$ so that it cannot diverge, e.g. the appearance of $\sin^{-1}(Y)$.  
Note that in the marginally bound case for LTB coordinates and a compatible LTB condition with $g_{\Delta}=1$, we have
\begin{align}\label{eq:XY_in_R}
    Y =\frac{\dot{R}}{R}, \quad X = 2 Y + \frac{R Y'}{R'},
\end{align}
The curvature scalars, which are functions of $R$ and its derivatives, can be rewritten as a function of $X,Y$ and their derivatives. More specifically, the Ricci scalar and Kretschmann scalar are given by
\begin{align}
    \mathcal{R} =& \frac{2 \dot{Y}' (X-2 Y)}{Y'}+8 X Y+6 \dot{Y}-4 Y^2, \\
    \mathcal{K} =&\frac{4\left(Y' \left(2 X Y+\dot{Y}-3 Y^2\right)+\dot{Y}' (X-2 Y)\right)^2}{Y'{}^2}\\
        &\quad+8 \left(Y^2 \left(-2 X Y+X^2+2 Y^2\right) +2 \dot{Y} Y^2+\dot{Y}^2\right)+4Y^4 . \nonumber
\end{align}
 Since there is no condition on $X$ coming from  the mimetic potential, the curvature scalars can still diverge if $X$ diverges. However, as can be seen from \eqref{eq:XY_in_R}, if $Y$ is bounded the divergence of $X$ can only happen at $R'=0$. This is nothing else than the shell-crossing singularities for the LTB coordinates. If $X$ diverges causing shell-crossing singularities, it means that shell-crossings are no coordinate artifacts but true singularities. Shell-crossing singularities also appear in classical general relativity and unlike to the central singularity they are weak singularities that can be removed by weak solutions \cite{Nolan:2003wp,Husain:2022gwp,Fazzini:2023ova}.  Note that for generic models in extended mimetic gravity with conditions on both $X$ and $Y$, e.g. like the models considered in \cite{Han:2022rsx}, it is possible to remove both the central and the shell-crossing singularities. This implies that the models we present here in the subclass II $\cap$ III can remove the central singularity due to a  condition on $Y$ in the reconstructed mimetic potential, but do not resolve the weak shell-crossing singularities. Usually a bounded polymerization function $\tilde{F}$ leads to a condition on $Y$, e.g. bounded $\tilde{F}$ usually implies bounded $\tilde{f}^{(2)}$. However, this is not necessary, since e.g. an unbounded polymerization $\tilde{F}$ can still have a bounded $\tilde{f}^{(2)}=\tilde{F}'/2$ and thus put a condition on $Y$. A concrete example is the Hayward solution presented in section \ref{sec:example_hayward}.
 ~\\
 ~\\
 Finally, we note that for the polymerized vacuum solution, which we will discuss in the following sections, one has $\dot{R} = - R'$. Therefore, for bouncing solutions, where $\dot{R} = 0$ is possible, one must carefully check whether or not $X$ diverges at the bounce. In contrast for non-bouncing solutions, e.g. Bardeen and Hayward solutions, if we have a bounded $|Y|$, this is enough to imply the existence of a regular black hole solution.  We will discuss in detail in section \ref{sec:examples} where we will discuss specific examples, that in the case of standard LQC with a symmetric bounce, Bardeen and Hayward no shell-crossing singularity occur for the corresponding polymerized vacuum solution.  In contrary, in the solution with an asymmetric bounce, a shell-crossing singularity occurs.

\section{Embedding regular black hole solutions: examples}\label{sec:examples}
In this section we will give several examples of regular black hole solutions, including the Bardeen \cite{Bardeen68} and Hayward \cite{Hayward:2005gi} metrics, as well as LQC-inspired bouncing models \cite{Ashtekar:2006rx,Ashtekar:2006wn,Yang:2009fp,Dapor:2017rwv,Han:2019vpw}. This allows to show more clearly the relationship between the Schwarzschild-like polymerized vacuum solution and the related Birkhoff-like theorem, the reconstruction of the corresponding effective LTB model, and its extension to the general marginally bound LTB solution. The latter gives the possibility to consider effective models describing the collapse of arbitrary inhomogeneous dust profiles. Since our formalism is able to analyze rather general regular black hole models, we will further discuss the properties and differences between bounded and unbounded polymerization functions. The bounded ones are usually chosen in LQC-inspired models, which leads to a minimal radius and the bounce, while the unbounded ones do not have a minimal radius and can be related to regular black hole solutions without a central singularity, corresponding to the Bardeen and Hayward metrics.
~\\
~\\
With the help of the lemma \ref{coro:reconstruction} we can give explicit examples of embedding regular black hole solutions in mimetic models. Moreover, we will show that the truncation of geometric series from Bardeen and Hayward solutions in Schwarzschild-like coordinates generates a series of bouncing solutions as truncated binomial series { in the static Schwarzschild-like coordinate}, where  { the solution corresponds to standard LQC in the Schwarzschild-like coordinate} can be related to the lowest order truncation of this series if we identify $\alpha$ with $\alpha_{\Delta}$ relating to the minimal area used in LQC. { The polymerization functions for the truncated series are still untruncated as nonpolynomial functions of $b$. It is their expansion coefficients that are identified up to $n+1$ for the truncated series at truncation order $n$. }

\subsection{Bounded polymerizations: bouncing solutions}
In this subsection we want to analyze two solutions that have a bounce. One example is the standard LQC solution with a symmetric bounce in the polymerized vacuum \cite{Ashtekar:2006rx,Ashtekar:2006wn}, which corresponds to a choice where the polymerization function is just the sine function. Even in this case we have to consider two monotonic segments of the phase space trajectory for the reconstruction, which will give exactly the same vacuum solution in Schwarzschild-like coordinates. Another example is the asymmetric bounce model \cite{Yang:2009fp,Dapor:2017rwv,Li:2018opr,Han:2019vpw} motivated by Thiemann regularized LQC \cite{Thiemann:1996aw,Giesel:2007wn}. In this case we have different Schwarzschild-like solutions, which are related to each other as described in the corollary  \ref{coro:g_i_extension}.
\subsubsection{Symmetric bounce}
\label{sec:Ex:SymmBounce}
To specialize our generic polymerized model to the standard LQC solution, we consider a decoupled Hamiltonian defined by the function $\tilde{F}$  \cite{Giesel:2023hys}
\begin{equation}\label{eq:F_LQC}
    \Tilde{F}(b)= \frac{\sin^2 \left(\alpha_\Delta b \right) }{\alpha^2_\Delta}\,,\quad \alpha_\Delta:=\gamma\sqrt{\Delta},\quad  \Delta:=4\pi l_P,
\end{equation}
with no inverse triad corrections and using the classical LTB condition $ {g}_{(\alpha)} =g_\Delta=1$. The detailed analysis of the corresponding spherically symmetric model can be found in \cite{Giesel:2023tsj}. 
From this we can compute according to \eqref{eq:gr_def_in_mimetic} the associated polymerized vacuum solution in Schwarzschild-like coordinates which results in
\begin{eqnarray}\label{eq:LQC_G}
    \mathcal{G}(r)^2 = \frac{r_s}{r} - \frac{\alpha^2_\Delta r_s^2}{r^4}\, .
\end{eqnarray}
Note that the requirement here that $\mathcal{G}(r)$ is a real function implies a minimum radius $r_{\min} = (r_s \alpha_{\Delta}^2)^{\frac{1}{3}}$. We will see later that this is the minimum radius at which the bounce occurs. 
This coordinate is defined only for the range of $b\in \left(0, \frac{\pi}{2}\right)$ or $b\in \left(\frac{\pi}{2}, \pi\right)$, which are exactly the two monotone segments of $\tilde{F}$. The segment of $b\in \left(0, \frac{\pi}{2}\right)$ is connected to the classical solution in the semiclassical limit. This case gives the example that one has a unique Schwarzschild-like polymerized vacuum solution even if the trajectory has two segments. 
\noindent
Now we will perform the reconstruction from this Schwarzschild-like metric following the procedure described in Fig \ref{fig:reconstruction}.
We can integrate the metric function in \eqref{eq:LQC_G} to derive with the help of \eqref{eq:recon_f} (step 1 in Fig \ref{fig:reconstruction}) the inverse function
\begin{eqnarray}
 \tilde{F}^{-1} = \frac{2 \cot^{-1} \left( \frac{ \sqrt{\frac{\alpha^2_\Delta r_s}{r^3}}}{1-\sqrt{1- \frac{\alpha^2_\Delta r_s}{r^3}}}\right)}{\alpha_\Delta} \quad\Rightarrow\quad   \tilde{F}(b)= \frac{\sin^2 \left(\alpha_\Delta b \right) }{\alpha^2_\Delta},
\end{eqnarray}
 and by taking the inverse function again we recover the standard polymerization function. We remark that the range of $\tilde{F}$, seen as the inverse function of $\tilde{F}^{-1}$, can be extended from $b \in \left(0, \frac{\pi}{2}\right)$ to the whole real domain. The fact that $\tilde{F}$ is a bound function with maximum $\frac{1}{\alpha_{\Delta}^2}$ implies the minimal of radius $r_{\min} = (r_s \alpha_{\Delta}^2)^{\frac{1}{3}}$ again from the conserved quantity relation $r_s = r^3\tilde{F} $ shown in section \ref{sec:BirkhofflikeThm}. The step 2 in Fig \ref{fig:reconstruction} then gives
 \begin{align}
     \tilde{f}^{(1)}(b) = \frac{3 \sin ^2(\alpha_{\Delta} b)-\alpha_{\Delta} b \sin (2 \alpha b)}{\alpha_{\Delta}^2}, \qquad \tilde{f}^{(2)}(b) = \frac{\sin (2 \alpha_{\Delta} b)}{2\alpha_{\Delta}} ,
 \end{align}
 thus the effective Hamiltonian with
 \begin{align}
   C^{\Delta} &=- \frac{E^{\phi} \sqrt{E^x}}{2G} \bigg[\frac{3\sin ^2\left(\alpha_{\Delta} b\right)}{\alpha_{\Delta}^2}  +\left(\frac{ 2 \sqrt{{E^x}} K_x}{{E^{\phi}}} - b \right) \frac{\sin \left(\alpha_{\Delta} b\right)}{\alpha_{\Delta}}+\frac{1 -\qty(\frac{  {{E^x}}'}{2{{E^{\phi}}} })^2}{E^x}  - \frac{2}{E^\phi}\Big(\frac{  {E^x}'}{2{{E^{\phi}}} }\Big)'\bigg] \,.
 \end{align}
 Following step 3, the corresponding mimetic potential $L_{\phi}(X,Y)$ is given by 
 \begin{align}
     L_{\phi}(X,Y) = 2 X Y - Y^2 + \frac{3 \sin ^2(\alpha_{\Delta} b)-\alpha_{\Delta} b \sin (2 \alpha_{\Delta} b)}{\alpha_{\Delta}^2} + 2 X b, \quad b \equiv \sin^{-1}_m(-Y)
 \end{align}
 where $\sin^{-1}_m(-Y)$ is defined in the cover space of $Y$ as $\sin^{-1}_{(m)}(\eta)=(-1)^m\arcsin(\eta)+m\pi\,\,\in \lt[m\pi-\frac{\pi}{2},m\pi+\frac{\pi}{2}\rt],\ \eta\in[-1,1]$. This completes the reconstruction of the theory.

Then by performing either step 5 or step 7 we can obtain the full solution in the marginally bound case in terms of the LTB coordinates. Here we directly use step 7 to construct the marginally bound solution from the Schwarzschild-like metric \eqref{eq:LQC_G}. According to \eqref{eq:marginal_sol_reconstruction} (step 7) we have
\begin{eqnarray}
    R(t,x) =  \left(2 G M(x) \right)^\frac{1}{3}\left( \alpha^2_\Delta + \frac{9}{4} z^2 \right)^\frac{1}{3}, \qquad z = s(x) - t\,.
\end{eqnarray}
We can see that this is a bouncing solution with minimal radius $r_{\min} = (2 G M(x) \alpha_{\Delta}^2)^{\frac{1}{3}}$ corresponding to the bounce $z=0$. This bound of minimal values of $r$ can be directly read out from $F$ as a bounded function with maximal value $1/\alpha^2$, and reminding ourselves that due to \eqref{eq:ham_in_mimetic} we have $\tilde{F} r^3 = 2G M$ is a conserved quantity. As a consistency check, for the polymerized solution in the Schwarzschild-like coordinates the zero of $\mathcal{G}(r)^2$ gives the same minimal value for radius, where in this case we have a constant mass $r_s=2GM(x)=2Gm$.

The two curvature scalars, i.e. the Ricci and the Kretschmann scalar of the general marginally bound solution above  with any mass function $M(x)$ can be directly computed to be
\begin{align}
    \mathcal{R}=\frac{\mathcal{A}}{\left(9 z^2+4 \alpha^2_\Delta\right)^2 \mathcal{S}}, \qquad
    \mathcal{K}=\frac{\mathcal{B} }{\left(9 z^2+4 \alpha^2_\Delta\right)^4 \mathcal{S}^2},    
\end{align}
with $\mathcal{S} = M'(x) \left(9 z^2+4 \alpha^2_\Delta\right)+18 M(x) s'(x) z$ and the detailed form of $\mathcal{A},\mathcal{B}$ can be read off in Appendix \ref{app:curvature}.
First we can see that the curvature scalars are well-defined at the bounce $z\to 0$ due to $\alpha_{\Delta}$ and thus we do not have the central or the Schwarzschild singularity. We can check with the full form of the curvature scalars given in appendix that this is also true in the polymerized vacuum case. However, in the non-vacuum case a shell crossing singularity will appear when the inequality $9 M(x)^2 s'(x)^2-4 \alpha^2_\Delta M'(x)^2 \geq 0$ holds as in such case $\mathcal{S}$ always has real roots in $z$. In this case a shock wave (weak) solution will appear as pointed out in \cite{Fazzini:2023ova}.

We remark that although the minimal radius of the bounce is related to the mass function $M(x)$, the curvature scalars at the bounce $z=0$ does not depend on $M(x)$. Actually they only depend on $\alpha_{\Delta}$ as in standard LQC with improved dynamics:
\begin{eqnarray}
   &M'(x) \neq 0&: \quad \mathcal{R}|_{z=0} = \frac{9}{\alpha_{\Delta}^2}, \qquad \mathcal{K}|_{z=0} = \frac{27}{\alpha_{\Delta}^4} ;\\
   &M'(x) = 0&: \quad \mathcal{R}|_{z=0} = - \frac{6}{\alpha_{\Delta}^2}, \qquad \mathcal{K}|_{z=0} = \frac{360}{\alpha_{\Delta}^4} . \label{eq:ricci_const_z}
\end{eqnarray}
For the polymerized vacuum solution $M'(x)=0$, this can be seen from the fact that the curvature scalars are only functions of $\frac{r_s}{r^3}$, with
\begin{align}
\mathcal{R}_{M(x)=m} = -6 \alpha_{\Delta}^2 \left( \frac{r_s}{r^3} \right)^2 , \qquad \mathcal{K}_{M(x)=m} = 12 \left( \frac{r_s}{r^3} \right)^2 -120 \alpha_{\Delta}^2 \left( \frac{r_s}{r^3} \right)^3 + 468\alpha_{\Delta}^4 \left( \frac{r_s}{r^3} \right)^4    .
\end{align}
At the bounce where $r=r_{\min}$ we have $\frac{r_s}{r^3}=\frac{1}{\alpha_{\Delta}^2}$ thus we recover \eqref{eq:ricci_const_z}. The fact that the curvature scalars are only functions of $\frac{r_s}{r^3}$ is related to the $\bar{\mu}$-scheme used in the polymerization function $\tilde{F}$ which is required by the underlying covariant mimetic lagrangian. This is true for all the examples in class II $\cap$ III presented in this section. 

\subsubsection{Asymmetric bounce}
\label{sec:ExAsymmBpunce}
\begin{figure}[!ht]
    \centering
    \includegraphics[width=0.42\textwidth]{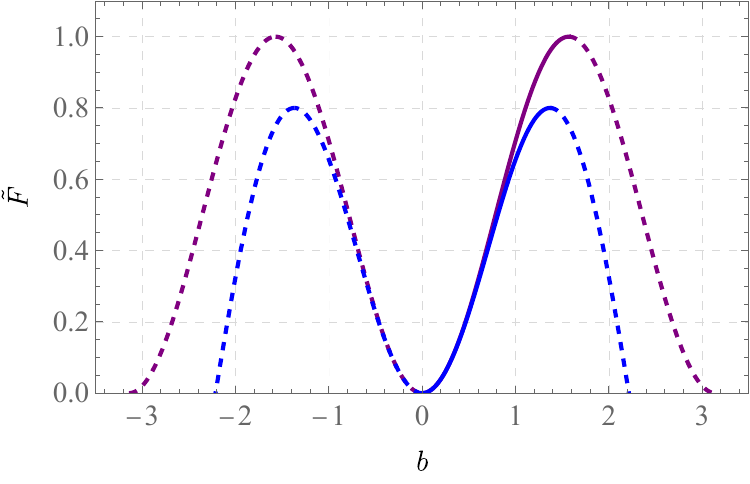}
    \includegraphics[width=0.42\textwidth]{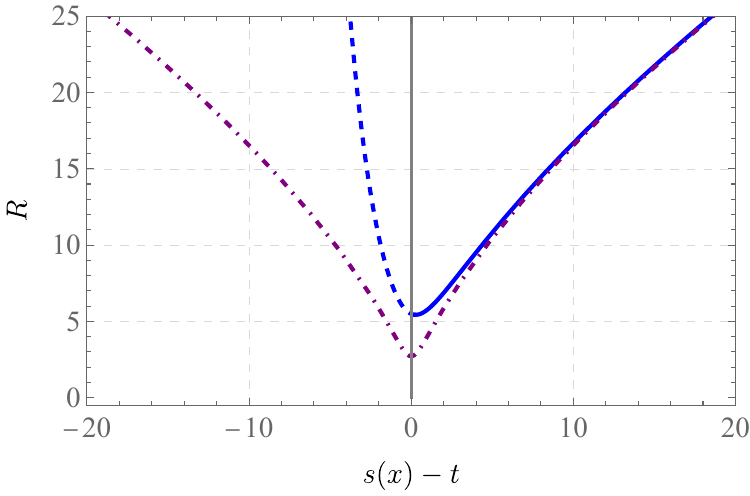}
    \caption{The plot of the function $\Tilde{F}=\tfrac{2Gm}{R^3}$ (left) and LTB solution (right) for standard LQC (purple) and Thiemann regularized LQC (blue), where the dashed line shows the extension from the original monotonic segment ( $\mathcal{G}_{-}(r)$ for Thiemann regularized LQC) in the reconstruction. For Thiemann regularized LQC, the asymmetry of $\mathcal{G}_{\pm}(r)$ results in an asymmetric bounce in LTB coordinates.}
    \label{fig:unsym_LQC}
\end{figure}

As a second example we want to consider an asymmetrically bouncing model inspired from Thiemann regularized LQC which can be generated by the $\tilde{F}$ function \cite{Yang:2009fp,Dapor:2017rwv,Li:2018opr,Han:2019vpw}
\begin{equation}\label{eq:F_unsym}
    \tilde{F}(b) = \frac{\sin ^2(\alpha_\Delta b \gamma) \left(1-\left(\gamma^2+1\right) \sin ^2(\alpha_\Delta b \gamma)\right)}{(\alpha_\Delta \gamma)^2} 
\end{equation}
where $\gamma$ is the real Barbero-Immirzi parameter of LQG. The plot of this function $\tilde{F}$ is shown in Fig. \ref{fig:unsym_LQC}. It is clear that the trajectory is composed of two monotonic segments $c_{-}$ and $c_{+}$, as $\tilde{F}$ is a quadratic function in $\sin(\alpha_\Delta b \gamma)$.

According to \eqref{eq:Gr_def_F} and using the Killing symmetry, we can write down the modified Friedmann equation of this model as
\begin{align}
    \dot{R}(t,x) &= \mathcal{G}_{\pm}(r)|_{r=R(t,x)} = \frac{R(t,x) \sqrt{\frac{1}{2}-\gamma^2 x_0} \sqrt{x_0 \pm \sqrt{1-2 \gamma^2 x_0 }+1}}{\alpha_\Delta \left(\gamma^2+1\right)},
\end{align}
with $x_0 := 2 \alpha^2_\Delta \left(\gamma^2+1\right) \left( \frac{r_s}{R(t,x)^3}\right)$, and the $\pm$ are related to the segments $c_{\pm}$. This gives us a concrete example of the situation described in the corollary \ref{Cor:Birkhoff_Schwarzschild}, where we have two different Schwarzschild-like solutions $\mathcal{G}_{\pm}(r)$ as the polymerized vacuum solution. The trajectory $c_{-}$ with the solution $\mathcal{G}_{-}(r)$ has the correct classical limit and is asymptotically connected to the classical Schwarzschild solution. The fact that $\mathcal{G}_{\pm}(r)$ are real functions gives a minimum radius which is given by $r_{\min} = 2^{2/3} \sqrt[3]{\alpha^2_\Delta \gamma^2 \left(\gamma^2+1\right) M(x)}$.

For example, let us choose the segment $c_{-}$ to perform the reconstruction of the polymerization functions. For the branch $c_{-}$, the integral defined in \eqref{eq:recon_f} (step 1 in Fig. \ref{fig:reconstruction}) can be solved analytically to be
\begin{align}
    \tilde{F}^{-1}_{(-)}\left(\frac{r_s}{r^3} \right) = \frac{i}{2 \alpha_\Delta \gamma} \log \left[\frac{\gamma^2+\sqrt{1-2\gamma^2 x_0}-i \gamma \sqrt{-2 \sqrt{1-2\gamma^2 x_0}+2x_0+2}}{1+\gamma^2} \right] .
\end{align}
Constructing the inverse function of this expression results in 
\begin{align}
    \frac{r_s}{r^3} = \frac{x_0}{ 2 \alpha^2_\Delta \left(\gamma^2+1\right)} &= -\frac{(x_1-1)^2 \left(x_1^2+2 x_1+4 \gamma^2+1\right)}{8 \alpha^2_\Delta \left(\gamma^2+1\right) \gamma^2 \left(x_1+\gamma^2\right)^2}\bigg|_{x_1 = (1+\gamma^2) e^{-i 2 \alpha_\Delta \gamma b} - \gamma^2} \\
    &= \frac{\sin ^2(\alpha_\Delta b \gamma) \left(1-\left(\gamma^2+1\right) \sin ^2(\alpha_\Delta b \gamma)\right)}{(\alpha_\Delta \gamma)^2}\, 
\end{align}
which recovers exactly \eqref{eq:F_unsym}. This computation can be seen as a concrete example of the lemma \ref{coro:reconstruction} and the corollary \ref{coro:g_i_extension}, since we can reconstruct the full function $\tilde{F}$ from a given $\mathcal{G}_{i}(r)$ for a piecewise monotone segment of $F$ and extend it to the entire domain. The reconstruction from $\mathcal{G}_{+}(r)$ gives exactly the same result. Moreover, it is also possible to derive $\mathcal{G}_{+}(r)$ from $\mathcal{G}_{-}(r)$ as described in the corollary \ref{coro:g_i_extension}. The reconstruction of the effective Hamiltonian $C^{(\alpha)}$ and the mimetic Lagrangian is straightforward, following steps 2 and 3 in Fig. \ref{fig:reconstruction}. $\tilde{f}^{(1)}(b)$ and $\tilde{f}^{(2)}(b)$ read
\begin{align}
    \tilde{f}^{(1)}(b) &= \frac{\sin (\alpha_{\Delta} b \gamma) \left(\sin (\alpha_{\Delta} b \gamma) \left(\left( \gamma^2+1\right) \left(2 \alpha_{\Delta} b \gamma \sin (2 \alpha_{\Delta} b \gamma)-3 \sin ^2(\alpha_{\Delta} b \gamma)\right)+3\right)-2 \alpha_{\Delta} b \gamma \cos (\alpha_{\Delta} b \gamma)\right)}{\alpha_{\Delta}^2  \gamma^2} \nonumber \\
    \tilde{f}^{(2)}(b) &= \frac{\sin (2 \alpha_{\Delta} b \gamma) \left(\left( \gamma^2+1\right) \cos (2 \alpha_{\Delta} b \gamma)- \gamma^2\right)}{2 \alpha_{\Delta}  \gamma},
\end{align}
Insert this into \eqref{eq:Ham_for_x} we obtain the corresponding $C^{\Delta}$. The reconstructed mimetic Lagrangian for this case is given by step 3 in Fig \ref{fig:reconstruction} with the provided $\tilde{f}^{(1)}(b)$ and $\tilde{f}^{(2)}(b)$ which reads
\begin{align}
    L_{\phi}(X,Y) =2 X Y - Y^2 + \tilde{f}^{(1)} \qty(b) + 2 X b, \quad b: \;\; -Y = \frac{\sin (2 \alpha_{\Delta} b \gamma) \left(\left( \gamma^2+1\right) \cos (2 \alpha_{\Delta} b \gamma)- \gamma^2\right)}{2 \alpha_{\Delta}  \gamma}
\end{align}
The full solution in the marginally bound case in terms of the LTB coordinates according to \eqref{eq:marginal_sol_reconstruction} is given as
\begin{eqnarray}
    R(t,x) =  \sqrt[3]{\frac{2 G M(x) \left(4 \alpha^2_\Delta \gamma^2+ 9 \eta ^2\right)^2}{18 \eta ^2-8 \alpha^2_\Delta \gamma^4}} ,\quad \,\, s(x) - t = \eta -\frac{2}{3} \alpha_\Delta \left(\gamma^2+1\right) \tanh ^{-1}\left(\frac{2 \alpha_\Delta \gamma^2}{3 \eta }\right)\,,
\end{eqnarray}
 Note that although the modified Friedmann equation is defined only for each of the segments $c_{\pm}$, the solution will cover the entire trajectory, while for $\eta \geq \eta_0 $ we are in trajectory $c_{-}$ and for $\frac{2}{3}\alpha_\Delta \gamma^2 < \eta \leq \eta_0 $ we are in trajectory $c_{+}$.
The solution has a bounce which occurs at $\eta = \eta_0 \equiv \frac{2}{3}\gamma {\alpha_\Delta \sqrt{1+2 \gamma^2}}$ where we have the minimum radius $r_{\min} = 2^{2/3} \sqrt[3]{a^2 \gamma^2 \left(\gamma^2+1\right) M(x)}$.  This minimum radius again coincides with the minimum radius obtained directly from the conserved quantity $r_s=\tilde{F} r^3$ or $\mathcal{G}_{\pm}(r)=0$ as explicitly shown in the first example. Taking $M(x)$ as constant $M(x)=m$ gives the unique stationary polymerized vacuum solution in LTB coordinates.

The curvature scalars of this model are of the form
\begin{align}
        \mathcal{R} = \frac{\mathcal{A}}{\left(9 \eta ^2+4 \alpha^2_\Delta \gamma^2\right)^4 \mathcal{S}}, \qquad \mathcal{K} = \frac{\mathcal{B}}{\left(9 \eta ^2+4 \alpha^2_\Delta \gamma^2\right)^8\mathcal{S}^2}
\end{align}
where the function $\mathcal{S}$ is given by
\begin{align}
   \mathcal{ S }= M'(x) \left(9 \eta ^2+4 \alpha^2_\Delta \gamma^2\right)^2+18 M(x) s'(x) \eta  \left(4 \alpha^2_\Delta \gamma^2 \left(2 \gamma^2+1\right)-9 \eta ^2\right)\,.
\end{align}
The detailed form of the curvature scalars can be found in the Appendix \ref{app:curvature}.
Unlike in the standard LQC case, we can check that for the vacuum solution where we have $M'(x) = 0$, we can write
\begin{align}
   \mathcal{R} \sim \frac{9 \gamma}{4 \left(\alpha_\Delta \left(\gamma^2+1\right) \sqrt{2 \gamma^2+1}\right) \left(\eta-\eta_0\right)}, \quad \mathcal{K} \sim \frac{81 \gamma^2}{16 \alpha^2_\Delta \left(\gamma^2+1\right)^2 \left(2 \gamma^2+1\right) \left(\eta-\eta_0\right)^2}\,.
\end{align}
Hence we still have a singularity present at the minimal radius $r_{min}$. We remark that this singularity is a shell crossing singularity which scales as $1/\eta^2$ for the Kretschmann invariants, while in comparison the Schwarzschild singularity scales as $1/\eta^4$. As a result, in order to avoid this singularity we have to use weak solutions for the vacuum solution as well, which means in other words that shock solutions are present. The detailed study of such a case goes beyond the scope of this article and will be investigated in future work.

\subsection{Unbounded polymerizations: solutions without a bounce -- Bardeen and Hayward}\label{sec:example_hayward}
In this subsection we want analyze two more regular solutions that are this time without a bounce. One example is the class of black holes with a regular center \cite{Bronnikov:2006fu,Ansoldi:2008jw}, where the center $r=0$ is contained in the trajectory. In such case the polymerization function $\tilde{F}(b)$ is unbounded from the conservation law $v \tilde{F}(b) = {2 G M}$ analyzed in section \ref{sec:BirkhofflikeThm}. The well-known examples in this class contains the Bardeen \cite{Bardeen68} and Hayward metric \cite{Hayward:2005gi}. 

It has been shown that this class of solutions can be obtained as magnetic monopole solutions of general relativity coupled to nonlinear electrodynamics, e.g. in \cite{Bronnikov:2000vy}. However, we will show that these solutions can be obtained directly from the reconstructed class II $\cap$ III Hamiltonian and the corresponding mimetic theory with unbounded polymerization functions. In this context, they can be seen as explicit examples of embedding a regular black hole solution in an underlying covariant mimetic model. Moreover, we are able to extend them in the sense that we are able to obtain an analytic form of the general marginally bound LTB solutions obtained from the corresponding mimetic theory for this class, which allows to describe the collapse of arbitrary inhomogeneous dust profiles.

Moreover, in these two examples, unlike what happens in the bouncing case, the polymerization function is monotonic. From corollary \ref{Cor:Birkhoff_Schwarzschild}, the Bardeen (or Hayward) metric is the unique polymerized vacuum Schwarzschild-like solution of the corresponding effective Hamiltonian and the mimetic theory.

The Bardeen metric \cite{Bardeen68} is a regular black hole metric which can be written in the coordinates of \eqref{metrictz3} as 
\begin{eqnarray}\label{eq:bardeen_m}
    \mathcal{G}(r)^2=\frac{r_s r^2}{(r^2 + l^2)^{\frac{3}{2}}} ,
\end{eqnarray}
with $l$ being a parameter of the dimension of a length.
The Hayward metric \cite{Hayward:2005gi} is another well known example of a regular black hole metric which written in the way of \eqref{metrictz3} takes the form
\begin{align}\label{eq:Hayward_G}
   \mathcal{G}(r)^2 =  \frac{r_s r^2}{r^3 + l^2 r_s}\,.
\end{align}
Note that in order to have Bardeen \eqref{eq:bardeen_m} or Hayward \eqref{eq:Hayward_G} as a consistent solution derived from the reconstructed Hamiltonian and the corresponding mimetic model, we require that $r_s=2G m$ appears in the metric related to the integration constant $m$. According to \eqref{eq:recon_f} this requires that $\frac{r}{2 \mathcal{G}(r)}$ is a function of $\frac{r_s}{r^3}$. Consequently, for the Bardeen case \eqref{eq:bardeen_m} we must identify $l = \alpha^{\frac{2}{3}} r_s{}^{\frac{1}{3}}$ with $\alpha$ the polymerization parameter. Note that with this identification $l$ still has the dimension of a length. For the Hayward case we can directly identify $l = \alpha$. Moreover, in contrast to the bouncing solutions, for both cases $\mathcal{G}(r)$ is well defined on $r 
\in [0, \infty)$, so there is no minimum radius and one can reach the center $r=0$.

We now can reconstruct an effective Hamiltonian, the corresponding mimetic Lagrangian and corresponding marginally bound solution from the Schwarzschild-like metric of Bardeen \eqref{eq:bardeen_m} and Hayward \eqref{eq:Hayward_G} model. Following the procedure described in Fig. \ref{fig:reconstruction}, the result is summarized in Table \ref{tab:bardeen_hayward}.
\begingroup

\setlength{\tabcolsep}{10pt} %
\renewcommand{\arraystretch}{1.5} %
\begin{table}[H]
    \centering
    \small
    \begin{tabular}{c|c|c}
            & Bardeen 
            & Hayward \\ 
            
    \noalign{\smallskip} 
    \hline 
    \noalign{\smallskip}
    
    metric  & $\small \begin{aligned}
    \mathcal{G}(r)^2=\frac{r_s r^2}{(r^2 + \alpha^{\frac{4}{3}} r_s{}^{\frac{2}{3}})^{\frac{3}{2}}}\end{aligned}$ 
    & $\small \begin{aligned}
    \mathcal{G}(r)^2=\frac{r_s r^2}{(r^3 + \alpha^{2} r_s)}\end{aligned}$ \\

    \noalign{\smallskip} 
    \hline 
    \noalign{\smallskip}
    
    \makecell{Polymerization \\ function} & $\footnotesize \begin{aligned}
        {\tilde{F}}^{-1}=& 
       \frac{\alpha r_s}{2 r^3}\, {}_2F_{1}\left(-\frac{3}{2},-\frac{3}{4}, -\frac{1}{2}, - \left(\alpha^2 \frac{r_s}{r^3} \right)^{-\frac{2}{3}}\right) \\
       &+\frac{\sqrt{\pi} \Gamma\qty(\frac{3}{4})}{\alpha \Gamma\qty(-\frac{3}{4})} 
    \end{aligned}$ 
    & {$\small \begin{aligned}
    {\tilde{F}}^{-1} =\frac{2 \eta \alpha + \sinh(2 \alpha \eta)}{4 \alpha}, \\
    \alpha \eta = \sinh^{-1} \sqrt{\frac{\alpha^2 r_s}{r^3}}  \end{aligned}$}\\

    \noalign{\smallskip} 
    \hline 
    \noalign{\smallskip}
    
    \makecell{Marginally bound \\ solution} & $\small \begin{aligned}
        R(t,x)=&(2G M(x))^{\frac{1}{3}}\sqrt{\eta^\frac{4}{3}-\alpha^{\frac{4}{3}}}, \quad \eta \geq \alpha\\
    s(x) -t =& \frac{2}{3} \eta + \alpha \tan^{-1}\big(\eta^{\frac{1}{3}} \alpha^{-\frac{1}{3}} \big)\\
    &- \alpha \Re\tanh^{-1}\big(\eta^{\frac{1}{3}} \alpha^{-\frac{1}{3}} \big)\ 
    \end{aligned}$ 
    & {$\small \begin{aligned}
    R(t,x) =& \left(\frac{2 G M(x) \alpha^2}{ \sinh^2( \alpha \eta)} \right)^{\frac{1}{3}}, \quad \eta \geq 0\\
    s(x) - t=& \frac{2}{3} \alpha \big( \coth( \alpha \eta)-\alpha \eta  \big)  \end{aligned}$}\\

    \noalign{\smallskip} 
    \hline 
    \noalign{\smallskip}
    
    \makecell{Curvature scalars\\{\footnotesize (see App. \ref{app:curvature})}} & $\small \begin{aligned}
        &\mathcal{R}=\frac{\mathcal{A}}{\eta^{14/3} \mathcal{S} } , \qquad
    \mathcal{K}=\frac{ \mathcal{B}}{\eta^{28/3} \mathcal{S}^2}  \\
    &\mathcal{S} = M'(x) \eta + 3 M(x) s'(x)
    \end{aligned}$
    & {$\small \begin{aligned}
    &\mathcal{R}=\frac{\mathcal{A} }{{4 \alpha^3 \mathcal{S}}}, \qquad
    \mathcal{K}=\frac{\mathcal{B}}{16 \alpha^6 \mathcal{S}^2}\\
    &\mathcal{S} =  M'(x) + 3 M(x) s'(x) \frac{\tanh (\alpha \eta)}{\alpha}
    \end{aligned}$}
    
    \end{tabular}
    \caption{The reconstruction from Bardeen and Hayward metric}
    \label{tab:bardeen_hayward}
\end{table}
\endgroup

\begin{figure}
    \centering
    \includegraphics[width=0.43\textwidth]{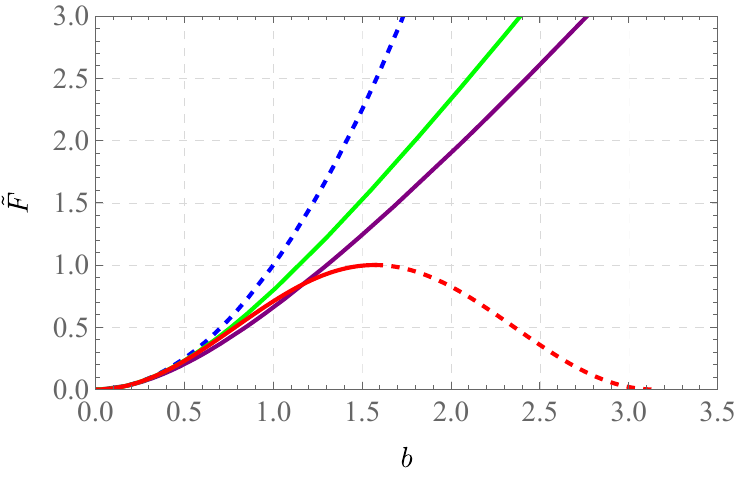}
    \includegraphics[width=0.43\textwidth]{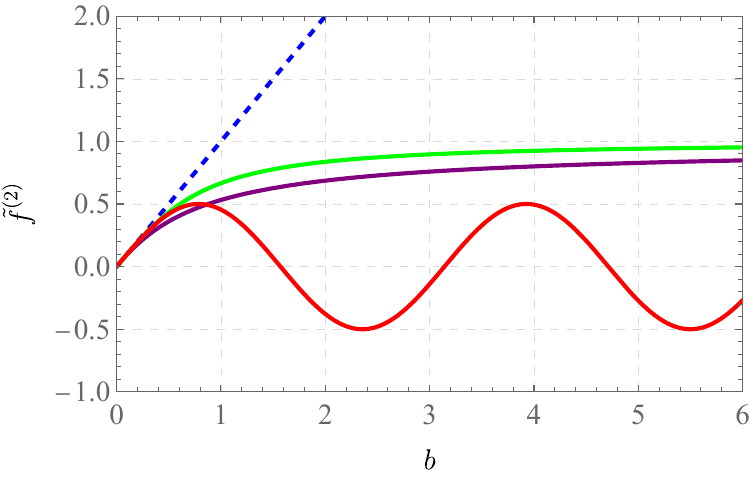}  
    \includegraphics[width=0.43\textwidth]{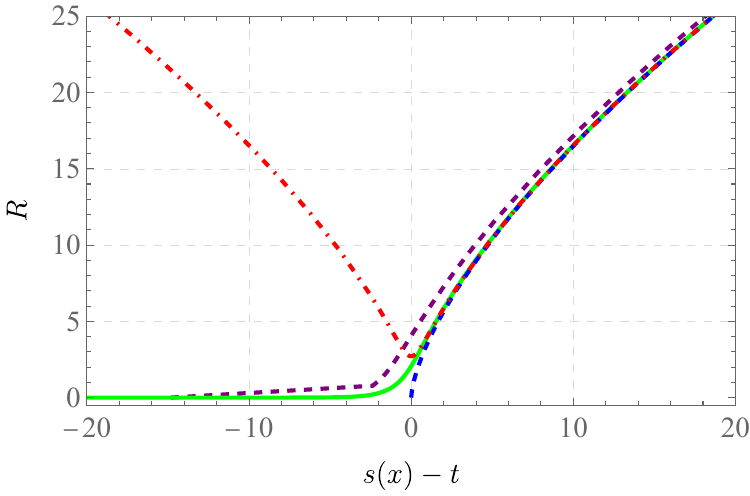}    
    \caption{Upper left:
    Left: the $\tilde{F}$ function, or equivalently the trajectory in phase space; Right: the $\tilde{f}^{(2)}$ function; Bottom: The $R(t,x)$ solution in LTB coordinates as a function of $s(x)-t$;  The blue dashed line is the classical case, purple line is the Bardeen case while the green line is the Hayward case, the red line is the symmetric bounce case. For the upper left plot we use dashed red line to show the extension of $\tilde{F}$ function in the case of symmetric bounce beyond its original branch used to solve Friedmann equation. All plots are ploted with $m=10, \alpha=1$.}
    \label{fig:bardeen_hayward}
\end{figure}

We note that in both cases $\tilde{F}$ is a monotone unbounded function, as shown in Fig. \ref{fig:bardeen_hayward}. However, we cannot express the function $\tilde{F}$ as elementary functions from its inverse $\tilde{F}^{-1}$. Nevertheless, the function $\tilde{F}$ can be defined as a series expansion from the Lagrange inversion formula, and there is no branching problem since $\tilde{F}^{-1}$ is monotone for $r>0$. For example, in the Hayward case we can define the inverse of $\tilde{F}^{-1}$ as the Leal function $y(x)=\text{Lsinh}_2(x)$ which is the solution of $x = y+\sinh y$ (see \cite{VAZQUEZLEAL2020e05418} for the detailed analysis of Leal functions). Then we express $\tilde{F}$ as
\begin{align}\label{eq:F_Hayward}
    \tilde{F} = \frac{\sinh^2\left(\frac{\text{Lsinh}_2(4 \alpha b)}{2}\right)}{\alpha^2} .
\end{align}
From the reconstruction (step 2 in Fig \ref{fig:reconstruction}) we then have
\begin{align}
    \tilde{f}^{(2)}(b) = \frac{2\tanh\left(\frac{\text{Lsinh}_2(4 \alpha b)}{2}\right)}{\alpha}, \quad \tilde{f}^{(1)}(b) = \frac{(3 \sinh (\text{Lsinh}_2(4 \alpha b))-4 \alpha b) \tanh \left(\frac{1}{2} \text{Lsinh}_2(4 \alpha b)\right)}{2 \alpha^2} ,
\end{align}
and the reconstructed mimetic potential (step 3 in Fig \ref{fig:reconstruction}):
\begin{align}
    L_{\phi}(X,Y) =2 X Y - Y^2 + \tilde{f}^{(1)} \qty(b) + 2 X b, \qquad b = \frac{Y}{2 \alpha^2 Y^2-2}-\frac{\tanh ^{-1}(\alpha Y)}{2 \alpha}
\end{align}
where we use $\text{Lsinh}_2'(x) = \frac{1}{1+\cosh(\text{Lsinh}_2(x))}$. 
Note that even we have an unbounded polymerization function $\tilde{F}$, its derivative $\tilde{f}^{(2)}$ is bounded. This leads to a requirement on $Y$ and the central singularity resolution according to section \ref{sec:curvature_limiting}. Similar thing happens for the Bardeen case as shown in Fig. \ref{fig:bardeen_hayward}.

We also note that in the limit of the shrinking physical radius to zero, $R \to 0$, the Ricci and Kretschmann curvature scalars are bounded. They read
\begin{align}
    \text{Bardeen}: \quad \lim_{R \to 0}\mathcal{R} = \frac{12}{\alpha^2}, \quad \lim_{R \to 0} \mathcal{K}= \frac{24}{\alpha^4}, \\
    \text{Hayward}: \quad \lim_{R \to 0}\mathcal{R} = \frac{12}{\alpha^2}, \quad \lim_{R \to 0} \mathcal{K}= \frac{24}{\alpha^4} .
\end{align}
This is true not only for the polymerized vacuum solution, but also for the marginally bound solution with arbitrary mass function $M(x)$. This indicates that for both Bardeen and Hayward we have removed the Schwarzschild-like singularity at the center for the dust collapse. Note that the denominator $\mathcal{S}$ can still be $0$. When $\mathcal{S} =0$ we have the shell crossing singularity. However, unlike the bouncing solutions, where the shell-crossing singularity is generally unavoidable, in this case the form of $\mathcal{S}$ is similar to that in classical GR. Such shell-crossing singularities can in principle be avoided by choosing a good matter profile $M(x)$ and suitable $s(x)$, as in classical GR. For example, if $M'(x)\geq 0$ and $s'(x)\geq0$ we are free of the shell crossing singularity.

\subsection{Interplay between the bouncing and non-bouncing solutions}
\label{sec:BionExp}
An interesting observation that we want to point out in this subsection is that considering non-bouncing black hole solutions with a regular center, e.g. the Bardeen or Hayward metric, they can generate bouncing solutions from a (truncated) binomial series. This series converges to the corresponding non-bouncing solutions in the region before the bounce. This implies that it is possible to have no-bouncing solutions without a central singularity arbitrarily close to bouncing solutions in the region before the bounce. Moreover, for the Hayward metric written in the form of \eqref{eq:Hayward_G}, the first term exactly matches the LQC solution given in \eqref{eq:LQC_G} if we identify the polymerization parameter $\alpha$ with $\alpha_{\Delta}$, which captures the minimum area in LQC. Note that this expansion is not a truncation of the expansion of the polymerization function. The polymerization function for all truncated series is still nonpolynomial and contains all orders of $\alpha$ contributions. The truncated series at the level of \eqref{eq:Hayward_G} is reflected in the identification of the expansion parameters in the polymerization function, e.g. the first and second order expansions in $\alpha$ for standard LQC are the same as for Hayward if we identify $\alpha$ with $\alpha_{\Delta}$, the remaining ones are different.

\begin{figure}[h!]
    \centering
    \includegraphics[width=0.4\textwidth]{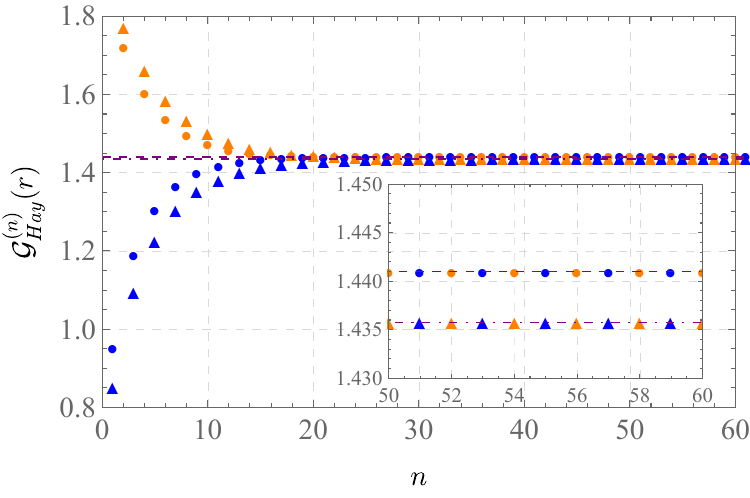}
\includegraphics[width=0.4\textwidth]{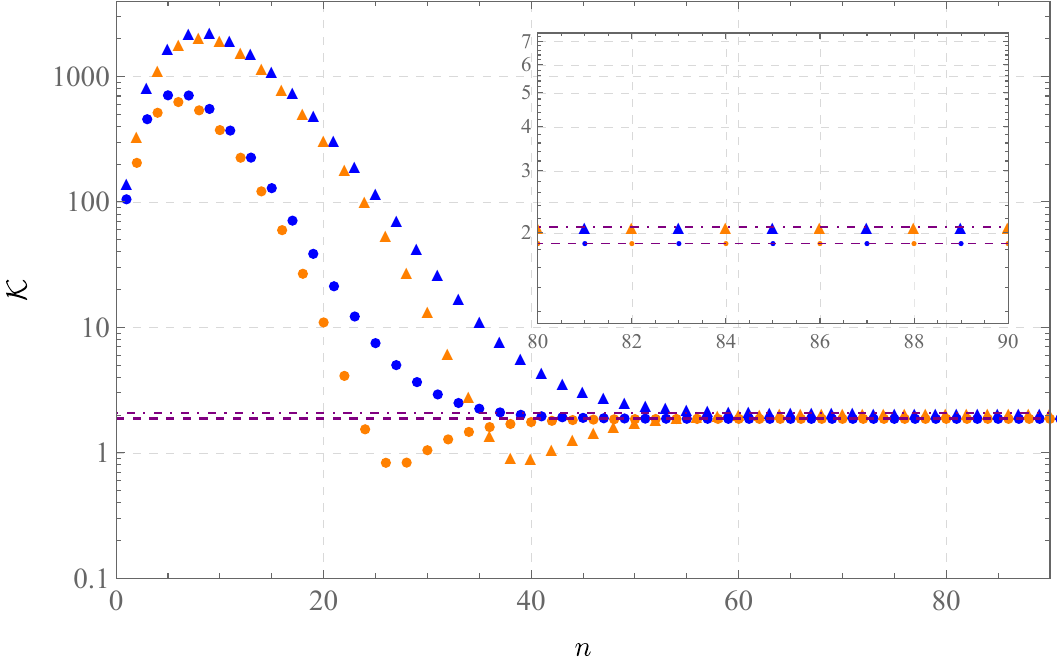}

    \caption{The convergence of $\mathcal{G}_{Hay}^{(n)}(r)$ (left) and the corresponding Kretschmann scalar (right) calculated from them. The blue points are with odd $n$ and the orange ones are even $n$. The purple dashed line indicates the original Hayward case. The plots are plotted with $\alpha = 1, r_s=8$ with the convergence $r > (\alpha^2 r_s)^{\frac{1}{3}} = 2$. The plots are evaluated at $r=2.2$ (circles) and $r=2.15$ (triangles). Close to $r=2$ the order $n$ increases after which the solutions converge to Hayward}
    \label{fig:kret_expan}
\end{figure}

\begin{figure}[h!]
    \centering\includegraphics[width=0.45\textwidth]{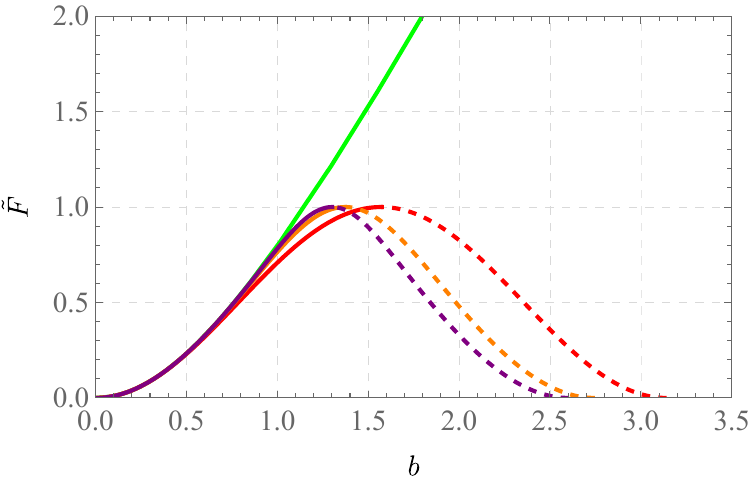}
    \includegraphics[width=0.45\textwidth]{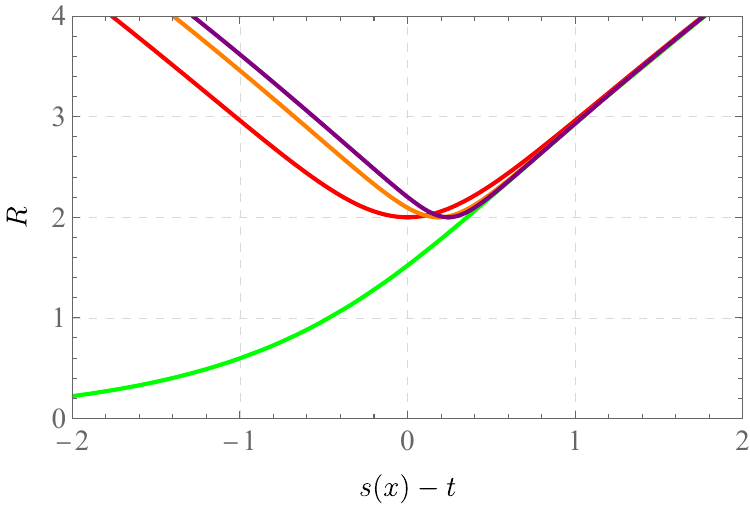}
    \caption{The $\tilde{F}=2Gm/r^3$ functions (left) and the marginally bound solutions corresponding to Hayward and its truncated geometric series $ \mathcal{G}_{Hay}^{(n)}(r)^2$ with $n=1,3,5$. The green line is Hayward, while the red line is $ \mathcal{G}_{Hay}^{(1)}(r)^2$, orange is $ \mathcal{G}_{Hay}^{(3)}(r)^2$ and purple is $ \mathcal{G}_{Hay}^{(5)}(r)^2$. The evaluation is done with $r_s=8, \alpha=1$ where we have $r_{\min}=2$. One see clearly both the $\tilde{F}$ function and the marginally bound solution $R(t,x)$ of $ \mathcal{G}_{Hay}^{(n)}(r)^2$ converging to the Hayward case with large $n$ before the bounce which happens at $r_{\min}$.}
    \label{fig:F_2nd_Hayward}
\end{figure}

We now use Hayward metric as an example. From the function $\mathcal{G}(r)^2$ of Hayward metric \eqref{eq:Hayward_G}, it admits a formal binomial expansion as geometric series:
\begin{eqnarray}
   \mathcal{G}_{Hay}(r)^2 = \frac{r_s }{r} \frac{1}{1+\frac{\alpha^2 r_s}{r^3}} 
= \sum_{k=0}^{\infty} (-1)^k \left( \frac{\alpha^2 r_s}{r^3}\right)^k 
\end{eqnarray}
This series is convergent when $r^3 > r_{\min}^2 \equiv \alpha^2 r_s $. Actually one can define the truncated summation 
\begin{align}
    \mathcal{G}_{Hay}^{(n)}(r)^2:=\frac{r_s }{r}\sum_{k=0}^{n}  \left( - \frac{\alpha^2 r_s}{r^3}\right)^k = \frac{r_s }{r} \frac{1-\left( - \frac{\alpha^2 r_s}{r^3}\right)^{n+1}}{1-\left( - \frac{\alpha^2 r_s}{r^3}\right)} , \qquad  n \in \mathbb{N}_{+} .
\end{align} 
With odd $n$, we have the expansion
\begin{align}
    \mathcal{G}_{Hay}^{(1)}(r)^2 =& \frac{r_s}{r} \left(1 - \frac{\alpha^2 r_s}{r^3} \right) ,\\
    \mathcal{G}_{Hay}^{(3)}(r)^2 =& \frac{r_s}{r} \left(1 - \frac{\alpha^2 r_s}{r^3} \right)\left(1 + \left(\frac{\alpha^2 r_s}{r^3} \right)^2 \right) \nonumber \\
    & \qquad \qquad  \cdots \nonumber\,\\
    \mathcal{G}_{Hay}^{(2 n +1)}(r)^2  =& \frac{r_s}{r} \left(1 - \frac{\alpha^2 r_s}{r^3} \right) \sum_{k=0}^{n} \left( \frac{\alpha^2 r_s}{r^3}\right)^{2k} .
\end{align}
The first term $\left(1 - \frac{a^2 r_s}{ r^3} \right) \geq 0$ that provides the strongest bound for the minimal value of $r$ in order to have a well defined function $\mathcal{G}_{Hay}(r)^2$. This gives bouncing solutions which is exactly in the convergent domain $r^3>a^2 r_s$ of the binomial expansion, where the bouncing point $r^3 = a^2 r_s$ is the boundary. We note that if we identify $\alpha$ with $\alpha_{\Delta}$, we obtain exactly the Schwarzschild-like metric for LQC from the 1st term: $\mathcal{G}_{LQC}(r)^2 = \mathcal{G}_{Hay}^{(1)}(r)^2$.
\noindent
However, this is no longer the case for even $n$, as in such case we have 
\begin{align}
     \mathcal{G}_{Hay}^{(2n)} = \frac{r_s }{r} \frac{1-\left( - \frac{\alpha^2 r_s}{r^3}\right)^{2n+1}}{1-\left( - \frac{\alpha^2 r_s}{r^3}\right)}= \frac{r_s }{r} \frac{1 + \left(\frac{\alpha^2 r_s}{r^3}\right)^{2n+1}}{1 + \frac{\alpha^2 r_s}{r^3}},
\end{align}
which clearly shows that we have a strictly positive function with no real zeros in $r$. This indicates that the whole binomial series with $n=\infty$ and the corresponding Hayward metric is not a bouncing solution, even if all of its odd $n$ truncated series are bouncing solutions.
The convergence of the geometric series and the corresponding Kretschmann scalar is shown in Fig. \ref{fig:kret_expan}. When $r$ approaching $r_{\min}$, the order n increases after which the solutions converge to Hayward.
\noindent
The binomial expansion is divergent for $r^3 \leq r_s \alpha^2$, thus fails to reproduce the original Hayward $\mathcal{G}(r)^2$. Actually for $r^3 < r_s \alpha^2$, we can use again the binomial expansion in the following form
\begin{eqnarray}
   \mathcal{G}_{Hay}(r)^2 = \frac{ r^2 }{\alpha^2} \frac{1}{1+\frac{r^3}{\alpha^2 r_s}} =   \frac{r^2 }{\alpha^2} \sum_{k=0}^{\infty} \left( - \frac{r^3}{\alpha^2 r_s}\right)^k .
\end{eqnarray}
This series does not correspond to any bouncing solution. Here as an example we can construct the model corresponding to $ \mathcal{G}_{Hay}^{(3)}(r)^2$. Following the reconstruction procedure given in Fig. \ref{fig:reconstruction} we get for the function $\tilde{F}$ the expression
\begin{equation}
    \tilde{F}^{(3)}(b)=\qty[\alpha^2 \left(\frac{1+i}{\text{sn}\left(\left.\sqrt{1+i} \,\alpha b\right|1+i\right)^2}-i\right)]^{-1} ,
\end{equation}
where $\text{sn}(u|m)$ is the Jacobi elliptic function. It is clear in such case we have a symmetric bouncing solution as shown in Fig. \ref{fig:F_2nd_Hayward}. In contrast, the reconstruction from $ \mathcal{G}_{Hay}^{(2)}(r)^2$ gives an unbounded polymerization and leads to a solution which still at singularity at the center $r=0$.
\noindent
By the reconstruction, this convergent series before $r_{\min}$ will lead to a convergent series of the corresponding polymerized Hamiltonian and the corresponding mimetic potential. The convergence can be seen already from the first three truncated series shown in Fig. \ref{fig:F_2nd_Hayward}. This implies that before the bounce, the bouncing solutions and the solutions with a regular center can be infinitely close to each other if the truncation is done at a sufficiently high order, thus in this case one can not distinguish them, while they have very different behavior at and after $r_{\min}$. This implies any perturbations before the bouncing radius $r_{\min}$, e,g. around the black hole horizon if it exists, can not distinguish bouncing and no-bouncing solutions. They can only be distinguished after the bounce happens in case it is a bouncing solution. For example, a shock wave may be present for a collapsing solution after the bounce as in \cite{Husain:2022gwp}, which generates observational effects. 

We finally remark that, as already seen from the form of $\tilde{F}^{(1)}$ and $\tilde{F}^{(3)}$, for all truncated series that they still have an truncated nonpolynomial polymerization. However, the polymerization functions will get closer and closer to each other and the untruncated Hayward polymerization if we increase the truncation order $n=2k+1$. This is reflected in the fact that the identification of the first $n+1$ coefficients of the infinite series expansion of the polymerization in terms of $\alpha$. For example, we have
\begin{align}
    \tilde{F}^{(1)}(b) &=b^2-\frac{1}{3}\alpha^2 b^4+\frac{2 }{45} \alpha^4 b^6-\frac{1}{315}\alpha^6 b^8+\frac{2 }{14175}\alpha^8 b^{10}+O\left(\alpha^9\right) , \\
    \tilde{F}^{(3)}(b) &=b^2-\frac{1}{3}\alpha^2 b^4+\frac{11}{45} \alpha^4 b^6-\frac{73 }{315}\alpha^6 b^8+\frac{1973 }{14175}\alpha^8 b^{10}+O\left(\alpha^9\right) ,\\
    & \qquad \qquad  \cdots \nonumber\, \\
        \tilde{F}_{Hay}(b) &= b^2-\frac{1}{3}\alpha^2 b^4+\frac{11 }{45}\alpha^4 b^6-\frac{73 }{315}\alpha^6 b^8+\frac{3548 }{14175}\alpha^8 b^{10}+O\left(\alpha^9\right) .
\end{align}
If we identify $\alpha$ to $\alpha_{\Delta}$, $\tilde{F}^{(1)}$ coincides with the standard LQC polymerization function \eqref{eq:F_LQC}.
\section{Conclusions}
\label{sec:Concl}
In this work we presented further applications of the formalism introduced in \cite{Giesel:2023tsj} with a particular focus on regular black hole models. For the purpose of our investigation we started with a definition and classification of polymerized effective models. In effective models one can encode the modifications compared to general relativity by involving
general polymerization functions into the Hamiltonian present in the classical theory. In general these functions can depend on the densitized triad as well as extrinsic curvature variables. We restricted to those models where no polymerization of the spatial diffeomorphism constraint is included.  In loop quantum gravity inspired models, these polymerization functions can be chosen such that the corrections to the classical theory can be understood as quantum gravity corrections of holonomies, which encode the discretized quantum geometry effects as well as inverse triad corrections,  which are  encoded in functions of the radial area variables. As analyzed in detail in \cite{Giesel:2023tsj} further restriction on the polymerization function apply if we require that with respect to the temporal coordinate gauge fixed canonical Hamiltonian is a conserved quantity and/or a compatible LTB condition exists.  Our analysis assumed validity of effective dynamics for the entire spacetime.
~\\
~\\
According to \cite{Giesel:2023hys}, the polymerized effective models presented here can be classified into three classes. Class I: these are models with an LTB reduction, that is a compatible LTB condition exists. 
Class II: covers those models that have an LTB reduction and in addition a conserved energy density. For those models the dynamics along the radial coordinate is completely decoupled. Furthermore, due to the conserved energy density, polymerized vacuum solution, where the modified Hamiltonian vanishes, are allowed. Class III: includes those models for which an underlying Lagrangian in the class of extended mimetic models exist. The mimetic field plays a dual role in our analysis. On  one hand it is  used as clock where the mimetic condition contributes to the action similar to non-rotational dust. On the other hand  it contributes as a scalar field whose higher order derivatives introduce the polymerization effects by choosing the mimetic  potential accordingly.
~\\
~\\
In this work we have mainly studied models in class II and further in its subclass II $\cap$ III. Due to the existence of the LTB reduction, we are able to derive an analytic marginally bound solution in LTB coordinates for a general dust collapse. 
With help of the marginally bound solution we are able to prove a Birkhoff-like theorem for models in class II (see Lemma \ref{thm:birkhoff-like}). As our results show the polymerized vacuum solution of models in class II is uniquely determined by an integration constant, and there exists a Killing vector field next to the one from spherical symmetry. The Killing vector field is asymptotically timelike, so that the spacetime is stationary.  
~\\
~\\ 
As in the classical case the result of Birkhoff's theorem can either be derived in LTB or Schwarzschild coordinates, we further explored the relationship between the polymerized vacuum solution in LTB coordinates and Schwarzschild-like coordinates (see Corollary \ref{Cor:Birkhoff_Schwarzschild}). It turns out that at the level of effective models the situation is more subtle than in the classical theory and here we could relate those differences to certain properties of the polymerization functions. Our results show that for non-monotonic polymerization functions, even if the polymerized vacuum solution is unique in LTB coordinates, it is not unique in Schwarzschild-like coordinates. The number of solutions is related to the number of monotonic segments of the polymerization functions. This is exactly the situation for bouncing solutions that typically occur with polymerization functions of loop quantum gravity inspired models. Among those a special case is a polymerization function that yields a symmetric bounce, since then the solution in Schwarzschild-like coordinates is still unique. Therefore, analysing the Birkhoff theorem in the context of effective models is another example similar to the discussion of shock solutions in \cite{Giesel:2023hys}, where a discussion of the results for different choices of coordinates is instructive. 
~\\
The existing results on Birkhoff-like theorems in certain loop quantum gravity inspired models in \cite{Giesel:2023tsj,Cafaro:2024vrw} 
can be embedded into the framework presented in this work as special choices of the polymerization functions. Furthermore, the formalism directly allows to compare the results in generalized LTB and Schwarzschild-like coordinates. 
\noindent
As we have proven a Birkhoff-like theorem for effective models in class II  in Sec. \ref{sec:Birkhoff_eff} this obviously involves the subclass of models included in II $\cap$ III. For this subclass a corresponding Lagrangian of an extended mimetic models exists, therefore the presented proof of the Birkhoff-like theorem in this work also applies to the corresponding extended mimetic gravity models.
~\\
~\\
The presentation in Sec. \ref{sec:Reconstruction_Detail} shows for effective models in the subclass II $\cap$ III, we can establish the correspondence between Schwarzschild-like solutions, the polymerized Hamiltonian and a mimetic Lagrangian as well as the marginally bound solutions with a generic mass function via a reconstruction algorithm (see Fig. \ref{fig:reconstruction}). With this algorithm, we are able to embed any Schwarzschild-like solution that is expressed as a function of $\frac{r_s}{r^3}$, where $r_s$ denotes the Schwarzschild radius and $r$ the coordinate in Schwarzschild-like coordinates, into a mimetic model. This embedding can be established by the reconstruction of its corresponding mimetic potential, such that the corresponding Schwarzschild-like solution is the solution of the model and satisfies the Birkhoff-like theorem. Moreover, this also implies that for a mimetic model with potential in the subclass  II $\cap$ III, we are able to obtain its possibly implicit analytic spherically symmetric solution. 
~\\
~\\
Our embedding  established in the framework presented here has several advantages: first, in our formalism, the mass of the static solutions appears naturally as an integration constant of the equations of motion, while the underlying covariant theory is characterized only by one constant parameter $\alpha$, which can be related to quantum gravity effects. In this sense, we have a consistent embedding, which differs from models where the mass is not an integration constant but part of the definition of the model, e.g. \cite{Nojiri:2023qgd}. Second, we embed not only the static solutions of the model, but also its associated (inhomogeneous) dust collapse models with the help of a compatible LTB condition. For instance, the Oppenheimer-Snyder extension with the standard junction condition as in classical GR is a solution of the corresponding underlying covariant model. In other words, the embedding fixes the underlying 1+1-dimensional field theory, in contrast to the situation where only the static part of the solution is embedded, which does not contain any field-theoretic degrees of freedom. As a consequence, the underlying dynamics, which is restricted to the spherically symmetric case as a 1+1-dimensional field theory, is unambiguous from the point of view of the equations of motion, since they are completely determined by the inhomogeneous dust collapse model.
~\\
~\\
To demonstrate how the formalism works, we applied it to four examples in section \ref{sec:examples}. For cases with a bounded polymerization function we consider the symmetric bounce in Subsec. \ref{sec:Ex:SymmBounce} and asymmetric one in Subsec. \ref{sec:ExAsymmBpunce}.  In both cases we choose the corresponding polymerization function and then start with the Schwarzschild-like polymerized vacuum solution from which we can reconstruct the effective Hamiltonian as well as the corresponding extended mimetic Lagrangian by determining the mimetic potential. Given that we can finally construct the generic marginally bound solutions for which we compute the curvature invariants. In section \ref{sec:example_hayward} we consider as two prominent examples for unbounded polymerization functions the reconstruction of the  well-known Bardeen and Hayward solution. We provide the corresponding extended mimetic model in Lagrangian form which gives these solutions and further discuss its generic marginally bound solution. 
~\\
~\\
With help of the extended mimetic model, we are able to provide a limiting curvature mechanism for regular black holes that we presented in Subsec. \ref{sec:LimCurvature}. According to our results for models in class II $\cap$ III, a bounded derivative of the polymerization function will impose an additional constraint on the higher derivative terms at the level of the mimetic potential, which successfully removes the center (or Schwarzschild) singularity. However, it will not remove the weak shell-crossing singularities, and it may transfer the center (or Schwarzschild) singularity into these weak shell-crossing singularities. Since the constraint is depending on the derivative instead of the polymerization function itself, the limiting curvature mechanism is not restricted to bounded polymerization or bouncing solutions. We can recover these results in the examples discussed in Sec. \ref{sec:examples}. For the class of solutions with a regular center such as  Bardeen and Hayward solutions, which are regular solutions with unbounded polymerization, we obtain exactly this behavior. Moreover, we have shown with the examples of the symmetric and asymmetric bounce that with a bounded polymerization, i.e. bouncing solutions, the weak shell-crossing singularities are unavoidable for the general dust collapse, thus one need weak solutions. This covers the previous finding for loop quantum gravity inspired symmetric bounce model in \cite{Fazzini:2023ova}. For models with a regular center, e.g. Bardeen and Hayward solutions, the weak shell-crossing singularities can be avoided similar to classical general relativity by choosing the matter profiles appropriately.
~\\
~\\
Since we obtained different properties of models with bounded and unbounded polymerization functions, we discussed in Subsec. \ref{sec:BionExp} the interplay between bouncing and non-bouncing solution. For this purpose we used the fact that it is possible to generate a series of models with bouncing solutions from models with a regular center, e.g. via a binomial series of the static metric in Schwarzschild like coordinate. As an explicit example we consider the the Hayward solution where the associated geometric series is an expansion in powers of $\frac{\alpha^2 r_s}{r^3}$, where $\alpha$ denotes the polymerization parameter. The geometric series converges to the Hayward solution before a minimum radius $r_{\min}$. The odd-order truncation of this geometric series gives a series of bouncing solutions, where the bounce occurs at $r_{\min}$. In this context we realize that the lowest order of the geometric series coincides with the model of standard LQC with a symmetric bounce, if we identify the parameter $\alpha=\alpha_{\Delta}$ for loop quantum gravity inspired models. Note that this truncation does not mean that we are truncating the polymerization function. For all truncated geometric series in this case, they have untruncated nonpolynomial polymerization functions. However, as we increase the truncation orders, these polymerization functions become closer and closer to each other before the bounce. This can be seen in the infinite perturbative expansion of the polymerization functions, where they have more and more identified coefficients. This implies that before the bounce if we consider a truncation at a sufficiently high order, the bouncing solutions and the solutions with a regular center can be infinitely close to each other, so that they cannot be distinguished. This also happens in the corresponding Hamiltonian and the mimetic potential via the reconstruction, implying that any perturbations before the bounce, e.g. around the black hole horizon if it exists, cannot distinguish non-bouncing solution with its corresponding bouncing solutions. Note that this does not apply to the standard LQC solutions because this would correspond to a truncation at the lowest order.
This suggests that to get the Page curve of the Hawking radiation \cite{Page:1993wv,Page:2013dx}, the details deep inside the  interior of the black hole are irrelevant.
~\\
~\\
Finally, the formalism and results presented in this work are only first steps towards a more general framework. Since we mainly focus on the marginally bound case in this article, it would be interesting to study the more general non-marginally bound case and compare with \cite{Cipriani:2024nhx} as well as the one with more general matter coupling, e.g. matter fields with pressure. With the help of the underlying extended mimetic model given, it is also possible to study the null dust collapse and, more importantly, the null evaporation. Moreover, for models in class II $\cap$ III theories, since we have a unique reconstruction of the 1+1 dimensional  dynamics and the (inhomogeneous) dust collapse from the static or cosmological solution by the LTB condition, it is possible to define an equivalent class of four dimensional Lagrangians that all lead to the same equations of motion in the case of effective spherically symmetric spacetime as an 1+1 dimensional field theory. In this sense, we can extend the correspondence to various modified gravity theories in spherically symmetric spacetime, e.g. more general DHOST theories \cite{Langlois:2018dxi}. 
\noindent
Moreover, the extension of the framework beyond spherical symmetry, e.g. to rotating spacetime, is also interesting topic for future research and could in principle follow a similar but generalized formalism with rotating generalised LTB coordinates \cite{Zaslavskii:2018lbb,Sorge:2021nqm}. { The exact spherically symmetric solutions presented here can be used as seed solutions for more general hairy solutions in the class of DHOST theories with solution generating methods, e.g. \cite{Babichev:2020qpr,BenAchour:2020wiw, BenAchour:2020fgy,Baake:2021jzv}. Another interesting topic is the cosmological and black hole perturbation theory and the computation of quasi normal modes \cite{Langlois:2021aji,Roussille:2023sdr}, especially for regular black hole solutions.} 
\section*{Acknowledgements}
This work is supported by the DFG-NSF grants PHY-1912274 and 425333893 and NSF grant PHY-2110207. H.L. is supported
by research grants provided by the Blaumann Foundation. H.L. also acknowledges Beijing Normal University, Zhejiang University of Technology and Hangzhou Normal University for the hospitality during his visits. K.G. acknowledges Louisiana State University for the hospitality during her visit in the final stage of the project.

\begin{appendix}
\section{Mimetic gravity as an underlying Lagrangian}
\label{app:MimeticIntro}
The (extended) mimetic Lagrangian is given by
\begin{align}
S[g_{\mu\nu},\phi,\lambda] =\frac{1}{8\pi G} \int_{\mathscr{M}_4} \rmd^4x \, \sqrt{-g} \, \left[ \frac{1}{2} \, {\mathcal{R}}^{(4)} + 
\, L_\phi(\phi,\chi_1,\cdots,\chi_p) \, + \, \frac{1}{2}\lambda(\mathscr{X} + 1)\right]\, ,
\end{align}
where
\begin{align}
\mathscr{X}=\phi_\mu\phi^\mu,\quad\chi_n \equiv \sum_{\mu_1,\cdots,\mu_n} \phi_{\mu_1}^{\mu_2} \, \phi_{\mu_2}^{\mu_3} \cdots \phi_{\mu_{n-1}}^{\mu_n} \, \phi_{\mu_n}^{\mu_1},\quad\phi_\mu=\nabla_\mu\phi,\quad  \phi_{\mu\nu}=\nabla_\mu\nabla_\nu\phi.
\end{align}
In the spherically symmetric case, we will only consider $\chi_1,\chi_2$, such a model can be rewrite as a two-dimensional (2D) action of the form
\begin{eqnarray}\label{eq:cov_action2}
    S_2&=&\frac{1}{4G}\int_{\mathscr{M}_2} \rmd^2 x\, \det(e) \big\{ e^{2\psi}\left(R_{h}+2h^{ij}\partial_{i}\psi\partial_{j}\psi\right)+2 
    +\, e^{2\psi}\big[ L_{\phi}\left(X,Y\right)+\frac{1}{2}\lambda\left(\partial_j \phi \partial^j \phi+1\right)\big]\big\}\,, \nonumber
\end{eqnarray}
where $h_{ij}$ is the 2D metric, $\det(e)$ denotes the determinant of 2D triad field and $R_h$ is 2D scalar curvature. The quantities $X,Y$ consist of couplings between the metric $h_{ij}$ and derivatives of $\phi,\psi$ and are given by 
\be\label{eq:choiceofXandY2}
X = - \Box_{h}\phi + Y \ , \quad  Y = - h^{ij}\partial_{i}\psi\partial_{j}\phi\ .
\ee
$X,Y$ are related to $\chi_1, \chi_2$ by
\be
\Box_{h}\phi=\frac{1}{3} \left(\chi_1-\sqrt{2} \sqrt{3 \chi_2-\chi_1^2}\right),\quad h^{ij}\partial_{i}\psi\partial_{j}\phi=\frac{1}{6} \left(2 \chi_1+\sqrt{2} \sqrt{3 \chi_2-\chi_1^2}\right),\label{lift1}\\
\text{or}\quad \Box_{h}\phi=\frac{1}{3} \left(\chi_1+\sqrt{2} \sqrt{3 \chi_2-\chi_1^2}\right),\quad h^{ij}\partial_{i}\psi\partial_{j}\phi=\frac{1}{6} \left(2 \chi_1-\sqrt{2} \sqrt{3 \chi_2-\chi_1^2}\right). \label{lift2}
\ee
We note that the space of $X,Y$ is the cover space of $\chi_1$ and $\chi_2$, as a result, we lift the potential of $L_{\phi}(\chi_1,\chi_2)$ in the four dimensional mimetic model to the cover space of $\chi_1,\chi_2$, namely $L_{\phi}(X,Y)$.

The equations of motion of the mimetic model are given by the pendant of Einstein's equations and the mimetic condition, which can be written as
\begin{eqnarray}\label{eq:cov_eq2}
   G^{(\alpha)}_{\mu\nu} := G_{\mu\nu}- T^{\phi}_{\mu\nu}=-\lambda\partial_{\mu} \phi\partial_{\nu} \phi, \quad \partial_{\mu} \phi \partial^{\mu} \phi = -1 \,.
\end{eqnarray}
The equation of motion for the mimetic field $\phi$ can be derived from the divergence free condition of the Einstein tensor $\nabla_{\mu} G^{\mu\nu} = 0$.
The right hand side of the modified Einstein equation is the energy momentum tensor of non-rotational dust, where $\lambda$ is the dust energy density. We call the case $\lambda = 0$ polymerized vacuum. In this case we still have a non-trivial mimetic field $\phi$ which leads to a non-trivial $T^{\phi}_{\mu\nu}$ and thus we keep the quantum gravity or polymerization effects. 

The mimetic potential encoded in the function $L_{\phi}$ corresponding to the case I $\cap$ III considered in this paper is given by
\begin{eqnarray}\label{eq:mimetic_L_from_F_app}
    L_{\phi}(X,Y) =2 X Y - Y^2 + \tilde{f}^{(1)} \qty[(\tilde{f}^{(2)})^{-1}\qty(-{Y})] + 2 X (\tilde{f}^{(2)})^{-1}\qty(-{Y}) .
\end{eqnarray}
Or equivalently
\be
L_{\phi}(X,Y,K_x,K_\phi) = && 2 XY -Y^2 + \tilde{f}^{(1)}(b) + 2 X b\,.
\ee
In case $f^{(2)}$ is not a monotonic function where $(f^{(2)})^{-1}$ is not unique, this implies that the function  $L_{\phi}$ is actually defined in the cover space of $Y$. 

Now we will show that in the unitary gauge $\phi  = t$ we recover exactly the Hamiltonian \eqref{eq:defpolyhamiltonianconstraintwithoutKx} by following the procedure given in \cite{BenAchour:2017ivq,Han:2022rsx}. As already studied in \cite{Han:2022rsx}, the gauge fixing of $\phi$ is exchangeable with the variation, thus we can directly work the gauge fixed action $S_2 |_{\phi = t, N =1}$.
In this gauge, the variables $X$ and $Y$ are given by
\begin{eqnarray}
    X = \frac{\partial_t E^{\phi} + \partial_x \left( N^x E^{\phi} \right)}{ E^{\phi}}  , \qquad Y = \frac{ \partial_t E^{x} + N^x \partial_x E^x}{ 2 E^{x}}\,,
\end{eqnarray}
which relate to the extrinsic curvatures of the constant-$\phi$ slice.
In total the corresponding Lagrangian up to boundary terms is given by
\begin{eqnarray}
    \mathcal{L}_2 &=&\frac{1}{2} E^{\phi} \sqrt{E^x} \left( \tilde{f}^{(1)} \qty[(\tilde{f}^{(2)})^{-1}\qty(-{Y})] + 2 X (\tilde{f}^{(2)})^{-1}\qty(-{Y}) + \frac{1}{2} R^{(3)} \right),\\
&& R^{(3)} =\frac{2}{E^x} +\frac{2(\partial_x E^{\phi})(\partial_x E^{x})}{2  (E^{\phi})^2 {E^x}} -\frac{(\partial_x E^x)^2}{2  (E^{\phi})^2 {E^x}} -\frac{\partial_x^2 E^x}{2  (E^{\phi})^2 }\,.\nonumber 
\end{eqnarray}
\noindent The generalized momenta of $E^{x}$ and $E^{\phi}$ are defined as
\begin{align}
    K_x &\coloneqq \frac{\partial \mathcal{L}_2}{ \partial 
(\partial_t E^{x})} = \frac{E^{\phi} \left(2X +  \tilde{f}^{(1)}{}' \qty[(\tilde{f}^{(2)})^{-1}\qty(-{Y})]\right)}{4 \sqrt{E^x} \tilde{f}^{(2)}{}'\qty[(\tilde{f}^{(2)})^{-1}\qty(-{Y})]} , \\
K_{\phi} &\coloneqq \frac{\partial \mathcal{L}_2}{ \partial 
(\partial_t E^{\phi})} = \sqrt{E^x}(\tilde{f}^{(2)})^{-1}\qty(-{Y}) \,. \nonumber
\end{align}
The inverse function then leads to
\begin{align}
    X = -\frac{1}{2} \tilde{f}^{(1)}{}' \qty(b) - \frac{2 \sqrt{E^x} K_x }{E^{\phi}} \tilde{f}^{(2)}{}' \qty(b) , \qquad Y = - \tilde{f}^{(2)}{}\qty(b)
\end{align}
where we define $b = \frac{K_{\phi}}{\sqrt{E^x}}$.
Using this to perform the Legendre transformation of the Lagrangian we end up with
\begin{align}
    C^{(\alpha)}(x)= \frac{ E^{\phi}}{2 G \sqrt{{{E^x}}}}\Bigg[ -{ E^x}\bigg(&\frac{4 \sqrt{E^x} K_x {f}^{(2)}(b)}{E^{\phi}} +  {{f}^{(1)}(b)} \bigg)+
\qty(\qty(\frac{  {{E^x}}'}{2{{E^{\phi}}} })^2 - 1   ) 
+2\frac{E^x}{E^\phi} \qty(\frac{   {{E^x}}'}{2{{E^{\phi}}} })'\Bigg](x)\,.\nonumber
\end{align}
which reproduces exactly the Hamiltonian in \eqref{eq:defpolyhamiltonianconstraintwithoutKx}.

\section{Curvature invariants for the examples}\label{app:curvature}
In this appendix we list the Ricci scalar and Kreschmann scalar for all examples in Sec. \ref{sec:examples}.
\begin{enumerate}
    \item Symmetric LQC
    \begin{align}
    \mathcal{R}=&\frac{36 \left(48 \alpha^2_\Delta M(x) s'(x) z+M'(x) \left(48 \alpha^2_\Delta z^2+27 z^4+16 \alpha^4_\Delta\right)\right)}{\left(9 z^2+4 \alpha^2_\Delta\right)^2 \mathcal{S}}\\[1em] 
    \mathcal{K}=&\frac{432 }{\left(9 z^2+4 \alpha^2_\Delta\right)^4 \mathcal{S}^2}\Big(384 \alpha^2_\Delta M(x) M'(x) s'(x) z \big(6 a^2 z^2-27 z^4+8 \alpha^4_\Delta\big)+M'(x)^2 \left(9 z^2+4 \alpha^2_\Delta\right)^2 \nonumber\\
    & \big(-24 \alpha^2_\Delta z^2+45 z^4+16 \alpha^4_\Delta\big)+432 M(x)^2 s'(x)^2 z^2 \left(-96 \alpha^2_\Delta z^2+27 z^4+160 \alpha^4_\Delta\right)\Big),    
\end{align}
    \item Asymmetric LQC
    \begin{align}
    \mathcal{R} = \frac{\mathcal{A}}{\left(9 \eta ^2+4 \alpha^2_\Delta \gamma^2\right)^4 \mathcal{S}}, \qquad \mathcal{K} = \frac{\mathcal{B}}{\left(9 \eta ^2+4 \alpha^2_\Delta \gamma^2\right)^8\mathcal{S}^2}
\end{align}
with  
    \begin{align*}
    \mathcal{ S }=& M'(x) \left(9 \eta ^2+4 \alpha^2_\Delta \gamma^2\right)^2+18 M(x) s'(x) \eta  \left(4 \alpha^2_\Delta \gamma^2 \left(2 \gamma^2+1\right)-9 \eta ^2\right)\,, \\
        \mathcal{A}=&36 \Big(M'(x) \left(9 \eta^2+4 \alpha_\Delta^2 \gamma ^2\right)^2 \Big(-48 \alpha_\Delta^4 \left(2 \gamma ^4+8 \gamma ^2-1\right) \gamma ^4 \eta^2 \\
        &+540 \alpha_\Delta^2 \left(\gamma ^2+2\right) \gamma ^2 \eta^4+243 \eta^6+64 \alpha_\Delta^6 \left(2 \gamma ^2+1\right) \gamma ^8\Big)\\
        &-48 \alpha_\Delta^2 \gamma ^2 M(x) s '(x) \eta \Big(-144 \alpha_\Delta^4 \left(\gamma ^4+18 \gamma ^2+14\right) \gamma ^6 \eta^2+324 \alpha_\Delta^2 \left(2 \gamma ^4+9 \gamma ^2+4\right) \gamma ^2 \eta^4 \\
        &-729 \left(3 \gamma ^2+4\right) \eta^6+64 \alpha_\Delta^6 \left(7 \gamma ^4+4 \gamma ^2-2\right) \gamma ^8\Big)\Big) , \\
        \mathcal{B}=&432 \Big(M'(x)^2 \Big(-6144 \alpha_\Delta^{10} \left(20 \gamma^6+42 \gamma^4+18 \gamma^2+1\right) \gamma^{12} \eta^2\\
        &+2304 \alpha_\Delta^8 \left(164 \gamma^8+604 \gamma^6+618 \gamma^4+148 \gamma^2+5\right) \gamma^8 \eta^4 \\
        &+81 \eta^6 (432 \alpha_\Delta^4 \left(95 \gamma^4+164 \gamma^2+74\right) \gamma^4 \eta^2-648 \alpha_\Delta^2 \left(29 \gamma^2+22\right) \gamma^2 \eta^4 \\
        &+3645 \eta^6-256 \alpha_\Delta^6 \left(97 \gamma^6+225 \gamma^4+144 \gamma^2+11\right) \gamma^6)\\
        &+4096 \alpha_\Delta^{12} \left(2 \gamma^2+1\right)^2 \gamma^{16}\Big) \left(9 \eta^2+4 \alpha_\Delta^2 \gamma^2\right)^4\\
        &-384 M(x) M'(x) s '(x) \eta \left(9 \alpha_\Delta \gamma \eta^2+4 \alpha_\Delta^3 \gamma^3\right)^2 \Big(\\
        &-768 \alpha_\Delta^{10} \left(172 \gamma^8+515 \gamma^6+381 \gamma^4+65 \gamma^2+1\right) \gamma^{12} \eta^2\\
        &+1728 \alpha_\Delta^8 \left(82 \gamma^8+743 \gamma^6+1299 \gamma^4+641 \gamma^2+43\right) \gamma^{10} \eta^4\\
        &+243 \eta^6 \big(36 \alpha_\Delta^4 \left(291 \gamma^6+890 \gamma^4+733 \gamma^2+144\right) \gamma^4 \eta^2\\
        &-162 \alpha_\Delta^2 \left(64 \gamma^4+143 \gamma^2+72\right) \gamma^2 \eta^4+729 \left(3 \gamma^2+4\right) \eta^6\\
        &-16 \alpha_\Delta^6 \left(261 \gamma^8+1211 \gamma^6+1479 \gamma^4+525 \gamma^2+16\right) \gamma^6\big)\\
        &+1024 \alpha_\Delta^{12} \left(22 \gamma^6+27 \gamma^4+6 \gamma^2-1\right) \gamma^{16}\Big)\\
        &+432 M(x)^2 s '(x)^2 \eta^2 \Big(-98304 \alpha_\Delta^{14} \left(605 \gamma^{10}+2762 \gamma^8+3122 \gamma^6+1087 \gamma^4+37 \gamma^2+1\right) \gamma^{16} \eta^2\\
        &+27 \eta^4 \big(-18432 \alpha_\Delta^{10} \left(2 \left(521 \gamma^8+3066 \gamma^6+5275 \gamma^4+3202 \gamma^2+603\right) \gamma^2+19\right) \gamma^{10} \eta^2\\
        &+81 \eta^4 \big(9 \eta^2 (144 \alpha_\Delta^4 \left(526 \gamma^4+1152 \gamma^2+623\right) \gamma^4 \eta^2-648 \alpha_\Delta^2 \left(18 \gamma^2+19\right) \gamma^2 \eta^4+729 \eta^6\\
        &-128 \alpha_\Delta^6 \left(951 \gamma^6+2844 \gamma^4+2519 \gamma^2+586\right) \gamma^6\big)\\
        &+256 \alpha_\Delta^8 \left(2934 \gamma^8+11868 \gamma^6+14852 \gamma^4+6236 \gamma^2+623\right) \gamma^8)\\
        &+4096 \alpha_\Delta^{12} \left(563 \gamma^{12}+6192 \gamma^{10}+15494 \gamma^8+12852 \gamma^6+3489 \gamma^4+100 \gamma^2+1\right) \gamma^{12}\big)\\
        &+65536 \alpha_\Delta^{16} \left(241 \gamma^8+392 \gamma^6+168 \gamma^4+8 \gamma^2+1\right) \gamma^{20}\Big)\Big)
    \end{align*}
   \item Bardeen
    \begin{align}
    \mathcal{R}=&\frac{3 M'(x) \eta^{7/3} \left(\eta^{4/3}+3 \alpha^{4/3}\right)+9 \alpha^{4/3} M(x) s'(x) \left(\eta^{4/3}-5 \alpha^{4/3}\right)}{\eta^{14/3} \left(M'(x) \eta + 3 M(x) s'(x)\right)} ,\\[1em] 
    \mathcal{K}=&\frac{ 3}{\eta^{28/3} \left(M'(x) \eta-3 M(x) s'(x)\right)^2} \Big(-6 \alpha^{4/3} M(x) M'(x) s'(x) \eta^{7/3} \big(-14 \alpha^{4/3} \eta^{4/3}\nonumber\\
    &+7 \eta^{8/3}+15 \alpha^{8/3}\big) +M'(x)^2 \eta^{14/3} \left(-6 \alpha^{4/3} \eta^{4/3}+5 \eta^{8/3}+9 \alpha^{8/3}\right)\\
    &+9 M(x)^2 s'(x)^2 \left(-28 \alpha^{4/3} \eta^4+107 \alpha^{8/3} \eta^{8/3}-150 \alpha^4 \eta^{4/3}+4 \eta^{16/3}+75 \alpha^{16/3}\right)\Big) . \nonumber
\end{align}
    \item Hayward
    \begin{align}
    \mathcal{R}=&\frac{ 3 \tanh ^4(a \eta ) \text{sech}^2(a \eta ) }{a^2 \left(a M'(x) \coth (a \eta )-3 M(x) \beta '(x)\right)} \times\\
    &\quad \Big(a M'(x) (2 \cosh (2 a \eta )-1) \coth ^3(a \eta )-6 M(x) \beta '(x) (\cosh (2 a \eta )-2)\Big) \nonumber\\[1em] 
    \mathcal{K}=&\frac{3 \tanh ^4(a \eta ) \text{sech}^4(a \eta ) }{16 a^4 \left(a M'(x)-3 M(x) \beta '(x) \tanh (a \eta )\right)^2} \times \nonumber\\
    &\quad \Big(16 a^2 M'(x)^2 (-2 \cosh (2 a \eta )+\cosh (4 a \eta )+6) \nonumber\\
    &\quad -96 a M(x) M'(x) \beta '(x) (-5 \cosh (2 a \eta )+\cosh (4 a \eta )+12) \tanh ^3(a \eta )\nonumber\\
    &\quad +9 M(x)^2 \beta '(x)^2 \big(-820 \cosh (2 a \eta )+196 \cosh (4 a \eta ) \\
    &-12 \cosh (6 a \eta )+\cosh (8 a \eta )+699\big) \tanh ^2(a \eta ) \text{sech}^4(a \eta )\Big).\nonumber
\end{align}
\end{enumerate}
\end{appendix}

\bibliographystyle{jhep}
\bibliography{refs}

\newpage
\appendix

\end{document}